\documentclass[12pt,preprint]{aastex}
\usepackage{url}
\usepackage[usenames]{color}
\usepackage{amsmath}
\newcommand{\logg} {\log g}
\newcommand{\Te} {T_{\rm eff}}
\newcommand{\mv} {$M_V$}
\newcommand{\msun} {$M_\odot$}
\newcommand{\mbol} {M_{\rm bol}}
\newcommand\gta{\lower 0.5ex\hbox{$\buildrel > \over \sim\ $}} 
\newcommand\lta{\lower 0.5ex\hbox{$\buildrel < \over \sim\ $}} 

\newcommand{\ha} {$\rm{H}{\alpha}$}
\newcommand{\hb} {$\rm{H}{\beta}$}

\begin{document}

\title{PHYSICAL PROPERTIES OF THE CURRENT CENSUS OF NORTHERN WHITE DWARFS WITHIN 40 pc OF THE SUN}

\author{M.-M. Limoges,$^1$\altaffilmark{,3} P. Bergeron,$^1$ and S.
L\'epine$^2$\altaffilmark{,4}}
\affil{$^1$D\'epartement de Physique, Universit\'e de Montr\'eal,
  C.P.~6128, Succ.~Centre-Ville, Montr\'eal, Qu\'ebec H3C 3J7, Canada}
\affil{$^2$Department of Physics and Astronomy, Georgia State
  University, Atlanta, GA 30302-4106}
\email{limoges@astro.umontreal.ca, bergeron@astro.umontreal.ca,
  slepine@chara.gsu.edu}
\altaffiltext{3}{Visiting Astronomer, Kitt Peak
    National Observatory, National Optical Astronomy Observatory,
    which is operated by the Association of Universities for Research
    in Astronomy (AURA) under cooperative agreement with the National
    Science Foundation.}
\altaffiltext{4}{Department of Astrophysics, American Museum of Natural
    History, New York, NY 10024}

\begin{abstract}
We present a detailed description of the physical properties of our
current census of white dwarfs within 40 pc of the Sun, based on an
exhaustive spectroscopic survey of northern hemisphere candidates from
the SUPERBLINK proper motion database. Our method for selecting white
dwarf candidates is based on a combination of theoretical
color-magnitude relations and reduced proper motion diagrams. We
reported in an earlier publication the discovery of nearly 200 new
white dwarfs, and we present here the discovery of an additional 133
new white dwarfs, among which we identify 96 DA, 3 DB, 24 DC, 3 DQ,
and 7 DZ stars. We further identify 178 white dwarfs that lie within
40 pc of the Sun, representing a 40\% increase of the current census,
which now includes 492 objects. We estimate the completeness of our
survey at between 66 and 78\%, allowing for uncertainties in the
distance estimates. We also perform a homogeneous model atmosphere
analysis of this 40 pc sample and find a large fraction of massive
white dwarfs, indicating that we are successfully recovering the more
massive, and less luminous objects often missed in other surveys. We
also show that the 40 pc sample is dominated by cool and old white
dwarfs, which populate the faint end of the luminosity function,
although trigonometric parallaxes will be needed to shape this part of
the luminosity function more accurately. Finally, we identify 4
probable members of the 20 pc sample, 4 suspected double degenerate
binaries, and we also report the discovery of two new ZZ Ceti
pulsators.
\end{abstract}


\section{INTRODUCTION}

The determination of the atmospheric parameters of individual white
dwarf stars --- effective temperature ($\Te$), surface gravity
($\logg$), and atmospheric composition --- has now reached an
unprecedented level of accuracy, thanks to significant progress on
both the observational and theoretical fronts. On the observational
side, large samples of high signal-to-noise optical spectra can now be
routinely obtained and analyzed using the so-called {\em spectroscopic
  technique} where observed line profiles are compared to the
predictions of model atmospheres \citep[see, e.g.,][]{bergeron92},
reaching a precision as high as 1.2\% in $\Te$ measurements and 0.038
dex in $\logg$ for the DA stars \citep{liebert05}. This technique has
been applied successfully to large samples of various spectral types
including DA stars from the ESO SN Ia Progenitor Survey
\citep{koester09}, from the Villanova Catalogue of Spectroscopically
Identified White Dwarfs \citep{gianninas2011}, and in particular from
the Sloan Digital Sky Survey (SDSS; \citealt{kepler07},
\citealt{tremblay11a}). The same spectroscopic technique has also been
used for DB stars \citep{voss07,kepler07,bergeron2011}. Similarly,
optical and infrared photometry can be combined and compared with
synthetic photometry to measure effective temperatures, as well as
stellar radii when trigonometric parallaxes are available. This {\em
  photometric technique}, first applied to large photometric data sets
by \citet{BRL,bergeron01}, is particularly useful to study cool white
dwarfs that lack the presence of strong absorption lines required by
the spectroscopic method. Particularly interesting in this context is
the large set of optical $ugriz$ photometry available for white dwarfs
in the SDSS, combined with independent $JHK$ photometry (see, e.g.,
\citealt{kilic2010,gianninas2015}). The photometric approach, however, is more
sensitive to issues related to the calibration of the synthetic
photometry \citep{HB06}, unlike the spectroscopic approach.

Several independent model atmosphere grids have been widely used for
the analysis of white dwarf stars (e.g., \citealt{bergeron92},
\citealt{vennes92}, \citealt{koester2001}), and the results obtained
from these models are reassuringly comparable (see Figure 9 of
\citealt{liebert05}). Despite this agreement between models,
significant improvements on the theoretical front are still being
achieved.  For instance, new calculations for the Stark broadening of
hydrogen lines that include nonideal effects directly inside the line
profile calculations have recently become available
\citep{tb09}. These models yield systematically higher $\Te$ (up to
1000 K) and $\logg$ (up to 0.1 dex) values, and a mean mass for DA
stars shifted by $+0.034$ \msun.  Similarly, \citet{KS2006} have
successfully modeled the opacity from the red wing of Ly$\alpha$, an
important absorption process that affects the flux in the ultraviolet
region of the energy distribution of cool, hydrogen-atmosphere white
dwarfs. More importantly perhaps, \citet[][see also
  \citealt{tremblay11a,tremblay13a}]{tremblay13b} have produced
realistic 3D hydrodynamical model atmospheres of hydrogen-rich white
dwarfs and successfully showed that the so-called high-$\logg$ problem
--- the apparent increase of spectroscopic $\logg$ values below
$\Te\sim13,000$~K (see \citealt{tremblay2010} and references therein)
--- was related to the limitations of the mixing-length theory used to
describe the convective energy transport in previous model atmosphere
calculations.

With the availability of large data sets and improved model
atmospheres, the statistical properties of white dwarf stars can now
be studied in greater detail, including the luminosity function, space
density, mass distribution, age distribution, and space
kinematics. This can be achieved not only for the local population of
white dwarfs, but also for various components of the Galaxy, including
open and globular clusters (see, e.g., \citealt{tremblay12},
\citealt{woodley12}). Since white dwarfs represent the endpoint of
over 97\% of the stars in the Galaxy, they are a powerful tool to
study the overall evolutionary history of the Galaxy.  However, the
determinations of these global properties are always confronted with
the problem of defining statistically complete samples, minimally
affected by selection biases. For instance, ultraviolet color excess
surveys such as the Palomar-Green (PG) survey \citep{green86} or the
Kiso Schmidt (KUV) ultraviolet excess survey \citep{kondo84} are
restricted to the detection of blue and thus hot white
dwarfs. Consequently, the luminosity functions derived from these
surveys \citep{liebert05,lim10,bergeron2011} do not sample the faint
end of the distribution where the majority of white dwarf stars are
located. Even the luminosity function determined by \citet{harris06}
using the magnitude-limited SDSS sample, which covers the entire range
of bolometric magnitudes ($7\lesssim \mbol \lesssim16$), includes {\it
  several} corrections for completeness and contamination to
counterbalance important selection effects, and these corrections
critically determine the faint end of the luminosity function.  It is
however possible, as shown by \cite{kilic06}, to refine the selection
criteria by combining the SDSS photometry and astrometry with the
USNO-B plate astrometry to build a reduced proper motion diagram,
which helps to identify cool white dwarf candidates in the SDSS
imaging area and thus recover the faint end of the luminosity
function.  White dwarfs identified in proper motion surveys are indeed
much better suited to identify cool white dwarfs at the faint end of
the luminosity function.  For many years, one of the most commonly
used observational luminosity function had been that published by
Liebert, Dahn, \& Monet (1988), and revised by \citet{leggett98},
based on the Luyten Half-Second Catalog (Luyten 1979), but the major
drawback is that the sample contains only 43 spectroscopically
confirmed white dwarfs.

Other types of statistical biases occur in determining the
distribution of mass as a function of effective temperature ($M$
versus $\Te$), or the cumulative mass distribution ($N$ versus
$M$). For instance, color excess surveys (PG, KUV, and even SDSS) are
also magnitude-limited, and as such the samples drawn from these
surveys suffer from a bias in mass where low-mass white dwarfs with
their large radii and high luminosities are over represented, while
high-mass white dwarfs are undersampled (see, e.g., Section 3.2 of
\citealt{liebert05}). Another important issue is that, until recently,
spectroscopic masses at low effective temperatures could not be
trusted due to the high-$\logg$ problem discussed above. Thus, most
analyses restricted their determination of the mass distribution of DA
stars to temperatures higher than 13,000 K. Alternatively,
\citet{noemi2012} applied an empirical correction to the $\logg$
distribution, based on the DA white dwarfs from the Data Release 4 of
the SDSS analyzed by \citet{tremblay11a}.  More recently,
\citet{tremblay13b} produced a more accurate set of $\Te$ and $\logg$
corrections to be applied to spectroscopic determinations, based on a
comparison of detailed 3D hydrodynamical simulations with 1D model
atmospheres calculated within the mixing length theory. Another
problem arises at low effective temperatures when spectroscopic lines
can no longer be used efficiently. This occurs below $\Te\sim13,000$~K
and $\sim$6500~K for helium- and hydrogen-atmosphere white dwarfs,
respectively, i.e.~near the peak of the luminosity function, in which
cases one must rely on the photometric technique to measure the
atmospheric parameters. With the photometric technique, unfortunately,
stellar radii or masses can only be determined for white dwarfs with
trigonometric parallax measurements, which are only available for 200
stars or so. This situation will of course change dramatically when
the Gaia mission is completed.

Also of interest is the study of the spectral evolution, which
describes the various physical mechanisms (gravitational settling,
convective mixing, convective dredge-up from the core, accretion from
the interstellar medium or circumstellar material, radiative
acceleration, stellar winds, etc.) that affect the surface composition
of white dwarfs as they evolve along the cooling sequence. Of
particular interest is the spectral evolution of white dwarfs at low
$\Te$, where convective mixing of a thin superficial hydrogen
convective layer with the deeper helium convection zone is believed to
occur (see \citealt{tremblay08} and references
therein). \citet{BRL,bergeron01} also suggested the presence of a
non-DA gap (or deficiency) between $\Te\sim5000$~K and 6000~K where
most stars appear to have hydrogen-rich compositions, while
helium-atmosphere white dwarfs exist above and below this temperature
range. On the other hand, \citet[][see also
  \citealt{noemi2012}]{KS2006} suggested that most, if not all, cool
DC stars probably have hydrogen-rich atmospheres, based on a
reanalysis of the \citet{BRL,bergeron01} photometry with their
improved atmospheric models, which include the previously missing red
wing opacity from Ly$\alpha$. Unfortunately, the white dwarf samples
analyzed by \citet{BRL,bergeron01} are not complete in any statistical
sense. For instance, \citet{kilic2010} analyzed 126 cool white dwarfs
identified in the SDSS and uncovered several helium-atmosphere white
dwarfs in the $\Te$ range of the gap. To complicate matters,
\citet{chen11,chen12} showed that the evolution of cool white dwarfs
cannot be interpreted monotonically as a function of $\Te$, and that
upon mixing the $\Te$ of a white dwarf can actually {\it
  increase}. Hence our understanding of the spectral evolution of cool
white dwarfs is at best sketchy, a situation that can only be improved
by studying better-defined, large, statistically complete samples.

The best way around the completeness problems discussed above is the
use of a volume-limited sample. Efforts to identify white dwarfs in
the immediate solar neighborhood, within 20 or 25 pc of the Sun, have
been summarized in \citet[][hereafter Paper I]{limoges2013}.
\citet{noemi2012} performed a detailed photometric and/or
spectroscopic analysis of every white dwarf suspected to lie within 20
parsecs of the Sun. Although \citet{holberg08} and \citet{noemi2012}
have established the completeness of the 20 parsec sample at $80\%$
and $90\%$, respectively, one is confronted with small number
statistics since this sample contains only 130 objects or so. Hence
some results reported by Giammichele et al.~may not be statistically
significant. For example, while the luminosity function shown in their
Figure 22 agrees well with previous investigations at low temperatures
and luminosities, space densities in the brighter luminosity bins
(above $\Te\sim12,000$~K) are larger by a factor of $\sim$2. As
mentioned by the authors, one likely explanation for this apparent
overdensity is the small number of white dwarfs in the brightest
luminosity bins, which contain only a few objects ($\sim$2 to 8). The
only way out of this situation is to significantly increase the volume
sampled by these surveys. For instance, \citet{holberg2011} are
working on defining the sample of white dwarfs to 25 parsecs of the
Sun, nearly doubling the number of objects analyzed by
\citet{noemi2012}.

It is with this idea in mind that we embarked in a large effort (see
Paper I) to increase the census of white dwarfs to the larger distance
range of 40 pc from the Sun (corresponding to a volume {\em 8 times}
that of the 20 pc sample).  Given the space density of
$4.39\times10^{-3}$ pc$^{-3}$ derived by \citet{noemi2012}, the
expected number of white dwarfs within 40 pc is $\sim$1200, or
$\sim$600 if we restrict our search to the northern hemisphere, a
sample size that would markedly improve the statistical significance
of previous analyses. The SUPERBLINK survey is an all-sky search for
high proper motion stars ($\mu>40$ mas yr$^{-1}$) based on a
re-analysis of the Digitized Sky Surveys, with its 20-45 yr baseline
\citep{lspm05,LG2011}. The SUPERBLINK catalog is at least 95\%
complete for the entire northern sky down to $V=19.0$, with a very low
rate of spurious detection. As discussed in Paper I, the SUPERBLINK
catalog should contain all white dwarfs down to the luminosity
function turnover (which occurs at $L/L_{\odot}\simeq10^{-4}$) to a
distance of 56.7 pc from the Sun. Our method for selecting white dwarf
candidates, described in detail in Paper I, is based on reduced proper
motion diagrams of the 1.6 million stars in SUPERBLINK with
$\delta>0$, combined with distances estimated from theoretical
color-magnitude relations. A list of 1341 candidates with {\it
  photometric distance} estimates of $D<40$ pc, to within the
uncertainties, was established for follow-up spectroscopic
observations, excluding the objects with an already known spectral
type. We successfully confirmed 193 new white dwarfs, among which 93
had {\it spectroscopic distances} placing them within 40 pc. Only DA
stars with strong enough Balmer lines were analyzed in Paper I, using
the spectroscopic method described above.

The specific goal of this work is to obtain a complete sample of white
dwarfs within 40 pc of the Sun in the northern hemisphere. We report
in this paper the outcome of our survey, and present a detailed
photometric and spectroscopic model atmosphere analysis of all the new
white dwarfs that were identified. We also provide a comprehensive
analysis of the mass distribution and the chemical distribution of
white dwarf stars in this volume-limited sample.  In particular, since
only about a third of the white dwarfs in our sample have trigonometric parallax
measurements available, we develop a robust method to derive distances
from spectroscopic and photometric data alone.

We first present in Section \ref{sec:update} an update of our census
of white dwarfs within 40 pc of the Sun, which includes a summary of
our earlier work as well as a detailed description of the follow-up
spectroscopic observations of our list of white dwarf candidates. We
then perform in Section \ref{sec:atm_param} a detailed photometric and
spectroscopic analysis of all objects in our sample with
state-of-the-art model atmospheres, where we also include all known
white dwarfs in the SUPERBLINK catalog and from the literature,
suspected to belong to the 40 pc sample. The resulting distance
estimates are then used to build a complete sample of white dwarfs
within 40 pc of the Sun, which we analyze in further detail in Section
\ref{sec:phys}. In particular, we discuss several physical properties
of this sample, including its completeness, kinematics, mass
distribution, spectral evolution, and luminosity function.  We then
offer some concluding remarks in Section \ref{sec:concl}.
 
\section{UPDATE ON OUR CENSUS OF WHITE DWARFS WITHIN 40 PC OF THE SUN}\label{sec:update}

\subsection{Selection of the Candidates Based on Reduced Proper Motion Diagrams}

Our method for selecting white dwarf candidates from the SUPERBLINK
catalog using reduced proper motion diagrams is discussed at length in
Paper I. We briefly summarize here, for completeness, the various
steps involved.

The white dwarf candidates are identified from the SUPERBLINK catalog
of stars with proper motions $\mu> 40$ mas yr$^{-1}$
\citep{lepine02,lspm05,LG2011}. Our selection method takes advantage
of the coordinates and proper motions provided by SUPERBLINK for
1,567,461 stars in the northern hemisphere ($\delta>0$). White dwarf
candidates are selected on the basis of their particular location at
the bottom left of reduced proper motion diagrams. Since the
construction of such diagrams requires, in addition to proper motion
measurements, a set of photometric color indices for each star, we
cross-correlate SUPERBLINK with other catalogs to obtain photometric
data covering a large portion of the electromagnetic spectrum, and in
some cases, we also obtain improved coordinates and proper motion
measurements. Our version of SUPERBLINK used in Paper I includes ---
in the northern hemisphere only --- 1,472,666 counterparts in the Two
Micron All Sky Survey (2MASS) Point Source Catalog \citep{skrut06},
345,958 in the Sloan Digital Sky Survey \citep[SDSS,][]{SDSS_DR6},
118,475 in the \emph{Hipparcos} and \emph{Tycho-2} catalogs
\citep{hog}, 1,567,461 in the USNO-B1.0 database \citep{monet03}, and
143,096 in the sixth data release (GR6) of the {\it{GALEX}} database
\citep{gil09}. Each object in SUPERBLINK with $\delta>0$ is then
placed in all corresponding ($H_m$, color-index) diagrams, depending
on the available photometry, where $H_m$ represents the reduced proper
motion defined as $H_m=m+5\log\mu+5$, $m$ is the apparent magnitude in
some bandpass, and $\mu$ is the proper motion measured in arcseconds
per year. More specifically, we rely on ($H_g$, $g-z$) based on
$ugriz$ photometry, ($H_V$, $NUV-V$) based on UV {\it{GALEX}}
photometry, ($H_V$, $V-J$) based on 2MASS $JHK_S$ photometry, and
($H_V$, $V-I_N$) based on USNO-B1 photographic magnitudes. We also
restrict our search to stars with $V<19$, since SUPERBLINK has an
estimated false detection level of less than $1\%$ down to $V=19$, but
the false detection rate increases significantly for fainter sources.

The limit that defines the white dwarf region in each diagram is
determined from the location of SUPERBLINK objects with known white
dwarf counterparts in the 2008 May electronic version of the Catalogue
of Spectroscopically Identified White
Dwarfs\footnote{http://www.astronomy.villanova.edu/WDCatalog/index.html}
\citep[][hereafter WD Catalog]{mccook99}.  To be selected as a white
dwarf candidate, a SUPERBLINK object must be identified in the
expected white dwarf region of the diagram with the most accurate
photometry. The highest priority is thus given to the reduced proper
motion diagram based on SDSS magnitudes. If $ugriz$ photometry is not
available, the second priority is given to UV {\it{GALEX}} photometry,
the third priority to 2MASS $JHK_S$ photometry, and if no other
photometric system is available, we use USNO-B1 photographic
magnitudes. Finally, a criterion in $V-J$ is applied to the stars
identified in ($H_V$, $V-J$) diagrams to exclude bright, red, main
sequence contaminants.

Since we want to restrain our survey to a distance less than $D=40$ pc
from the Sun, and in the absence of trigonometric parallax
measurements for most white dwarf candidates in our sample, we must
rely on {\it photometric distances} estimated from the distance
modulus, $m-M=5\log D-5$, where the absolute magnitude $M$ of each
object is determined from theoretical color-magnitude relations
combined with a measured color index in some specified photometric
system.  These theoretical relations at constant mass values (see
Figures 5 to 8 of Paper I) are based on synthetic photometry from
white dwarf model atmospheres, following the procedure described in
\citet{HB06}. Hence, absolute magnitudes for each white dwarf
candidate are determined from color-magnitude relations in ($M_g$,
$g-z$), ($M_V$, $NUV-V$), ($M_V$, $V-J$) or ($M_V$, $V-I_N$), for a
0.6 $M_{\odot}$ hydrogen-atmosphere sequence, and for a
helium-atmosphere sequence at the same mass in order to evaluate the
distance uncertainty resulting from the unknown atmospheric
composition of our candidates.  A comparison of these photometric
distances with those obtained from parallax measurements show a
1$\sigma$ dispersion of 8.5 pc (see Figure 9 of Paper I), and we thus
adopt in our survey a conservative buffer of 15 pc to include all
white dwarfs that could potentially lie within 40 pc of the Sun, that
is with $D_{\rm phot}<55$ pc.

The method described above led to the identification of 193 new
spectroscopically identified white dwarfs in Paper I. However, even
though 14594+3618 (we omit here and below the letters ``PM I'' from
the designation) was identified as a white dwarf in Paper I, a new
spectrum at a significantly improved S/N shows that the star is a main
sequence F star, hence reducing the total of new white dwarfs
identified to 192. We also established that our survey is efficient at
identifying nearby white dwarfs distributed uniformly across the
northern sky, and estimated the ratio of new to known white dwarfs to
be around $77\%$ within 40 pc. More importantly, our survey has
identified a large amount of cool white dwarfs that could possibly
refine the determination of the faint end of the luminosity
function. We describe in the next section the update of our
spectroscopic survey.

\subsection[Spectroscopic Follow-up of White Dwarf Candidates]{Spectroscopic Follow-up of White Dwarf Candidates Within 40 pc of the Sun} \label{sec:follow-up}

In Paper I, we compiled a list of 1341 white dwarf candidates within
40 pc of the Sun --- but likely extending to 55 pc given the
uncertainties --- and reported spectroscopic observations for 422
objects from this target list, including 192 new white dwarf
identifications. We also reported that our selection criteria
recovered 499 known nearby white dwarfs from the literature (see
Figure 11 of Paper I). However, a thorough search indicated that while
several of these objects have a WD spectral classification in Simbad,
they have never been {\it confirmed spectroscopically} (and indeed,
some of these turned out not to be white dwarfs after all). We also
improved the cross-correlation of objects in our target list with
known white dwarfs from the literature, and discovered several
improper matches. Hence the number of previously known, actual white
dwarfs identified from our selection criteria is 416. Finally, we have
recently determined that our criterion based on $V-J$ color used to
exclude bright, red, main sequence contaminants (see above) had not in
fact been applied to our list of white dwarf candidates. Taking all
these changes into account, our list of white dwarf candidates now
numbers 1180 objects.

The spectroscopic observing log of our earlier survey is presented in
Table 2 of Paper I. Since then, additional optical spectra have been
obtained with the Steward Observatory 2.3-m telescope and the NOAO
Mayall 4-m and 2.1-m telescopes during 7 different observing campaigns
between 2011 January and 2013 October. A few of the brightest
candidates were also observed in spectroscopy with the 1.6-m
Mont-M\'egantic Observatory (OMM), while $\sim$60 hours with the
Gemini North and South 8-m telescopes were used to observe our
faintest candidates ($V\simeq17-18$). The adopted configurations allow
a spectral coverage of $\lambda\sim3200-5300$ \AA\ and $\sim3800-6700$
\AA, at an intermediate resolution of $\sim6$~\AA\ FWHM. Spectra were
first obtained at low signal-to-noise ratio (S/N $\sim25$), which is
sufficient to identify main sequence objects, but also represents the
lower limit required to obtain reliable model fits to the spectral
lines. Some stars were however reobserved at higher S/N and
resolution, whenever required. Table \ref{tab_obs} summarizes our
spectroscopic observation campaigns carried out since Paper I, along
with their instrumental setups.

We have now secured spectra for 588 objects from our list of 1180
candidates, thus adding 163 spectra to the list of 422 objects
reported in Paper I. The content of our complete spectroscopic data
set, described in the next subsection, includes 325 new white dwarf
identifications (192 reported in Paper I), and 263 spectra from
contaminants, mainly main sequence stars and quasars. We note that 3
of the newly identified white dwarfs are included in the SDSS Data
Release 7 but were not classified as such; we rely on the
corresponding SDSS spectra for these stars.

The current state of our survey is summarized in Figure \ref{Mv_d}
(upper panel), which plots the estimated absolute visual magnitudes
(from the calculated $V$ magnitudes and photometric distances --- see
Paper I) as a function of the photometric distance for the 325 new
white dwarfs (filled symbols) and the 416 previously known white
dwarfs (open symbols). We note that the subset of new white dwarfs is
dominated by objects fainter than $V=16$, and that most of them are
found at photometric distances larger than 20 pc. The 592 white dwarf
candidates on our list still without spectroscopic confirmation are
displayed separately in the lower panel of Figure \ref{Mv_d}; objects
selected on the basis of the less reliable USNO photographic
magnitudes (crosses) are considered second priority targets because of
their higher probability of being main-sequence contaminants. If we
exclude these second priority targets, we are left with $\sim$330
first-priority candidates for the spectroscopic follow-up
survey. However, only 4 candidates with $D<30$ pc and $V<18$ remain to
be observed, and every candidate with a high probability of being a
white dwarf with an estimated distance less than 20 pc has been
observed.

\subsection{Spectroscopic Content of our Updated Survey} \label{sec:SPEC}

Our spectroscopic follow-up observations from 2011 January to 2013
October (Table \ref{tab_obs}) have led to the identification of the
133 new white dwarfs listed in Tables \ref{tab_names} and
\ref{tab_phot} (including the 3 SDSS white dwarfs discussed above).
Table \ref{tab_names} provides astrometric data as well as NLTT and
SDSS designations, when available, while Table \ref{tab_phot} lists
the available photometry and adopted spectral types for the same
objects. Since Paper I, the SUPERBLINK catalog has been updated with
optical magnitudes from the 7th Data Release of the SDSS catalog, and
ultraviolet magnitudes from the DR4-5 data release of {\it{GALEX}}. In
particular, the number of stars with {\it{GALEX}} counterparts has
increased to 258,076 objects, while the number of stars with SDSS
counterparts now reaches 740,826. This improved photometry is
essential for the analysis of the energy distribution of the
SUPERBLINK white dwarfs, as well as for the estimation of their
distances. For this reason, we also include in Table \ref{tab_phot}
revised photometry for the 192 new white dwarfs identified in Paper
I. In summary, the new white dwarfs reported in this paper comprise 96
DA (including 1 DAZ, 3 DA+dM, and 50 magnetics), 3 DB, 24 DC, 3 DQ, and
7 DZ (including 1 DZA) stars.

Our new DA spectra are displayed in Figures \ref{DA_hot} and
\ref{DA_cool}; only \ha\ is displayed in
Figure \ref{DA_cool} since the bluer portion of the spectrum is either
featureless or too noisy to be of any use in the coolest DA stars. We
also show at the bottom of the last panel of Figure \ref{DA_hot}, 3
new spectra of stars already presented in Paper I (see Table 4), that
are double degenerate binary candidates and are further analyzed
below.  Note that 05431+3637 is actually a DAZ star. Two of the DA
stars discovered in our survey --- 06018+2751 (GD 258) and 18435+2740
(GD 381) --- were flagged as WD candidates in \citet{giclas80}, but to
our knowledge these had not been spectroscopically confirmed,
suggesting that many of the Giclas objects may still be white dwarfs
awaiting spectroscopic confirmation. The magnetic DA white dwarfs, or
magnetic candidates, in our updated sample are displayed in Figure
\ref{DAH}. The bottom five objects are new identifications, with the
Zeeman triplets clearly visible despite their low S/N. The top object,
04523+2519, was initially classified non-magnetic in Paper I, but the
flat bottom line cores observed in our spectroscopic fit (see Figure
17 of Paper I) led us to reobserve this star at \ha, where the weak
Zeeman splitting is now just barely detected, making us believe this
star is magnetic as well. Finally, our survey also led to the
discovery of 3 new DA + M dwarf binary systems, shown separately in
Figure \ref{DA+dM}.

Our subsample of new DZ (DZA), DB (DBZ), and DQ stars are displayed
together in Figure \ref{DZDQDB}. The first object in the figure,
00050+4003, is the star GD 1, another WD candidate listed in
\citet{giclas80}. Also shown are 5 DZ stars already identified in
Paper I but for which new optical spectra in the blue have been
secured: 01216+3440, 03196+3630, 16477+2636, 21420+2252, and
23003+2204. Indeed, the observational setup used with the 2.1-m and
4-m NOAO telescopes does not allow simultaneous coverage of
wavelengths shorter than $\sim$3900 \AA\ and of \ha\ in the red. In
order to better constrain the metal abundances in these objects,
additional spectra covering the blue portion of the spectral energy
distribution were thus obtained with the Steward Observatory 2.3-m Bok
telescope.  The 3 new DB white dwarfs displayed in Figure \ref{DZDQDB}
also include 02236+4816, also known as GD 27, another WD candidate
from \citet{giclas80} that was also lacking spectroscopic confirmation
to this date. Finally, our 3 DQ spectra, some of which are easily
recognizable from their strong C$_2$ Swan bands, are shown in the
bottom right section of Figure \ref{DZDQDB}; 16142+1729 is actually
one of those peculiar DQ stars with shifted C$_2$ Swan bands referred
to as DQpec white dwarfs, a phenomenon explained by
\citet{kowalski2010} as a result of pressure shifts of the carbon
bands that occur in very cool, helium-dominated
atmospheres. 05449+2602 has a DC spectral type in the WD Catalog (WD
0541+260), but this spectral classification was erroneously taken from
Table 2 of \citet{lim10}, where it was confused by McCook \& Sion with LSPM
J0021+2640. Our spectrum displayed here actually shows a weak
unidentified absorption feature near 5200 \AA. Since it does not show
any of the calcium features usually observed in DZ stars, it was classified as a
DQ? star in Table \ref{tab_phot}, although it could also be magnetic.

Finally, our featureless DC spectra are displayed in Figure \ref{DC}
in order of right ascension. All spectra cover the
$\lambda\sim3900-6700$ \AA\ range, except for 05462+1115 for which only
the blue part of the spectrum ($\lambda<5200$ \AA) is available, and
we notice a few cases where the blue portion of the spectrum is
particularly noisy, preventing us from detecting the possible presence
of calcium lines. These noisier spectra actually come from Gemini
North and South, where the integration times were calculated for the
central wavelengths near 5000 \AA, but the gratings used with the
GMOS-N and -S instruments (B600 G5307), chosen for their spectral
coverage, have a quantum efficiency that falls below 43\% blueward of
3937 \AA, compared with 83\% at 4983 \AA.

The progress of our spectroscopic survey can also be summarized in the
color-color diagram shown in Figure \ref{color_gmr}, where we display
the subset of 151 spectroscopically confirmed white dwarfs in our
sample that also have available $ugriz$ photometry from SDSS. White
dwarf candidates without spectroscopic confirmation are shown in red,
and most of these have estimated distances larger than 30 pc, as can
be seen from Figure \ref{Mv_d}. This figure reveals that our sample of
new white dwarfs is composed mainly of DA stars, but also contains a
significant number of cool ($\Te<5000$~K) white dwarfs, with objects
as cool as $\Te\sim4000$~K.

Finally, the entire white dwarf population detected in SUPERBLINK is
presented in Figure \ref{plot_equal}, where we display its
distribution on the sky. In the upper panel, the 325 new
identifications are shown with solid dots, while the white dwarfs from
the literature are represented with open circles. In the bottom panel,
we plot the sky density as a function of right ascension for the `old'
white dwarf population (dotted line), and compare it to that of the
325 new identifications (dashed line) and to the sum of the old and
new white dwarfs (solid line). In Paper I, Figure 11 showed that the
white dwarf candidates in our survey had the potential to fill the
void left in the galactic plane by earlier surveys. Here we notice
that the density of new identifications, especially near
$\rm{RA}=100$, suggests we are on our way to identify the missing
white dwarfs in this particular region of the sky.

\section{ATMOSPHERIC PARAMETER DETERMINATION}\label{sec:atm_param}

\subsection{Theoretical Framework}

Our model atmospheres and synthetic spectra for hydrogen-atmosphere
white dwarfs are built from the model atmosphere code originally
described in \citet{BSW95} and references therein, with recent
improvements discussed in \citet{tb09}. These are pure hydrogen,
plane-parallel model atmospheres, with non-local thermodynamic
equilibrium effects explicitly taken into account above
$\Te=30,000$~K, and energy transport by convection included in cooler
models following the ML2/$\alpha=0.7$ prescription of the
mixing-length theory. The theoretical spectra are calculated within
the occupation formalism of \citet{HM88}, which provides a detailed
treatment of the level populations as well as a consistent description
of bound-bound and bound-free opacities. We also include the improved
calculations for the Stark broadening of hydrogen lines from
\citet{tb09}, which take into account nonideal perturbations from
protons and electrons directly inside the line profile calculations,
as well as the opacity from the red wing of Ly$\alpha$ calculated by
\citet{KS2006}, which is known to affect the flux in the ultraviolet
region of the energy distribution, and in particular the FUV, NUV, and
$u$ magnitudes used in our analysis. Our model grid covers a range of
effective temperatures between $\Te = 1500$ K and 120,000 K in steps
of 250 K for $\Te < 5500$ K, 500 K up to $\Te= 17,000$ K, and 5000 K
above. The $\logg$ ranges from 6.0 to 9.5 by steps of 0.25 dex.
We also calculated cooler models with mixed hydrogen and helium
compositions \citep[see][]{gianninas2015} for the analysis of white dwarfs
in our sample that show evidence for collision-induced absorptions by
molecular hydrogen due to collisions with helium.

Our model atmospheres and synthetic spectra for helium-atmosphere
stars are described at length in \citet{bergeron2011}. These include
the Stark profiles of neutral helium from \citet{beauchamp97} as well
as van der Waals broadening. The synthetic spectra are calculated
using the occupation probability formalism of \citet{HM88} for helium
populations and corresponding bound-bound, bound-free, and
pseudo-continuum opacities. Our model grid covers a range of effective
temperatures between $\Te = 3000$ K and $40,000$ K in steps of $1000$
K, while the $\logg$ ranges from 7.0 to 9.0 in steps of 0.5 dex. In
addition to pure helium models, we also calculated models above
$11,000$ K with $\log\rm{H/He} = -6.5$ to $-2.0$ by steps of 0.5 dex.

The photometric analyses of DQ and DZ white dwarfs rely on the LTE
model atmosphere calculations developed by \citet{dufour05,dufour07}
for the study of DQ and DZ stars, respectively. Both are based on a
modified version of the code described in \citet{BSW95}. The main
addition to the models is the inclusion of metals and molecules in the
equation of state and opacity calculations. These heavier elements
provide enough free electrons to affect the atmospheric structures and
predicted energy distributions of cool, helium-rich white dwarfs.

\subsection{Photometric Analysis}

\subsubsection{General Procedure}\label{sec:GP}

The photometric technique developed by \citet{BRL} makes use of the
apparent magnitudes in any photometric system in order to measure the
atmospheric parameters ($\Te$ and $\logg$) and the chemical
composition. This method is particularly useful for the analysis of
cool white dwarfs when spectral features are either too weak or
completely absent.  The magnitudes in each bandpass are first
converted into a set of average fluxes $f^m_{\lambda}$ following the
procedure described in \citet{HB06}, which is mainly based on the Vega
fluxes from \citet{bohlin2004}, but also includes $ugriz$ photometry
in the AB magnitude system.  In particular here, we make use of the
transmission functions described in \citet{Morrissey2004} and
available from the {\it{GALEX}} Web
site\footnote{http://{\it{GALEX}}gi.gsfc.nasa.gov/docs/{\it{GALEX}}/Documents/PostLaunchResponseCurveData.html}
for the FUV and NUV filters, while the bandpasses for the $ugriz$
system \citep{Fukugita1996} are taken from the SDSS Web
site\footnote{http://www.sdss.org/dr6/instruments/imager/\#filters}. Similarly,
for the 2MASS filters described in \citet{Cohen2003}, we use the
transmission functions from the 2MASS survey Web
site\footnote{http://www.ipac.caltech.edu/2mass/releases/allsky/doc/sec6\_4a.html}.
Finally, the USNO-B1.0 $B_J$, $R_F$, and $I_N$ magnitudes are
described in \citet{monet03}, and the transmission functions are taken
from the Digitized Sky Survey
website\footnote{http://www3.cadc-ccda.hia-iha.nrc-cnrc.gc.ca/dss/}.

These observed average fluxes can be compared to the average
model fluxes $H^m_{\lambda}$ by the relation
\begin{equation}
f^m_{\lambda}=4\pi(R/D)^2H^m_{\lambda}
\end{equation}

\noindent
where $R/D$ defines the ratio of the radius of the star to its
distance from Earth.  The model fluxes $H^m_{\lambda}$ --- which
depend on $\Te$, $\logg$, and chemical composition --- are obtained
from averages over the transmission function of the corresponding
bandpass\footnote{This synthetic photometry is available at
  http://www.astro.umontreal.ca/\~{}bergeron/CoolingModels}.  We then
minimize the $\chi^2$ value defined in terms of the difference between
observed and model fluxes over all bandpasses, properly weighted by
the photometric uncertainties.  Our minimization procedure relies on
the nonlinear least-squares method of Levenberg-Marquardt
\citep{press86}, which is based on a steepest descent method. Only
$\Te$ and the solid angle $\pi(R/D)^2$ are considered free parameters
(for an assumed chemical composition), while the error of both
parameters are obtained directly from the covariance matrix of the
fit. For stars with known trigonometric parallax measurements, we
first assume a value of $\logg=8$ and determine $\Te$ and the solid
angle, which combined with the distance $D$ obtained from the
trigonometric parallax measurement, yields directly the radius of the
star $R$. The radius is then converted into mass using evolutionary
models similar to those described in \citet{fontaine01} but with C/O
cores, $q({\rm He})\equiv \log M_{\rm He}/M_{\star}=10^{-2}$ and
$q({\rm H})=10^{-4}$, which are representative of hydrogen-atmosphere
white dwarfs, and $q({\rm He})=10^{-2}$ and $q({\rm H})=10^{-10}$,
which are representative of helium-atmosphere white dwarfs. In
general, the $\logg$ value obtained from the inferred mass and radius
$(g=GM/R^2)$ will be different from our initial assumption of
$\logg=8$, and the fitting procedure is thus repeated until an
internal consistency in $\logg$ is reached. For white dwarfs with no
parallax measurement, we simply assume a value of $\logg=8$ and an
uncertainty of 0.25 dex, which corresponds approximately to a
$2\sigma$ dispersion of the surface gravity distribution of hot DA
stars \citep{gianninas2011}.

\subsubsection{Analysis with Hydrogen- and Helium-atmosphere Models}\label{sec:HHE}

We first perform a photometric analysis of all featureless DC stars in
our sample, and of all DA stars for which the Balmer lines are too
weak to be analyzed with the spectroscopic method. Sample fits are
shown in Figure \ref{ex_photo} for a subsample of 5 newly identified
white dwarfs.  Average observed fluxes are represented by error bars
in the left panels (with the photometric bandpasses used in the
fitting procedure indicated at the top of each panel), while our best
fits with pure hydrogen and pure helium models are shown as filled or
open circles, respectively. The corresponding atmospheric parameters
are given at the bottom of each panel. The USNO-B1.0 photographic
magnitudes have an error of 0.5 mag, which explains the large error
bars associated with their photometry. However, since the fit is
weighted by the photometric uncertainty, these less accurate
magnitudes will have little impact on the solution but they are still
useful when no other photometric information is available. Also, some
bandpasses had to be removed from the fitting procedure (shown in red
in the left panels) either because they are obviously incorrect, or
because they are contaminated by the presence of a red companion. In
the right panels we compare the spectroscopic observations near
\ha\ with the model predictions {\it assuming the pure hydrogen
  solution}; these only serve as an internal check of our photometric
solutions and are not used in the fitting procedure.  When \ha\ is
observed spectroscopically, we adopt the pure hydrogen solution, as is
the case for two objects in Figure \ref{ex_photo}, and even for
magnetic DA stars (one shown in Figure \ref{ex_photo}).  When \ha\ is
predicted by the pure hydrogen solution but is not observed
spectroscopically, we adopt the pure helium solution instead (see,
e.g., second object from the top in Figure \ref{ex_photo}).  In cases
where the star is too cool to show \ha\ ($\Te\leq 5000$ K), one has to
rely on the predicted energy distributions to decide which atmospheric
composition best fit the photometric data. However, according to
\citet{KS2006} based on their analysis of cool white dwarfs with
models including the Ly$\alpha$ opacity, almost all cool DC stars
appear to have hydrogen-rich atmospheres, a conclusion also reached by
\citet[][see their Figure 9 and their Section 4.2]{noemi2012}. We thus
assume here the pure hydrogen solution for all cool DC stars in our
sample (bottom object in Figure \ref{ex_photo} for instance).  Based
on a close inspection of these photometric fits and predicted
\ha\ features, we adopt the solution shown in red in the left panels.

The photometric fits for all 146 DC and cool DA stars in our sample
are displayed in Figure \ref{photoDADC}.  The particular case of the
DZ star 12145+7822 will be discussed in Section \ref{secDQDZ}.  Also
included here are the photometric fits to 3 DA + M dwarf systems
(03031+2317, 04032+2520E, and 08184+6606) hot enough to be analyzed
with the spectroscopic method, but for which the optical spectra are
so heavily contaminated by the presence of the companion that the
spectroscopic technique cannot be used reliably. In those cases, we
also had to omit from our photometric fits the $I_N$ and $JHK_s$
magnitudes for 03031+2317, the $R_F$, $I_N$, and $JHK_s$ for
04032+2520E, and we kept only the $B_J$, $R_F$, and $ugr$ magnitudes
for 08184+6606. Figure \ref{photoDADC} also includes the fits to 11
new magnetic white dwarfs identified in our survey. For these, the
photometric technique is adopted since the presence of such small
magnetic fields are not likely to affect the predicted energy
distributions.

A closer inspection of all the photometric fits shown in Figure
\ref{photoDADC} reveals that most solutions for the DA stars predict
an \ha\ feature that agrees remarkably well with observations (with
the glaring exception of 04263+4820, 11337+6243, 11598+0007, and
14278+0532, discussed further in Section \ref{sec:DD}), giving us
confidence in our photometric temperature scale, even for non-DA
stars.  Since we are often forced in our survey to rely on magnitudes
with large uncertainties, we need to worry about the overall accuracy
of the photometric method. But our results indicate that the lack of
accurate photometric measurements for some objects is compensated to
some extent by the large number of data points, and also by the fact
that the fit is weighted by the error on the photometry.

\subsubsection{Photometric Analysis of DQ and DZ Stars}{\label{secDQDZ}}

The photometric technique used to fit the energy distributions of DQ
and DZ stars is similar to that described above for the cool DA and DC
stars in our sample, with the exception that the abundance of heavy
elements (carbon or metals) is determined from fits to the optical
spectra \citep[see also][]{noemi2012}. Briefly, the energy
distribution is first fitted for an arbitrary abundance of heavy
elements to obtain an initial estimate of the effective temperature
($\logg$ is assumed or constrained from trigonometric parallax
measurements). The spectroscopic observations are then used to
determine the carbon abundance in the case of DQ stars --- fitting the
$\rm{C}_2$ Swan bands --- or the metal abundances in the case of DZ
stars --- fitting the Ca {\sc ii} H \& K doublet --- at these initial
values of $\Te$ and $\logg$.  This improved determination of heavy
element abundances is then used to obtain new estimates of the
atmospheric parameters, and this procedure is repeated until a
consistent photometric and spectroscopic solution is reached.

Results for the 5 new DQ stars in our sample (3 from Paper I and 2
from this work) are presented in Figure \ref{DQfits}, where we display
in the left panels the observed and model fluxes, as well as the
adopted $\Te$, $\logg$, and carbon abundance, and in the right panels,
the observed and model spectra (for the DQ? star 05449+2602, we
assumed a pure helium composition and our best fit is shown in Figure
\ref{photoDADC}). The predicted energy distributions and spectra agree
well for the two normal DQ stars shown at the top. The three other
objects are DQpec white dwarfs, with characteristic pressure-shifted
$\rm{C}_2$ Swan bands \citep{kowalski2010}. Since these pressure
shifts are not included in our models, the line strengths and shifts
are not properly reproduced. For these DQpec stars, we simply
fit/force the carbon abundance to reproduce the overall strength of
the $\rm{C}_2$ molecular bands, which is sufficient for our
purposes. Note also that the effective temperature for 12476+0646 has
been forced to $\Te=5000$~K, which corresponds to the coolest
temperature in our DQ model grid.

The results for the 13 new DZ stars in our sample (6 from Paper I and
7 from this work) are presented in Figure \ref{DZfits}, where the
spectroscopic observations used to determine the calcium abundances in
the fitting procedure are shown in the right panels. As mentioned
above, the calcium abundances are determined from the strength of the
Ca {\sc ii} H \& K doublet (see also \citealt{dufour07}), while the
abundance of other heavy elements, whether or not they are
spectroscopically detected, are assumed to have solar ratios. Because
hydrogen is often present in some of these DZ stars (DZA stars), we
rely on model grids calculated with hydrogen abundances of $\log
\rm{H/He} = -3$, $-4$, and $-5$. The insert in the right panels shows
the \ha\ absorption line used to measure or constrain the hydrogen
abundance in these stars.  The \ha\ line is particularly strong in
17574+1021, a new DZA star identified in our survey. For stars without
\ha\ or in the absence of spectroscopic data in this region sample,
the fits are performed at a fixed hydrogen abundance, determined from
the quality of the fit to the H \& K doublet, since the amount of
hydrogen present in the atmosphere influences the strength of these
lines. The predicted energy distributions and spectra agree well with
the observations for all 13 stars, except for 12145+7822, for which it
is impossible to reproduce the narrow calcium lines with
helium-atmosphere models at the low inferred temperature of
$\Te\sim4200$~K. This suggests that this star has a hydrogen-rich
atmosphere instead, which should produce much narrower absorption
lines due to lower atmospheric pressures. Because we do not have model
atmospheres that cover the appropriate range of parameters to fit this
star, we adopt a photometric solution based on pure hydrogen models
(see Figure \ref{photoDADC}), since the presence of metals are not
expected to affect the atmospheric structures of hydrogen-rich models.

\subsection{Spectroscopic Analysis}

\subsubsection{Spectroscopic Analysis of DA Stars}

The atmospheric parameters of DA stars with well-defined Balmer lines
($\Te$ $\gtrsim 6500$~K) can be accurately determined from the optical
spectra using the so-called spectroscopic technique pioneered by
\citet[][see also \citealt{liebert05}]{bergeron92}. The optical
spectrum of each star, as well as all model spectra (convolved with a
Gaussian instrumental profile), are first normalized to a continuum
set to unity. The calculation of $\chi ^2$ is then carried out in
terms of these normalized line profiles only. The atmospheric
parameters -- $\Te$ and $\logg$ -- are considered free parameters in
the fitting procedure. Since two solutions exist for a given star, one
on each side of the maximum strength of the Balmer lines, we take
advantage of our photometry to resolve the ambiguity. Also, since most
of our spectra cover \ha, we include this line in our fitting
procedure, allowing us to extend our spectroscopic fits down to
$\Te\sim6100$~K, when \ha\ is available. We however find that the
internal errors on $\logg$ increase significantly for stars cooler
than $\Te<6300$ K --- and in particular for spectra with low
signal-to-noise ratios (${\rm S/N}<40$) --- reaching a spread of
values as large as $\sigma_{\logg}\sim0.3$~dex, while for the best
spectra $\sigma_{\logg}$ can be as low as $0.04$. In such cases where
the internal errors become too large, $\logg$ is fixed at 8.0 and the
uncertainty is set at 0.25 dex, which corresponds to a $\sim2\sigma$
dispersion of the surface gravity distribution of hot DA stars
\citep{gianninas2011}.

Spectroscopic fits for 158 new DA stars identified in our survey,
which can be analyzed with the spectroscopic technique, are presented
in Figure \ref{DAspectro2}. It is worth mentioning that 05025+5401 and
07029+4406 have atmospheric parameters placing them within the ZZ Ceti
instability strip, and that periodic light variations are confirmed by
observations that are presented along with the dominant pulsation
periods in \citet{green2015}.  Special care needs to
be taken in the case of DA stars with unresolved M dwarf components,
in order to reduce the contamination from the companion.  When the
contamination affects only \hb\ and/or $\rm{H}{\gamma}$, we simply
exclude these lines from the fit, as indicated in Figure
\ref{DAspectro2} by the green lines.  At other times, emission lines
from the M dwarf are also observed in the center of some Balmer lines
(see, e.g., 13096+6933), in which case we simply exclude the line
centers from our fitting procedure.  A similar approach is adopted
when the contribution from the M dwarf is large enough to fill up the
Balmer line cores, resulting in predicted lines that are too deep
(see, e.g., 04586+6209). However, as discussed in Section
\ref{sec:HHE}, the white dwarf spectrum is sometimes too contaminated
by the M dwarf companion to be fitted with the spectroscopic technique
(03031+2317, 04032+2520E, and 08184+6606), and we must therefore rely
on the photometric technique alone for these stars.  Note that
04389+6351 was classified as a single DA star in Paper I, but we now
find from our fits that the predicted \hb\ is too deep, suggesting
that a red dwarf companion is filling the line (\hb\ is actually
excluded from our fit here). Our photometric fit for this star (not
shown here) shows a significant infrared excess at $I_N$ and $JHK_S$,
also suggesting the presence of a companion.  We thus reclassify this
star as a DA + M dwarf binary system.

There are also a few DAZ stars in our sample (including our new DAZ
identifications 05431+3637, 14106+0245, and 22276+1753), for which the
calcium H line (at 3968 \AA) is blended with $\rm{H}\epsilon$. Since
the upper Balmer lines are particularly sensitive to surface gravity,
it is important to model properly the calcium lines for these
stars. To do so, we relied on a small grid of synthetic spectra, based
on our grid of pure hydrogen models, where calcium lines have been
included in the calculation of the synthetic spectrum only (see
\citealt{gianninas2011} and references therein). This grid covers a
range in $\Te$ from 6000 to 9000 K in steps of 500 K, $\logg$ from 7.0
to 9.5 in steps of 0.5 dex, and $\log$ Ca/H from $-7.0$ to $-9.5$ in
steps of 0.5 dex.

\subsubsection{Spectroscopic Analysis of DB Stars}\label{sec:DB}

For the analysis of the DB and DBA white dwarfs in our sample, we rely
on the spectroscopic technique described at length in
\citet{bergeron2011}, which is similar to that used for DA stars but
modified to fit simultaneously $\Te$, $\logg$, and H/He. The first
step is to normalize the flux from individual predefined wavelength
segments, in both observed and model spectra, to a continuum set to
unity. The comparison with model spectra, which are convolved with the
appropriate Gaussian instrumental profile, is then carried out in
terms of this normalized spectrum only. However, as for DA stars, two
solutions exist for a given DB spectrum, one on each side of the
maximum strength of the neutral helium lines. Fortunately, all DB
stars in our sample are relatively cool and it is easy to distinguish
the cool and hot solutions from an examination of our best fits.
Furthermore, the hydrogen abundance in DBA stars is better constrained
if spectroscopic data near \ha\ are available, which is the case for
most of our DB stars.

The spectroscopic fits for the 4 DB white dwarfs identified in our
survey, of which 2 are DBA stars, are presented in Figure
\ref{DBfit}. Note that the hydrogen abundances in 12430+2057 represent
only upper limits based on the absence of H$\alpha$.

\subsection{Unresolved Double Degenerate Binaries}\label{sec:DD}

Four objects in Figure \ref{photoDADC} --- 04263+4820, 11337+6243,
11598+0007, and 14278+0532 --- show a predicted \ha\ absorption
feature significantly deeper than the observed profile; these are
plotted separately in Figure \ref{fitsDADC} for clarity.  Note that
14278+0532 (1425+057) was fitted as a helium-rich DC white dwarf in
\citet{sayres2012} based on a noisy SDSS spectrum, but we clearly
detect the \ha\ feature in our spectrum \citep[see also][]{lim10}. For
all stars, we achieve excellent fits of the model spectral energy
distributions to the photometry, but the fits to the observed spectra
are poor. We experimented with helium-rich models containing traces of
hydrogen instead of pure hydrogen models (see, e.g., L745-46A and Ross
640 shown in Figure 14 of \citealt{noemi2012}).  For 11337+6243 and
14278+0532, our best fits to these stars (not shown here) predict
\ha\ profiles that are much broader than the observed profiles due to
van der Waals broadening of hydrogen lines by neutral helium. The
sharp features observed here rather suggest that these correspond to
DA stars whose \ha\ absorption lines are diluted by the presence of an
unresolved DC white dwarf companion. While it was possible to
reproduce the \ha\ profile for 04263+4820 using helium-rich models,
our best solution predicts a steep Balmer decrement due to the
destruction of the high atomic levels of hydrogen, in sharp contrast
with the optical spectrum which shows the whole Balmer series all the
way to H$\epsilon$. Here again we suggest that we are rather dealing
with an unresolved DA+DC degenerate binary.

For 11598+0007, although the discrepancy observed in Figure
\ref{fitsDADC} is not as extreme as for the other three objects,
further evidence that we are also dealing with an unresolved binary is
provided by the spectroscopic fit, displayed in Figure
\ref{DA+DA}. Also shown for comparison is our best spectroscopic fit
to 04263+4820, discussed above, and SDSS 1257+5428, a double white
dwarf binary (DA + DA) discussed in \citet{badenes09},
\citet{kulkarni2010}, and \citet{marsh2011}, of which the optical
spectrum has been kindly provided to us by M.~H.~van Kerkwijk and
S.~R.~Kulkarni. Not only is the spectroscopic temperature for
11598+0007 ($\Te\sim7900$~K) significantly different from the
photometric temperature ($\sim9700$~K), but the quality of the fit is
poor, not unlike our best fit to SDSS 1257+5428 under the assumption
of a single DA star.  In addition, the hydrogen lines in 11598+0007
exhibit a strong asymmetry similar to that observed in Figure 1 of
\citet{kulkarni2010} for SDSS 1257+5428, attributed in this case to
orbital motion and differences in gravitational redshift from both
components of the binary system.  Kulkarni \& van Kerkwijk obtained
$\Te=6250$ K and $\logg=6.0$ for the primary, and $\Te=13,000$ K and
$\logg=8.5$ for the secondary; \citeauthor{marsh2011} obtained
slightly different parameters but the basic suggestion of a cool,
low-mass primary with a hotter, high-mass secondary remains the same.
We thus suggest that 11598+0007 also represents an unresolved DA + DA
double degenerate system.  As demonstrated by \citet{liebert1991}, it
is normally impossible to infer the presence of such DA + DA binary
systems using the spectroscopic technique alone since the coaddition
of synthetic spectra of two DA stars with different values of $\Te$
and $\logg$ can be reproduced almost perfectly by a single DA spectrum
{\it unless the surface gravities of both components differ
  significantly}, which is certainly the case for SDSS 1257+5428, and
it is thus most probably the case for 11598+0007 as well. Finally, the
similarity between our best spectroscopic fits to 11598+0007 and
04263+4820, both displayed in Figure \ref{DA+DA}, clearly suggests the
same interpretation, although in the latter case, there is no obvious
asymmetry in the line profiles, either because there is no velocity
difference between both components of the system, or perhaps
04263+4820 is composed of a DA + DC system.

We do not attempt here to deconvolve the individual components of
these double white dwarf binary candidates, and we simply adopt the
effective temperatures from the photometric fits, which are more
reliable than those derived from the line profiles. We further assume
that both components have identical effective temperatures and surface
gravities, and thus that they contribute equally to the total
luminosity of the system, resulting in distances that are a
factor of $\sqrt{2}$ larger than the values obtained under the
assumption of a single white dwarf. These binaries will also
contribute as two objects for the calculations of the luminosity
function and the space density.  A more detailed analysis of these
double degenerate binaries will be presented elsewhere.

\subsection{Known White Dwarfs within 40 pc of the Sun}\label{sec:ADD40}

In order to get a full picture of the physical properties of white
dwarfs within 40 pc of the Sun, we must also include, in addition to
the new white dwarfs identified in our survey, all previously known
white dwarfs suspected to lie within 40 pc of the Sun.  Our selection
criteria applied to the SUPERBLINK catalog recovered a total of 416
known white dwarfs with $D_{\rm phot}<55$ pc. We reexamine this
subset to exclude from the lot the known white dwarfs with distances
well beyond 40 pc. This we do based on (1) a more robust photometric
distance estimate described in Section \ref{sec:AtmP}, and (2)
reliable spectroscopic distances found in the literature. We find 116
relatively distant white dwarfs, which leaves 300 white dwarfs that
need to be included in our model atmosphere analysis, and for which
optical spectra are thus required. Fortunately we already had spectra
for 208 objects in this list, acquired over the years for various
independent projects. Spectra for another 12 stars were directly
available from the SDSS database. Also, optical spectra for 46
additional white dwarfs hot enough to be analyzed with the
spectroscopic technique were secured during the spectroscopic
observation campaigns listed in Table \ref{tab_obs}. Among these, GD
338 listed as a WD candidate by \citet{giclas80}, turned out to be a
main sequence star, thus reducing the number of known nearby white
dwarfs to 299. Finally, 34 white dwarfs on our list are classified in
the literature as DC or very cool DA stars, subtypes that can only be
analyzed using the photometric technique, and for which we did not
obtain new optical spectra because only a spectral type is sufficient
for our present purposes.

We also need to include all known white dwarfs suspected to lie within
$\sim$40 pc of the Sun but that were missed in our search of the
SUPERBLINK catalog, either because they failed our selection criteria
or are missing from the SUPERBLINK catalog itself for any reason. As
discussed in Paper I, about 20\% of the nearby white dwarfs are likely
to be missed in our search because of their unusual photometry, in
particular Sirius-like systems of white dwarfs companions to
main-sequence stars. With this limitation in mind, we searched the
following papers for objects with parallaxes larger than $0\farcs025$
yr$^{-1}$ or photometric/spectroscopic distances less than 40 pc of
the Sun: the WD Catalog, \citet{BRL}, \citet{bergeron01},
\citet{kawka2006}, \citet{kilic06}, \citet{gatewood09},
\citet{kilic2010}, \citet{kilic2012}, \citet{sayres2012},
\citet{noemi2012}, \citet{harris13}, \citet{gianninas2015}, as well as
the large spectroscopic samples of \citet{gianninas2011} and
\citet[][SDSS DR7]{genest2014} for DA stars, and \citet{bergeron2011}
for DB stars, and also \citet{holberg2013} for Sirius-like systems.
We also included individual objects such as the ultracool white dwarfs
LHS 3250 \citep{harris99,bergeron02}, WD 0343+247 \citep{hambly99},
and SDSS J1102+4113 \citep{hall08}.  Finally, we also searched nearby
DZ stars in the sample of \citet{dufour07} and nearby DQ stars in the
samples of \citet{dufour05} and \citet{kk2006}, although we did not do
a thorough search of the latest data releases of the SDSS.  As was the
case for the 20 pc sample, identifying all known white dwarfs with 40
pc of the Sun is a major endeavor, and we do not pretend that the
above list is complete, in particular given the large number of white
dwarfs continuously being identified in the SDSS.

All these {\it known} white dwarfs suspected to lie within 40 pc of
the Sun, within the uncertainties, have been combined with the 325
{\it new} white dwarfs identified in our survey of SUPERBLINK.  We
performed a spectroscopic or photometric analysis of each known white
dwarf following the same fitting procedures described above. Since
most of these stars have already been analyzed elsewhere in the
literature, we do not provide here the detailed fits for individual
stars, although our atmospheric parameters may reflect improved data
sets and/or model spectra. We discuss some specfic cases in turn.

As pointed out in \citet{bergeron2011}, several DB stars cooler than
$\Te\sim15,000$ K show masses in excess of 1 \msun, most likely
because these cool objects, with their weak and shallow line profiles,
lie at the limit of reliability for the spectroscopic technique. Three
of these DB stars (KUV 02499+3442, 21003+3426, and 21499+2816) are in
our sample of nearby candidates.  Since the spectroscopic $\logg$
values and inferred distances are uncertain, we adopt $\logg=8.0$ for
these 3 white dwarfs. Doing so, we obtain for 21499+2816 a distance of
$37.7\pm1.5$ pc, in better agreement with the value of $35.3\pm3.8$ pc
suggested by the trigonometric parallax. As mentioned in Bergeron et
al., a value of $\logg=8.2$ would actually reconcile both distance
estimates perfectly, which suggests that the spectroscopic $\logg$
values for these stars are likely overestimated.

An interesting object in our sample is 09487+2421 (0945+245,
LB 11146), an unresolved binary system composed of a DA star and a
magnetic white dwarf (DA + DAXP). We use here the values from
\citet{liebert1993} obtained by the deconvolution of both spectra:
$\Te=14,500\pm1000$ K, $\logg=8.4\pm0.1$ for the DA component, and
$\Te=16,000\pm2000$ K and $\logg=8.5\pm0.2$ for the magnetic
component.  Based on their distance estimate of $40\pm5$ pc, this
binary system is included in the 40 pc sample, and will count as 2
white dwarfs in the luminosity function.

Another interesting object in our sample is 01489+1902 (GD
16), a helium-rich DAZB star analyzed in detail by
\citet{koester2005}. They obtained $\Te=11,500\pm300$~K, $\log
\rm{H/He}=-2.89\pm0.3$, and $\log \rm{Ca/He}=-8.7\pm0.2$ under the
assumption of $\logg=8.0$. Since this last assumption will affect the
spectroscopic distance estimate, we analyzed this star ourselves using
the same spectroscopic technique for DB stars outlined in Section
\ref{sec:DB}, with mixed H/He model atmospheres that also include the
opacity from Ca {\sc ii} H \& K; the additional lines of Mg and Fe
visible in the spectrum are not included in our models. Our best fit
is displayed in Figure \ref{fit_01489} where we find $\Te=10,420$ K
and $\logg=7.71$; these are the only two free parameters in our
fit. The helium abundance is set according to the depth of the
He{\sc{I}} 5875 line reported by \citet{koester2005}, while the
calcium abundance is fixed at a value that reproduces the observed
strength of the calcium lines in our spectrum.  The values we obtain
for the helium abundance ($\log \rm{H/He}=-2.70$), and for the calcium
abundance ($\log \rm{Ca/He}=-8.5$), are in good agreement with the
values derived by \citet{koester2005}, while the effective temperature
is $\sim$1100 K lower, a difference that may be explained by the fact
that we do fit $\logg$ instead of simply assuming a value of 8.0. The
corresponding distance of $D=63.7\pm2.9$ pc is 13.7 pc further away
than the distance obtained by Koester et al., and significantly
outside our limit of 40 pc. Finally, we notice that our spectrum
displayed in Figure \ref{fit_01489} shows what appears to be a blue
component in the wings of \hb, $\rm{H}\gamma$, and $\rm{H}\delta$,
which could indicate that this object is in fact an unresolved double
degenerate binary.

\subsection{Adopted Atmospheric Parameters}\label{sec:AtmP}

As discussed in the Introduction, the spectroscopic $\logg$ values of
DA stars show a significant increase at low temperatures
($\Te\lesssim13,000$ K) with respect to the $\logg$ distribution of
hotter DA stars --- the so-called high-$\logg$ problem. Hydrodynamical
3D models \citep[][see also
  \citealt{tremblay11a,tremblay13a}]{tremblay13b} have now
successfully shown that this problem is related to the limitations of
the mixing-length theory used to describe the convective energy
transport in standard 1D model atmospheres calculations. These
spurious $\logg$ values prevent us from obtaining reliable mass and
distance estimates for DA stars in our sample in this particular range
of temperature.  Indeed, those higher $\logg$ values will yield
underestimated spectroscopic distances, biasing our census of white
dwarfs within 40 pc. To overcome this problem, \citet{noemi2012}
derived an empirical procedure (see their Section 5 and Figure 16) to
correct the $\logg$ values based on the DA stars in the SDSS (DR4),
analyzed by \citet{tremblay11a}; this was also our approach in Paper
I.  Here we make use of the latest results from \citet{tremblay13b}
who presented their first complete grid of 3D synthetic spectra for DA
white dwarfs based on 3D hydrodynamical model atmospheres. In
particular, they provided correction functions to be applied to both
$\Te$ and $\logg$ measurements determined using the spectroscopic
technique with model spectra calculated within the mixing-length
theory --- the ML2/$\alpha=0.7$ prescription in our case.  We thus
apply these corrections to our sample of DA stars whose atmospheric
parameters have been obtained using the spectroscopic method.

Because the distance to each white dwarf in our sample is a key issue
in our study, some care must be taken to reduce the uncertainty on the
distance estimates as much as possible. For stars with trigonometric
parallax measurements available (167 objects), we adopt the
corresponding distances directly. Out of these 167 parallax
measurements, 9 have uncertainties larger than 20\%, while Feige 4 and
HZ 9 have uncertainties of 31\% and 30\%, respectively. However, even
for these last two objects we prefer to use the parallax-based
distances because the spectroscopic distance to Feige 4 is largely
inconsistent with the parallax value \citep{bergeron2011}, and because
HZ 9 is a DA + M dwarf system whose spectral energy distribution is
highly contaminated by the M dwarf, which means we only have a single
data point to determine the distance from the photometric fit.  For
stars without parallaxes and fitted using the photometric method (DC,
DQ, DZ, and cool DA stars), the distance is determined from the
measured solid angle $\pi(R/D)^2$ combined with the stellar radius
corresponding to $\logg = 8.0 \pm 0.25$, as discussed in Section
\ref{sec:GP}. Finally, for white dwarfs in our sample only fitted with
the spectroscopic technique and without parallaxes, we fit all the
available photometric data but force the spectroscopic values of $\Te$
and $\logg$ (and the corresponding radius) and derive the distance
from the fitted solid angle \citep[see also][]{limoges2013}.  The
distance uncertainty in this case is obtained from the combination in
quadrature of the error on the spectroscopic $\logg$ and the solid
angle.  This approach for measuring spectroscopic distances has the
advantage of using all the photometric information rather than relying
on the distance modulus from a single bandpass.

Our final results for the 325 white dwarfs identified in our survey
are presented in Table \ref{tab_results}, while the results for the
492 white dwarfs, or white dwarf systems, identified within 40 pc of
the Sun with $\delta>0$ are presented in Table \ref{tab_40pcresults},
which includes 178 new objects from our survey. Note that Table
\ref{tab_40pcresults} includes several unresolved double degenerate
systems (noted in the table). We give for each object the PM I
designation if the white dwarf is a SUPERBLINK object, the spectral
type, effective temperature ($\Te$), surface gravity ($\logg$),
stellar mass ($M/M_\odot$), atmospheric composition (H- or
He-dominated), absolute visual magnitude (\mv), luminosity
($L/L_{\odot}$), distance ($D$), trigonometric parallax ($\pi$) if
available, white dwarf cooling time ($\log \tau$), and the method used
to determine the atmospheric parameters. As discussed above, the
spectroscopic solutions for both $\Te$ and $\logg$ for the DA stars
have been corrected for the high-$\logg$ problem and these differ from
the uncorrected values given in Figure \ref{DAspectro2}.
 
\section{PHYSICAL PROPERTIES OF THE 40 PC SAMPLE}\label{sec:phys}

The atmospheric parameters of white dwarfs within 40 pc of the Sun
given in Table \ref{tab_40pcresults} represent the end result of our
large spectroscopic survey undertaken in 2009, and the foundation of
the homogeneous study of the local white dwarf population presented
below. Although not yet complete, this sample of 492 white dwarfs ---
or white dwarf systems --- should be relatively free of kinematic
bias, allowing us to draw a picture of the local white dwarf
population (in the northern hemisphere) to a distance twice as large
as the 20 pc sample analyzed by \citet{holberg08}, \citet{noemi2012},
and their predecessors.  We provide in the following a detailed
analysis of the physical properties of this sample, but we first
attempt to evaluate the completeness of the sample in order to better
understand its limitations.

\subsection{Completeness of the 40 pc Sample}

We compile 492 white dwarfs (501 including the double degenerate
binaries) within 40 pc --- to within the distance uncertainties ---
among which 178 are new identifications, marked with a star symbol
in Table \ref{tab_40pcresults}. Only 167 of these stars have
trigonometric parallax measurements, however, and additional
measurements in the near future (e.g.~by the Gaia mission) will most
likely add or remove stars from this sample. Until then, we can still
evaluate the completeness of the 40 pc sample by calculating the
cumulative number of stars as a function of distance. The results are
displayed in Figure \ref{NvsD} where the cumulative number of stars in
our sample is compared to the expected number of white dwarfs assuming
a space density of $4.8\times10^{-3}$ pc$^{-3}$, which corresponds to
the value derived from the smaller 13 pc sample, which is however
believed to be complete \citep{holberg08}. The expected number of
stars is also divided by a factor of two since our survey is
restricted to the northern hemisphere. Also shown for comparison is
the expected number of stars for the whole celestial sphere.

We first notice that within 20 pc, the cumulative number of stars
reaches a value of 69, out of an expected number around 80 (or 85\%), but the
number of white dwarfs in our sample increases to 99 if we take into
account the formal uncertainties in our distance measurements,
stressing the importance of improving these distance estimates through precise
trigonometric parallax measurements.  Out of the 69 stars within 20 pc
in our sample, 4 are new identifications (13413+0500, 14456+2527,
16325+0851, 23098+5506E), among which only 2 are actually fitted with
the spectroscopic method allowing for more robust distance
measurements, while for the other 2 objects, the distances are only
estimated from the photometric method under the assumption of
$\logg=8$. We also find 4 additional known white dwarfs within 20 pc
with respect to the sample of \citet{noemi2012} --- 00222+4236,
11508+6831, 15350+1247, and 15425+2329.  Hence, these results suggest
that the 20 pc sample in the northern hemisphere is probably close to
completeness, with a density consistent with the 13 pc sample.

Similarly, we find 125 white dwarfs within 25 pc (possibly up to 172
if we consider the distance uncertainties), whereas \citet{sion2014}
report a total of 224 white dwarfs for the same volume but in both
hemispheres. Sion et al.~estimate the completeness of their
sample at 65\%, which implies an expected number of white dwarfs
around $\sim$172 in the northern hemisphere only, precisely the
number we are finding if we allow for distance
uncertainties. At 30 pc, the cumulative number of stars in Figure
\ref{NvsD} becomes significantly less than expected (assuming that the
space density of the 13 pc sample is valid everywhere), which suggests
that our sample is significantly incomplete beyond this range.

By taking only the subsample of 427 stars with $D<40$ pc from Table
\ref{tab_40pcresults}, we derive a space density of
$3.19\times10^{-3}$ pc$^{-3}$, with an upper estimate of
$3.74\times10^{-3}$ pc$^{-3}$ if we take into account the distance
uncertainties, which correspond to 66\% to 78\% of the density of the
13 pc sample ($4.8\times10^{-3}$ pc$^{-3}$). Our upper estimate is
thus consistent with the 77\% completeness value estimated in Paper I
for the present efficiency of our survey in detecting white dwarfs
using reduced proper motion diagrams. However, as mentioned in Section
\ref{sec:follow-up}, $\sim$330 high-priority objects still remain to
be observed out of our larger list of 1180 white dwarf
candidates. Results from Paper I suggested that our discovery rate
dropped to 54\% when stars that did not meet our best selection
criteria (but with distance estimates less than 30 pc) were observed
spectroscopically. Therefore, even if only 50\% of the remaining 330
high-priority candidates are confirmed as white dwarfs within 40 pc of
the Sun, the number of such stars could rise to more than 600, which
is the expected number of objects in Figure \ref{NvsD} based on the
space density at 13 pc.

We must also consider the various sources of incompleteness in our
survey that can prevent the identification of all white dwarfs within
40 pc of the Sun, especially those associated with our use of the
SUPERBLINK catalog as the primary source of candidates. First, our
selection excludes stars fainter than $V=19$ since the rate of
spurious detection in SUPERBLINK increases significantly at fainter
magnitudes. Stars are also excluded if they have non-standard
magnitudes for a white dwarf, a good example of which are Sirius-like
systems, which are completely overlooked in our selected sample. In
Sirius-like systems, as opposed to binary systems where the companion
is an M dwarf, the flux is completely dominated by the main sequence
star. In the 20 pc sample of \citet{noemi2012}, there are 7
Sirius-like systems, all of which are in the southern hemisphere, with
the exception of Procyon B (0736+053), while in \citet{holberg2013},
there are 6 Sirius-like systems with $D<40$ pc in the northern
hemisphere only. As expected, none of these were recovered by our
selection criteria. We note, however, that 5 of these 6 systems were
included in our analysis afterward; 0911+023 at 34.8 pc was let aside,
however, since it is unresolved with a B star. The Sirius-like systems
thus represent 3\% of the white dwarf population within 25 pc, and we
may expect that $\sim$13 of these systems are missed in the 40 pc
sample. Five other stars were missed because of
their inaccurate colors, or because of a mismatch with SDSS and/or
2MASS magnitudes.

Some white dwarfs were also missed simply because they are not in the
SUBERBLINK catalog to begin with. Stars with proper motions $<40$ mas
yr$^{-1}$ would of course not be part of our selection, but there are
other cases where stars might be missing from the catalog. To be
included in SUPERBLINK, a star must either be a TYCHO-2 or Hipparcos
catalog stars, or it must be detected as a high proper motion star on
the digitized POSS I and II plates. As discussed by \citet{lspm05},
SUPERBLINK is less efficient at recovering stars with bright saturated
cores on those plates, leaving some bright stars out. Thus we note
that some WD + M dwarf systems known to be within 40 pc were missed,
in particular HZ 9, which is brighter than $V=13$ at $I_N$. Sixteen
very bright white dwarfs could also not be found in the published
version of the SUPERBLINK catalog of \citet{lspm05}, and are thus also
missing from our candidate list.

Finally, some stars can also be overlooked because of missing or
inaccurate tabulated photometry in SUPERBLINK, most often due to
faulty optical magnitudes from the USNO-B1.0 catalog
counterpart. Because our selection criteria and distance estimates are
based on empirical $V$ magnitudes derived from USNO-B1.0 magnitudes,
the estimated distances can be severely affected if the $B_J$, $R_F$,
and $I_N$ are inaccurate by more than the adopted 0.5 mag value. We
actually identified 2 known white dwarfs that were not recovered in
the USNO-B1.0 catalog, and 11 white dwarfs were also excluded from our
candidate list because we estimated a distance larger than 55 pc. All
in all, about 3\% of the stars are missed in this manner.

Taking into account all the sources of incompleteness, we estimate
that, to the best of our knowledge, some 50 white dwarfs have probably
been overlooked by our survey, for a ratio of missed-to-found of 17\%,
a value consistent with the apparent 77\% efficiency of our survey at
detecting white dwarfs, as discussed above. With these limitations in
mind, we provide a detailed statistical analysis of our sample knowing
that $\sim$25\% of the white dwarfs might still be missing from our
sample. Nevertheless, even if our 40 pc sample cannot be analyzed as
thoroughly as the 20 pc sample, the small number statistics problem
inherent to the 20 pc sample discussed by \citet{noemi2012} has been
significantly improved.

\subsection{Kinematics of the 40 pc Sample}\label{sec:kin}

Before interpreting the global properties of the 40 pc sample, we must
first verify if the sample is relatively free of kinematic bias, which
we do by examining its velocity-space distribution. The velocity
components $(U,V,W)$ are determined from the photometric distances (or
trigonometric parallaxes when available) and proper motions for each
star. Since radial velocities are not available for most of the stars
in our sample, we assume $V_{\rm rad}=0$, but in order to obtain an
unbiased representation of the motion of the stars in our sample
despite this approximation, we use the method described in
\citet{lepine2013}. We first evaluate the $(X,Y,Z)$ positions of the
stars in the Galactic reference frame and then we use the fact that
$U\propto(X/D) V_{\rm rad}$, $V\propto(Y/D) V_{\rm rad}$ and
$W\propto(Z/D) V_{\rm rad}$ to obtain the $(U,V,W)$ velocity
components.  However, if $\left|X\right| > \left|Y\right|$ and
$\left|X\right|>\left|Z\right|$, then the position vector as well as
the radial velocity vector mainly points toward the $X$ direction, so
the radial velocity mainly contributes to the $U$ component of velocity,
but its contribution to the $V$ and $W$ components of velocity is small.
Then, the $V_{\rm rad}=0$ approximation is valid to obtain
\emph{estimates} for the velocity components $V$ and $W$, but not for
$U$. Likewise, stars with the largest $\pm Y$ components (or $\pm Z$
components) are good tracers of the velocity dispersion in $U$ and $W$
(or $U$ and $V$), even if their radial velocities are not known. We
can then estimate in this manner at least two velocity components for
each star, and the component that depends the most on the radial
velocity is excluded from any representation or statistical
calculation.

Our results are displayed in Figure \ref{uvw}.  For reasons outlined
in the previous paragraph, each star appears in a single panel
only. The mean values of the velocity components are
\begin{align*}
&\langle U \rangle=-9.82,~\sigma_U=41.00~\rm{km\ s}^{-1}  \\
&\langle V \rangle=-26.58,~\sigma_V=29.46~\rm{km\ s}^{-1}  \\
&\langle W \rangle=-8.26,~\sigma_V=17.37~\rm{km\ s}^{-1}  
\end{align*}
\noindent
which happen to be in relative agreement with those reported by
\citet{fuchs2009} for stars in the SDSS belonging to the thin disk:
$\langle U \rangle=-8.6$ and $\sigma_U=32.4$ km s$^{-1}$, $\langle V
\rangle=-20.0$ and $\sigma_V=23.0$ km s$^{-1}$, and $\langle W
\rangle=-7.1$ and $\sigma_V=18.1$ km s$^{-1}$. If we consider only the
new white dwarf identifications (shown as solid circles in Figure
\ref{uvw}), the space velocities we derive are of the same order,
$\langle U \rangle=-8.09$ and $\sigma_U=34.27$ km s$^{-1}$, $\langle V
\rangle=-23.98$ and $\sigma_V=27.87$ km s$^{-1}$, and $\langle W
\rangle=-6.85$ and $\sigma_V=19.31$ km s$^{-1}$, but we still notice
the presence of ``holes'' in the distribution of new white dwarfs near
$(U,W)=(0,0)$ and $(U,V)=(0,0)$ when compared to the velocity
distribution of known white dwarfs. The red circle in each panel of
Figure \ref{uvw} represents the presumed kinematic limit of our survey
due to the $\mu>0\farcs04$ yr$^{-1}$ proper motion limit and $D<40$ pc
distance range, which corresponds to a transverse motion $v_T=4.74\,
\mu D=7.6$ km s$^{-1}$.  In \emph{Hipparcos}, 2.3\% of the stars
fainter than $V\sim9$ with $\delta>0$ and $\pi>0\farcs25$ yr$^{-1}$
($D<40$ pc) have $\mu>0\farcs04$ yr$^{-1}$, implying that we are only
missing $\sim$11 white dwarfs within 40 pc because of the presumed
kinematic limit of our survey. We also notice that the holes are
larger than our presumed limit of detection, implying that some
low-velocity white dwarfs are probably still hiding in our list of
candidates, awaiting spectroscopic confirmation.

We find no white dwarfs in our 40 pc sample that appear reliably old
enough to belong to the halo population of old stars. This is
consistent with the \citet{sion2014} study of the 25 pc sample, where
no definite halo white dwarf was found either. Our only possible halo
candidates are the two DC stars cool enough to be very old, however we
cannot be certain of their age because we assumed $\logg=8.0$ by
default for these stars, and an accurate cooling age can only be
obtained when the stellar mass is known. Another possible candidate is
the star 19401+8348, seen in the upper right corner of the $(U,W)$
diagram in Figure \ref{uvw}, which has relatively high velocity
components compared to the other stars in the diagram. This object is
a DC star with $\Te\sim4800$~K and a distance of $D=38.8\pm12$ pc, for
which we also assumed $\logg=8.0$, which means its cooling age remains
uncertain. Even if we extend our search to stars with $D>40$ pc (not
shown in Figure \ref{uvw}), we do not find any white dwarf that would
reliably appear old enough to be a member of the halo population. This
is however not inconsistent with the single-point luminosity function
of \citet{fontaine01} based on the two halo white dwarfs 2316$-$064
and 1022+009, from which we estimate $n(L)=10^{-5.39}$
pc$^{-3}~M_{\rm{bol}}^{-1}$ at $\log L/L_{\odot}\sim-4.09$
($\Te\sim5000$~K), which predicts the existence of only a single halo
white dwarf within 40 pc of the Sun in this particular luminosity
range. Finally, we note a small asymmetry in the two bottoms panels of
Figure \ref{uvw} towards negative $V$ values, which might suggest a
small but non-negligible component of the thick disk population in our
sample.

\subsection{Mass Distribution}

The mass distribution for the white dwarfs in Table
\ref{tab_40pcresults} with $D\lesssim40$ pc is displayed in Figure
\ref{correltm} as a function of effective temperature. Only those with
a measured mass are shown here (288 out of 492 objects from Table
\ref{tab_40pcresults}); the objects with an assumed value of $\logg=8$
are shown at the bottom of the figure.  These include the coolest DA
stars with weak hydrogen lines analyzed spectroscopically, most of the
magnetic white dwarfs, and all non-DA stars analyzed photometrically
but without trigonometric parallax measurements.  Since most of our
new identifications do not have trigonometric parallaxes available
yet, the number of cool helium-rich white dwarfs with mass
measurements is rather small in this figure, in comparison for
instance with the mass distribution displayed in Figure 21 of
\citet{bergeron01}, which includes all cool white dwarfs with
trigonometric parallax measurements available at that time. There, 54
out of 150 white dwarfs (or 36\%) had helium-rich atmospheres,
compared to 39 out 288 (or 13\%) in Figure \ref{correltm}. In fact,
most cool helium-atmosphere white dwarfs with mass determinations come
from the 20 pc sample. The situation will of course change
significantly when the Gaia mission is completed.

The cumulative mass distribution for the same subset of white dwarfs,
regardless of their effective temperature, is displayed in Figure
\ref{histo_mass_spectro}, where the separate contributions of
hydrogen- and helium-rich stars are also shown. This distribution can
be contrasted with that shown in Figure 21 of \citet{noemi2012} for
the 20 pc sample. Here again we see that the helium-atmosphere white
dwarfs in the 40 pc sample are significantly undersampled in the
cumulative mass distribution. The relative number of helium- to
hydrogen-rich objects in our sample is small, but we observe that the
median masses and mass dispersions of both subsamples are generally
comparable. The {\it mean mass} of the hydrogen-rich sample, however,
is about 0.04 \msun\ larger than the helium-rich counterpart. This is
mostly due to the prominent high-mass tail observed in the
distribution of hydrogen-atmosphere white dwarfs. To better illustrate
this feature, we contrast in Figure \ref{histo_compare} the mass
distribution of the 40 pc sample with that of the 20 pc sample from
\citet{noemi2012}, which is based on both spectroscopic and
photometric mass measurements, and with the spectroscopic mass
distributions of DA stars from the SDSS \citep{tremblay11a} and from
the WD Catalog \citep{gianninas2011}. The excess of high mass white
dwarfs in the 40 pc sample is quite obvious, a result that strongly
suggests we are successfully recovering the high-mass, less luminous
white dwarfs, which are often missed in magnitude-limited surveys
(see, e.g., \citealt{liebert05} in the case of the PG survey, and in
particular their discussion of Figure 13). A higher fraction of
massive white dwarfs has also been identified in the analyses of hot
DA stars from ROSAT and EUVE \citep{vennes1996,vennes1997,vennes1998},
since such surveys catalog all sources, regardless of the brightness
of the object.

This excess of massive white dwarfs seems to be related to the
population of low temperature white dwarfs apparent in Figure
\ref{correltm}, between roughly $\Te=6000$~K and 7000 K. These are all
DA stars for which the atmospheric parameters --- and in particular
$\logg$ and thus $M$ --- have been determined spectroscopically. Note
that even though 3D hydrodynamical corrections, which are negligible
in this temperature range, have been properly applied to our
spectroscopic results, it is still legitimate to question the validity
of the spectroscopic masses in this particular range of
temperature. As discussed in \citet{tremblay2010} --- see in
particular their Figure 14 (panel b) and Section 5.1 --- the strength
of the hydrogen lines in this temperature regime is particularly
sensitive to the {\it neutral particle interactions} in the
description of the occupation probability formalism, and a slight
change in the hard sphere radius in this case may result in
significantly lower $\logg$ values. Indeed, \citet{bergeron91} found
that a direct implementation of the Hummer-Mihalas occupation
probability formalism yields $\logg$ values that are too low in the
regime where non-ideal effects become dominated by neutral
interactions ($\Te\lesssim 8000$~K), a problem that could be overcome
by simply dividing the hydrogen radius by a factor of two to reduce
the non-ideal effects for the higher lines of the Balmer series (this
is the factor actually used in our models). At the same time, we also
notice in Figure \ref{correltm} that the mass distribution near
$\sim$0.6 \msun\ in the same temperature range, and below, appears
perfectly normal, suggesting that the model spectra are properly
calibrated. Hence the high mass tail observed in Figure
\ref{histo_mass_spectro} might be real after all. Clearly, a detailed
comparison of mass measurements derived from spectroscopy and {\it
  precise} trigonometric parallaxes in this range of temperature
should shed some light on this result.  Finally, we also note in
Figure \ref{correltm} an abrupt cutoff of massive white dwarfs below
$\sim$6000~K, but this is certainly due to the fact that cooler
objects would appear as DC stars (or weak DA stars) whose masses can
only be determined from the photometric method using trigonometric
parallax measurements, which are currently unavailable. Hence it is
possible that this high mass tail extends to even lower temperatures.

Also superposed on the mass distribution shown in Figure
\ref{correltm} are the theoretical isochrones for our C/O core
evolutionary models with thick hydrogen layers, as well as the
corresponding isochrones with the main sequence (MS) lifetime taken
into account; here we assume for simplicity \citep[see][]{wood92}
$t_{\rm MS}=10(M_{\rm MS}/M_\odot)^{-2.5}$ Gyr and $M_{\rm
  MS}/M_\odot=8\ln[(M_{\rm WD}/M_\odot)/0.4]$.  In the 20 pc sample of
\citet{noemi2012}, the 5 stars older than 8 Gyr (among which the
oldest is 9.5 Gyr old) are all located in the southern hemisphere, and
they are thus not included in our sample.  The oldest white dwarf in
Figure \ref{correltm} is only $\sim$8 Gyr old, although we see that
there are plenty of objects without mass determinations that may be
significantly older.  The isochrones that include the MS lifetime
reveal that white dwarfs with $M\lesssim0.48$ \msun\ cannot have C/O
cores, and yet have been formed from single star evolution within the
lifetime of the Galaxy. Some, and probably all, of the low-mass white
dwarfs in Figure \ref{correltm} must be unresolved double degenerate
binaries with helium cores, i.e.~the result of common envelope
evolution. The known double degenerate systems are identified in
Figure \ref{correltm}. Three of the low-mass ($M<0.45$ \msun) white
dwarfs with $\delta>0$, 1345+238, 2048+263, and 2248+293, have
already been discussed in \citet{noemi2012}, while we identify here
01294+1023 (0126+101, DD, \citealt{bergeron01}), 03467+2456
(0343+247), 06026+0904, 09466+4354 (0943+441), 13309+3029 (1328+307,
DZ), 13455+4200, 15555+5025 (1554+505), 16540+6253 (1653+630; LHS
3250), 17055+4803W (1704+481.2; Sanduleak B), 18205+1239 (1818+126,
DD, \citealt{bergeron01}), 23253+1403 (2322+137), 22225+1221
(2220+121), and 23549+4027 (2352+401, DQ). Not surprisingly, most of
these were already known in the literature since these objects, with
their large radii and high luminosities, can easily be detected in
most surveys.

The mean mass of the 40 pc sample (with mass determinations) is
$\langle M\rangle=0.699$ \msun\ with a standard deviation of
$\sigma_M=0.185$ \msun\ (Figure \ref{histo_mass_spectro}). These
values are significantly larger than those reported by
\citet{noemi2012} for the 20 pc sample, $\langle M\rangle=0.650$
\msun\ and $\sigma_M=0.161$ \msun. Our higher mean mass is actually
due to the hydrogen-rich white dwarfs in the 40 pc sample, with
$\langle M_{\rm H}\rangle=0.705$ \msun, while Giammichele et
al.~obtained $\langle M_{\rm H}\rangle=0.650$ \msun. However, the mean
masses for the helium-atmosphere white dwarfs are identical, $\langle
M_{\rm He}\rangle=0.660$ \msun, which also compare really well with
the mean mass for DB stars determined spectroscopically by
\citet{bergeron2011}, $\langle M_{\rm H}\rangle=0.671$ \msun. As
mentioned above, the mass distribution of the 40 pc sample also peaks
at larger mass values than the 20 pc sample, although it is comparable
to the peak value obtained by \citet{gianninas2011} for the DA stars
in the WD catalog of McCook \& Sion, also reproduced in
Figure~\ref{histo_mass_spectro}.

\subsection{Evolution of Surface Composition}

Since the photometric data set for the 40 pc sample is not as accurate
as that of the 20 pc sample, it is not yet possible to study the
evolution of the surface composition as a function of temperature in
as much detail as in \citet{BRL,bergeron01}, for instance (see also
\citealt{BDG2013}). Also, since trigonometric parallax measurements
are not available for most of the new white dwarfs identified in the
40 pc sample, stellar masses cannot be determined for these objects
either. It is still possible, however, to examine the spectral
evolution of the white dwarfs in our sample by ignoring this second
parameter, and by remembering that our temperature scale remains
somewhat uncertain for some stars, in particular those that have only
USNO-B1.0 photographic magnitudes available.

The distribution of the main spectral types (DA, DC, DQ, DB, and DZ)
as a function of effective temperature is displayed in Figure
\ref{spec_evol}; the only 2 white dwarfs missing from this figure are
the DXP stars 17481+7052 (1748+708, G240-72) and 18303+5447 (1829+547,
G227-35). We can see that, as expected, the local sample is dominated
by cool white dwarfs (see also Figure \ref{correltm}), with the
typical rise in the number of cool DA and DC stars --- the dominant
spectral types here --- expected at the faint end of the luminosity
function. We notice that the sudden drop in the number of DA stars
below $\Te\sim5000$~K is largely compensated by the significant
increase in the number of DC stars in the same temperature range,
which suggests that several of these DC stars probably have
hydrogen-rich atmospheres. Many of them might even reveal the presence
of H$\alpha$ when observed at even larger signal-to-noise ratios using
4 to 10-m class telescopes, as demonstrated for instance by
\citet{greenstein86} or \citet{BRL}.

While no DB stars were reported in the 20 pc sample studied by
\citet{noemi2012}, the 40 pc sample now includes 7 DB stars ---
00051+7313 (0002+729), Feige 4 (0017+136), KUV 05034+1445 (0503+147),
GD 325 (1333+487), 16473+3228 (1645+325), 20123+2338 (2010+310) and
21499+2816 (2147+280) --- and none of these correspond to new
detections. DQ and DZ stars are rather rare in our sample,
representing only 5\% of all white dwarfs below $\Te\sim12,000$~K, but
still 40\% of all helium-atmosphere white dwarfs in the
$6000-12,000$~K temperature range (the spectral type alone below
6000~K is not sufficient to infer the chemical composition). While it
is difficult to misclassify a DQ star in our sample, in particular at
low temperatures where the C$_2$ Swan bands are usually the strongest,
a significant number of DZ stars might still be present among the 118
cool DC stars in our sample due to the lack of appropriate spectral
coverage, resolution, or signal-to-noise ratio of our spectroscopic
observations.

As discussed in the Introduction, it is now believed that most, if not
all, cool DC stars with $\Te\lesssim5000$~K probably have
hydrogen-rich atmospheres, a result based on a reanalysis of the
existing photometry with model atmospheres that include the red wing
opacity from Ly$\alpha$ (this excludes DQ and strong DZ stars in the
same temperature range, which obviously have helium-dominated
atmospheres). We have 4 DC white dwarfs below 5000~K in Figure
\ref{correltm} with helium atmospheres, 2 of which are known in the
literature and only one with trigonometric parallax, and thus mass
measurements.  For some of these objects, we do not even have a
spectrum near H$\alpha$, and in some other cases the photometry is
clearly suspicious, in particular at $JHK$. Hence for all 4 stars, the
pure hydrogen solution would be equally acceptable. We thus reaffirm
the conclusion first made by \citet{KS2006}, and supported by
\citet{noemi2012}, that most cool DC stars probably have hydrogen
atmospheres.

One of the most puzzling results of our analysis is displayed in
Figure \ref{ratio_H} where we show in the left panel the total number
of stars as a function of effective temperature per bin size of 2000
K, as well as the individual contribution of the hydrogen-atmosphere
white dwarfs, while we show in the right panel the ratio of
helium-atmosphere white dwarfs to the total number of stars. As
mentioned in the previous paragraph, the results below $\Te=5000$~K
should be considered with caution. The results above this temperature,
however, are fairly robust since most solutions are constrained by the
presence or absence of H$\alpha$ when the photometric method is used,
or in the case of hotter DA stars, the atmospheric parameters have
been obtained from the spectroscopic method. Above 15,000~K, the
fraction of helium-atmosphere white dwarfs is around 25\%, in
good agreement with the fraction of DB stars found in the PG
survey, as determined by \citet{bergeron2011}. In the
$13,000-15,000$~K temperature range, this fraction drops to only 5\%,
although the total number of helium-rich stars in this temperature bin
is admittedly small (see left panel). Below 13,000~K, the fraction of
helium-atmosphere white dwarfs gradually increases towards lower
temperatures, and keeps increasing to a ratio around $\sim$25\% in the
$7000-9000$~K temperature range. Even though this trend is entirely
consistent with the results reported by \citet[][see their Figure
  20]{noemi2012} for the 20 pc sample, the fraction of
helium-atmosphere white dwarfs reaches a much higher value around 40\%
in the 20 pc sample. To better understand this discrepancy, we show in
Figure \ref{ratio_D} similar results as those displayed in Figure
\ref{ratio_H} but only for temperatures in the range $7000<\Te<9000$~K
and this time as {\it a function of distance}. Below 20 pc, we recover
the results of \citet{noemi2012} almost perfectly, but beyond this
distance, the fraction of helium-atmosphere white dwarfs drops
abruptly. By looking at the results shown in the left panel of Figure
\ref{ratio_D}, we can see that the 20 pc sample probably suffers from
small number statistics, and that these statistics become more
significant at larger distances.  Otherwise, we cannot think of a
single bias in our survey that would produce such a trend, either in
favor of DA stars beyond 20 pc, or against non-DA stars. We thus
believe that the peak value near 40\% reported by Giammichele et
al.~was overestimated.

In \citet[][see also \citealt{tremblay08}]{noemi2012}, the increase in
the fraction of helium-atmosphere white dwarfs at lower temperatures
was interpreted as the result of convective mixing, where the thin
convective hydrogen atmosphere is mixed with the deeper and more
massive helium convection zone. Since the fraction of
helium-atmosphere white dwarfs at low temperatures in our sample is
now consistent with that of DB stars at higher temperatures (here and
in the PG sample), there appears to be little evidence that convective
mixing ever occurs in cool DA stars, at least in the temperature range
considered here. This in turn implies that DA stars have fairly thick
hydrogen layers of the order of $M_{\rm H}/M_{\rm tot}\sim 10^{-6}$
\citep[see Figure 1 of][]{tremblay08}. This revised conclusion is a
direct consequence of our analysis of a more statistically
significant, volume-limited sample.

Finally, as discussed in the Introduction, \citet{BRL,bergeron01}
suggested the presence of a non-DA gap (or deficiency) between
$\Te\sim5000$~K and 6000~K where most stars appear to have
hydrogen-rich compositions, while helium-atmosphere white dwarfs exist
above and below this temperature range (see also \citealt{BDG2013}).
While we have indeed very few ($\sim$3) helium-atmosphere white dwarfs
in Figure \ref{correltm} with a mass determination in this particular
range of temperature, our sample still contains a significant number
of such helium-rich stars without mass determinations. However, as
mentioned before, the photometric data for most of these objects are
not accurate enough to pinpoint their temperature with sufficient
precision, and we will thus refrain from further interpreting the
presence or absence of this non-DA gap in our sample at this stage.

\subsection{Luminosity Function}

As discussed in the Introduction, one of the goals of our
spectroscopic survey is to provide an improved determination of the
cool end of the white dwarf luminosity function (WDLF), as
statistically complete as possible, which can then be compared to
those obtained from magnitude-limited surveys, from large photometric
and spectroscopic surveys like the SDSS, or from the volume-limited
sample of white dwarfs within 20 pc of the Sun.  The WDLF is a measure
of the number of stars per pc$^3$ per unit of bolometric magnitude,
which can be obtained in our case using the bolometric magnitude of
each white dwarf within 40 pc of the Sun and with $\delta>0$ derived
from the spectroscopic or photometric results provided in Table
\ref{tab_40pcresults}. The bolometric magnitudes can be obtained from
the luminosity of each star given in the table ($\log L/L_{\odot}$)
and the simple relation $\mbol=-2.5\log L/L_{\odot}+M^{\odot}_{\rm
  bol}$, where $M^{\odot}_{\rm bol}=4.75$ is the bolometric magnitude
of the Sun. We present here an {\it observed luminosity function}, in
the sense that we do not attempt to apply any correction due to the
incompleteness of our survey.  Each object in the sample is then
simply added to the appropriate bolometric magnitude bin, and the
overall results are divided by the volume defined by a 40 pc
half-sphere.  Since the WDLF requires a proper account of the number
of individual stars in each magnitude bin, the confirmed and suspected
double degenerate binary systems are counted as two stars. Also, in
order to obtain the most accurate mass density as possible, we
deconvolve the individual masses of each system by using the procedure
described in Section 6.4 of \citet{noemi2012}. Doing so, our
luminosity function now includes a total of 501 individual white
dwarfs.

The luminosity function for the white dwarfs within 40 pc of the Sun
is presented in Figure \ref{lf} (the approximate temperature scale for
a 0.6 \msun\ evolutionary sequence is shown at the top of the
figure). Our results are also compared with those obtained by
\citet{noemi2012} for the 20 pc sample, by \citet{harris06} for white
dwarfs in the SDSS, and by \citet{bergeron2011} for the DA and DB
stars in the PG survey. In Figure \ref{lf2} we reproduce the same
luminosity function but in half-magnitude bins together with
theoretical luminosity functions from \citet{fontaine01} for a total
age of 10, 11, and 12 Gyr, normalized to our own observational results
between $\mbol=12-14$. These were obtained, as explained in
\citet{fontaine01}, with a constant star formation rate (SFR), a
classic Salpeter initial mass function (IMF, $\phi=M^{-2.35}$), an
initial-to-final mass relation (IFR) given by $M_{\rm
  WD}=0.4\,e^{0.125M}$, and a main sequence lifetime law given by
$t_{\rm MS}=10\,M^{-2.5}$ Gyr ($M$ and $M_{\rm WD}$ are in solar
units).  The half-magnitude bins have been selected to better match
the peak of the theoretical luminosity functions.

If the SFR is constant, we expect a monotonous rise of the luminosity
function, as shown by the theoretical curves in Figure
\ref{lf2}. Alternatively, bursts in star formation would show up as
bumps in the luminosity function \citep{noh90}. Our WDLF displayed
in both Figures \ref{lf} and \ref{lf2} show a definite bump near
$\mbol\sim10$. A similar bump was also observed in the WDLF determined
by \citet[][see their Figures 4 and 8]{harris06} for the SDSS sample
and by \citet[][see their Figure 22]{noemi2012} for the 20 pc sample
(both reproduced here in Figure \ref{lf}), but this peculiar result in
Giammichele et al.~had been attributed by the authors to small number
statistics. Here we have more than tripled the number of stars in the
magnitude bins of interest, and the result now appears to be
statistically significant. The brightest magnitude bins have error
bars that are still large enough to be consistent with the theoretical
expectations, but the points at $\mbol=10$ and 11 are solid
determinations. This particular feature in the WDLF can also be
observed {\it directly} in Figure \ref{correltm} and in the left panel
of Figure \ref{ratio_H} and could be explained by a sudden burst of
star formation in a recent past.  A direct comparison of our results
with the simulations shown in Figure 6 of \citet{noh90} suggests a
burst of star formation that occurred about 0.3 Gyr ago, a conclusion
also reached by \citet{harris06} based on the SDSS data. An
alternative, but less likely explanation would be that the luminosity
function for $M_V>10$ is still very incomplete. Indeed, the drop in
the space density near $\mbol=11$ is also observed in the luminosity
function of the PG sample, precisely in the region where the PG survey
becomes incomplete. However, the number of stars missing in our survey
would have to be enormous (Figures \ref{lf} and \ref{lf2} use a
logarithmic scale), an unlikely possibility given our estimate of the
completeness of this survey. In \citet{noemi2012}, this drop in the
number of stars near $\Te\sim14,000$~K was also tentatively explained
by the inaccurate treatment of convective energy transport in the
models, since this corresponds to the temperature where the
atmospheres of DA stars become convective. However, since we applied
here the appropriate 3D to 1D hydrodynamical corrections from
\citet{tremblay13b}, this explanation can be ruled out.

As expected, the WDLF displayed in Figure \ref{lf} shows that our
survey samples the cool end of the luminosity function really well,
with a significant number of stars in each magnitude bin, as opposed
to the color-excess PG survey for instance. Our derived space density
is consistent with that obtained by \citet{harris06} for white dwarfs
in the SDSS, except near the peak region, but again, our results have
not been corrected for incompleteness. At the same time, there are
significant corrections applied to the SDSS sample in this particular
range of luminosities, all of which remain extremely uncertain. Note
also that the number of stars in the fainter magnitude bins of our
WDLF must be considered with caution, since for most cool white dwarfs
in our sample, we had to assume a value of $\logg=8.0$ due to the lack
of trigonometric parallax measurements.  Hence a change in $\logg$
values could shift some stars from one bin to the other. As a matter
of fact, the mean mass of our sample is actually closer to 0.7
\msun\ rather than the 0.6 \msun\ assumed here for those stars. A
larger average mass would imply smaller radii and lower luminosities,
thus shifting the stars to fainter magnitude bins in both Figures
\ref{lf} and \ref{lf2}. Precise parallax measurements, like those that
will become available from the Gaia mission, will be required to
improve the shape of the WDLF at low luminosities, in particular since
only the last magnitude bin is sensitive to the age of the galactic
disk. Despite these uncertainties, we can still conclude that our
results are consistent with an age of the galactic disk around 11 Gyr.

Our WDLF displayed in Figure \ref{lf2} also fails to reproduce the
peak of the theoretical luminosity functions near $\mbol\sim15$,
despite the finer resolution used in this plot. This pronounced peak,
or bump, in the theoretical luminosity functions has been discussed in
detail by \citet[][see their Figures 5 and 6]{fontaine01} who
attribute this feature to the combined signatures of both convective
coupling and crystallization, although the contribution of the latter
process is significantly less than that of the former since the
release of latent heat operates over a wide range of
luminosities. Consequently, its effects tend to be averaged out over
that luminosity interval. Given the size of the error bars of our
observed WDLF in this particular region of bolometric magnitudes, it
is unlikely that our white dwarf sample misses so many objects. The
most likely explanation is that the assumptions built into the
construction of these theoretical luminosity functions mentioned above
(SFR, IMF, IFR, etc.) should perhaps be reexamined and explored
further, and in particular the simplistic initial-to-final mass
relation used in these calculations. Another aspect that would need to
be explored quantitatively is the effect of an old thick disk
component (see Section \ref{sec:kin}) on the predicted luminosity
function, with a different scale height appropriate for this
population.

Finally, by integrating the WDLF over all magnitude bins, it is
possible to obtain a measure of the {\it total space density} of white
dwarfs in our sample, which in turn can be used to evaluate the
completeness of our survey. \citet{holberg08} obtained for the 13 pc
sample, believed to be complete, a space density of
$(4.8\pm0.5)\times10^{-3}$ pc$^{-3}$ and a mass density of
$(3.2\pm0.3)\times10^{-3}$ \msun\ $\rm{pc}^{-3}$.  Using these numbers
as a reference point, \citet{noemi2012} concluded that the 20 pc
sample was $90\%$ complete. For the 40 pc sample, we derive a space
density of $3.74\times10^{-3}$ pc$^{-3}$ and a corresponding mass
density of $2.46\times10^{-3}$ \msun\ pc$^{-3}$, which would imply
that the 40 pc sample in the northern hemisphere is $78\%$ complete, in agreement with our
other estimates above.

\section{Concluding Remarks}\label{sec:concl}

Our spectroscopic survey of white dwarfs within 40 pc of the Sun is
not yet complete and there is still a significant amount of work to be
done. First of all, we have a few high-priority targets in our sample,
some of which are bright enough to be observed with 2-m class
telescopes, while some faint targets around $V\sim19$ will require
large aperture telescopes, such as Gemini North or South. Among the
brightest objects which could not be observed because of uncooperative
weather, several candidates with $D<30$ pc remain on our target list,
as well as 5 Giclas objects: 01309+5321 (GD 278), 02011+1212 (GD 21),
07544+6611 (GD 454), 14065+7418 (GD 492), and 22022+3848 (GD
399). Also, the southern hemisphere should eventually be dealt with,
but it is not clear how much effort should be put into this given the
work of \citet{sayres2012} for declinations close to $\delta=0$, as
well as the SOAR + SMARTS Southern White Dwarf SURVEY (SSSWDS) of
\citet{suba09}, which uncovered 100 new white dwarfs. The current
homogeneity of white dwarf surveys within 20 pc of the Sun in both
hemispheres is about to be extended to 25 pc in a near future, but
eventually to larger distances.  The ultimate confirmation of the
white dwarfs identified in our survey as members of the 40 pc sample
will of course come from precise trigonometric parallaxes from the
Gaia mission.  In addition to adding or removing stars from the 40 pc
sample, these measurements will also help to better define the faint
end of the luminosity function revealed by our survey.

\acknowledgements We are grateful to the referee, Hugh C. Harris, for his
detailed and constructive comments, which have greatly helped
improving the content and presentation of our results.  We would like
to thank the director and staff of Steward Observatory, Kitt Peak
National Observatory, and the Observatoire du Mont-M\'egantic for the
use of their facilities, as well as the Director and staff of Gemini
North and South Observatories for the remote observing. We would also
like to thank P.~Dufour for making his DQ and DZ models available to
us, G.~Fontaine for the theoretical luminosity functions, and
S.R.~Kulkarni and M.H.~van Kerkwijk for allowing us to use their
spectrum of SDSS 1257+5428.  This work was supported in part by the
NSERC Canada and by the Fund FRQ-NT (Qu\'ebec). S.L. was supported in
this research by NSF grants AST 06-07757 and AST 09-08419.  S.L also
acknowledges support from the {\it{GALEX}} Guest Investigator program
under NASA grant NNX09AF88G. This research made use of the SIMBAD
database and the VizieR catalog access tool, operated at CDS,
Strasbourg, France, and also made use of data products from the Two
Micron All Sky Survey, which is a joint project of the University of
Massachusetts and the Infrared Processing and Analysis
Center/California Institute of Technology, funded by the National
Aeronautics and Space Administration and the National Science
Foundation.

\bibliography{ms}{}
\bibliographystyle{apj}

\begin{deluxetable}{lllcccc}
\tabletypesize{\scriptsize}
\tablecolumns{12}
\tablewidth{0pt}
\tablecaption{Spectroscopic Observing Runs}
\tablehead{
\colhead{Date} &
\colhead{Telescope} &
\colhead{Spectrograph} &
\colhead{Grating} &
\colhead{Blaze} &
\colhead{Coverage}&
\colhead{Slit} \\

\colhead{ } &
\colhead{ } &
\colhead{ } &
\colhead{ (l mm$^{-1}$)} &
\colhead{ (\AA)} &
\colhead{ (\AA)}&
\colhead{ ($\arcsec$)}}

\startdata
2011 Jan&NOAO 2.1 m&Goldcam&500 &5500&3800--6700&2\\
2011 Mar&NOAO Mayall 4 m&RC&316 &5500&3900--6700&2\\
2011 Apr&Steward Observatory Bok 2.3 m&B$\&$C&600 &3568&3800--5600&4.5\\
2011 Apr&NOAO 2.1 m&Goldcam&500 &5500&3800--6700&2\\
2012 Sep&NOAO Mayall 4 m&RC&316 &5500&3900--6700&2\\
2011B &Gemini North &GMOS-N&600 &4610&3800--6700&1\\
2011B &Gemini South &GMOS-S&600 &4610&3800--6700&1\\
2012B &Gemini North&GMOS-N&600 &4610&3800--6700&1\\
2013 Jan&Observatoire du Mont-M\'egantic 1.6 m&B$\&$C&600 &4000&3800--6700&4.3\\
2013 Jun&Steward Observatory Bok 2.3 m&B$\&$C&600 &3568&3800--5600&4.5\\
2013 Oct & Steward Observatory Bok 2.3 m&B$\&$C&600 &3568&3800--5600&4.5\\
\enddata
\label{tab_obs}
\end{deluxetable}

\clearpage
\begin{deluxetable}{lcccccrr}
\tabletypesize{\scriptsize}
\tablecolumns{8}
\tablewidth{0pt}
\tablecaption{Spectroscopically Confirmed White Dwarfs from SUPERBLINK $-$ Astrometry}
\tablehead{
\colhead{PM I} &
\colhead{NLTT} &
\colhead{SDSS} &
\colhead{RA} &
\colhead{DEC} &
\colhead{$\mu_{\rm{tot}}$} &
\colhead{$\mu$RA} &
\colhead{$\mu$DE}\\ 
\colhead{  } &
\colhead{  } &
\colhead{  } &
\colhead{($J_{2000}$)} &
\colhead{($J_{2000}$)} &
\colhead{$(\arcsec \rm{yr}^{-1})$} &
\colhead{$(\arcsec \rm{yr}^{-1})$} &
\colhead{$(\arcsec \rm{yr}^{-1})$}}
\startdata
00050$+$4003 &133       &                &00:05:01.08 &  +40:03:34.6& 0.229  & 0.206  & 0.101  \\
00056$+$4825 &          &                &00:05:40.41 &  +48:25:05.9& 0.153  & 0.152  &$-$0.013\\
00074$+$3403 &301       &                &00:07:28.87 &  +34:03:39.9& 0.199  & 0.021  & 0.197  \\
01088$+$7600 &          &                &01:08:49.61 &  +76:00:18.4& 0.242  & 0.237  & 0.050  \\
01278$+$7328 &4799      &                &01:27:49.09 &  +73:28:47.7& 0.177  &$-$0.160& 0.076  \\
\\
01327$+$4604 &5100      &                &01:32:46.67 &  +46:04:58.8& 0.211  &$-$0.211&$-$0.002\\
01327$+$6635 &          &                &01:32:42.90 &  +66:35:46.0& 0.051  & 0.039  &$-$0.033\\
\enddata
\label{tab_names}
\end{deluxetable}

\clearpage
\begin{deluxetable}{lccccccccccccclc}
\tabletypesize{\scriptsize}
\tablecolumns{19}
\tablewidth{0pt}
\rotate
\tablecaption{Spectroscopically Confirmed White Dwarfs from SUPERBLINK $-$ Photometry}
\tablehead{
\colhead{PM I} &
\colhead{FUV} &
\colhead{NUV} &
\colhead{$B_F$} &
\colhead{$R_I$} &
\colhead{$I_N$} &
\colhead{$J$} &
\colhead{$H$} &
\colhead{$K_S$} &
\colhead{$u$} &
\colhead{$g$} &
\colhead{$r$} &
\colhead{$i$} &
\colhead{$z$} &
\colhead{ST} &
\colhead{Notes}}
\startdata
    00023$+$6357 &$-$  &$-$  &17.8&16.2&17.3&15.80&15.57&15.51&$-$  &$-$  &$-$  &$-$  &$-$  &DC   &\\
    00050$+$4003 &$-$  &18.14&16.3&16.2&16.5&$-$  &$-$  &$-$  &$-$  &$-$  &$-$  &$-$  &$-$  &DZA: &\\
    00056$+$4825 &$-$  &19.38&17.1&16.9&16.5&16.25&15.60&15.89&$-$  &$-$  &$-$  &$-$  &$-$  &DA   &\\
    00074$+$3403 &$-$  &21.07&17.7&17.3&16.8&16.39&16.25&15.83&$-$  &$-$  &$-$  &$-$  &$-$  &DC:  &\\
    00079$+$3947 &$-$  &$-$  &17.4&16.2&15.8&15.18&14.85&14.65&$-$  &$-$  &$-$  &$-$  &$-$  &DC   &\\
\\
    00217$+$2640 &$-$  &$-$  &17.9&17.2&16.9&16.18&15.91&15.81&$-$  &$-$  &$-$  &$-$  &$-$  &DC   &\\
    00276$+$0542 &$-$  &$-$  &17.2&15.6&15.4&14.97&14.67&14.57&$-$  &$-$  &$-$  &$-$  &$-$  &DA   &\\
\enddata
\tablecomments{
(1) WD in McCook and Sion. 
(2) Also in \citet{Schilbach2012}. 
(3) Also in \citet{vennes2011}. 
(4) SDSS spectrum. 
(5) Also in \citet{sayres2012}. 
(6) Also in \citet{kilic2010}. 
(7) DA+dM in \citet{Rebassa-Mansergas2010}. 
(8) Also in \citet{tonry2012}. 
(9) LSPM J1227+3150 wrongly identified as PM I12273+3150 in \citet{limoges2013}. }
\label{tab_phot}
\end{deluxetable}

\clearpage
\begin{deluxetable}{lllrccccrrcl}
\tabletypesize{\scriptsize}
\tablecolumns{12}
\tablewidth{0pt}
\rotate
\tablecaption{Atmospheric Parameters of New White Dwarfs from SUPERBLINK}
\tablehead{
\colhead{PM I} &
\colhead{ST} &
\colhead{$\Te$ } &
\colhead{log $g$} &
\colhead{$M/M_{\odot}$} &
\colhead{Comp.} &
\colhead{$M_V$} &
\colhead{log $L/L_{\odot}$} &
\colhead{$D $} &
\colhead{log $\tau$} &
\colhead{Fit$^a$} &
\colhead{Other name$^b$}  \\
\colhead{  } &
\colhead{  } &
\colhead{(K)} &
\colhead{ } &
\colhead{ } &
\colhead{ } &
\colhead{ } &
\colhead{ } &
\colhead{$({\rm pc})$} &
\colhead{ } &
\colhead{ } &
\colhead{ }}
\startdata
   00023+6357   &DC        &  4630 (563)   &8.00 (0.25)    &0.58 (0.15)  &H &14.77     &$-$4.18     & 26.7 (5.2) & 9.85          &2&                         \\
   00050+4003   &DZA:      &  8150 (549)   &8.00 (0.25)    &0.58 (0.15)  &He&12.50     &$-$3.20     & 52.2 (9.4) & 9.05          &2&GD 1                     \\
   00056+4825   &DA        &  6970 (105)   &7.95 (0.09)    &0.56 (0.05)  &H &13.47     &$-$3.43     & 48.4 (1.3) & 9.15          &1&                         \\
   00074+3403   &DC:       &  5830 (113)   &8.00 (0.25)    &0.57 (0.15)  &He&14.23     &$-$3.79     & 42.0 (5.2) & 9.46          &2&                         \\
   00079+3947   &DC        &  4890 (453)   &8.00 (0.25)    &0.58 (0.15)  &H &15.16     &$-$4.09     & 20.2 (3.0) & 9.79          &2&                         \\
\\
   00217+2640   &DC        &  5160 (645)   &8.00 (0.25)    &0.58 (0.15)  &H &14.67     &$-$3.99     & 35.2 (6.1) & 9.69          &2&                         \\
   00276+0542   &DA        &  5520 (855)   &8.00 (0.25)    &0.58 (0.15)  &H &14.65     &$-$3.87     & 21.5 (3.7) & 9.49          &2&                         \\
\enddata
\tablenotetext{a}{Fit: (1) spectroscopic, (2) photometric.}
\tablenotetext{b}{SDSS name unless noted otherwise.}
\label{tab_results}
\end{deluxetable}

\clearpage
\begin{deluxetable}{lllrccccrcrclc}
\tabletypesize{\scriptsize}
\tablecolumns{14}
\tablewidth{0pt}
\rotate
\tablecaption{Atmospheric Parameters of Northern White Dwarfs within 40 pc of the Sun}
\tablehead{
\colhead{PM I} &
\colhead{ST} &
\colhead{$\Te$} &
\colhead{log $g$} &
\colhead{$M/M_{\odot}$} &
\colhead{Comp.} &
\colhead{$M_V$} &
\colhead{log $L/L_{\odot}$} &
\colhead{$D$} &
\colhead{$\pi$} &
\colhead{log $\tau$} &
\colhead{Fit$^a$} &
\colhead{WDname} &
\colhead{Notes}\\
\colhead{  } &
\colhead{ } &
\colhead{ (K)} &
\colhead{ } &
\colhead{  } &
\colhead{ } &
\colhead{  } &
\colhead{ } &
\colhead{$({\rm pc})$} &
\colhead{$({\rm mas})$} &
\colhead{ } &
\colhead{ } &
\colhead{ } &
\colhead{ }}
\startdata
   00023+6357*  &DC        &  4630 (563)     &8.00 (0.25)    &0.58 (0.15)    &H &14.77          &$-$4.18        & 26.7 (7.2) &               & 9.85          &2&                         &          \\
   00051+7313   &DB        & 14490 (236)     &8.34 (0.12)    &0.80 (0.08)    &He&11.90          &$-$2.40        & 34.7 (5.8) & 28.8 (  4.7)  & 8.59          &1&               0002+729  &          \\
   00073+1230   &DC        &  5090 ( 65)     &8.00 (0.25)    &0.57 (0.15)    &He&14.62          &$-$4.02        & 21.0 (3.4) &               & 9.76          &2&               0004+122  &          \\
   00074+3403*  &DC:       &  5830 (113)     &8.00 (0.25)    &0.57 (0.15)    &He&14.23          &$-$3.79        & 42.0 (7.3) &               & 9.46          &2&                         &          \\
   00079+3947*  &DC        &  4890 (453)     &8.00 (0.25)    &0.58 (0.15)    &H &15.16          &$-$4.09        & 20.2 (4.2) &               & 9.79          &2&                         &          \\
\\
   00113+4240   &DA        &  7140 (106)     &8.24 (0.07)    &0.75 (0.05)    &H &13.80          &$-$3.56        & 19.3 (0.8) &               & 9.35          &1&               0008+424  &          \\
   00122+5025   &DAP       &  6420 ( 99)     &8.15 (0.10)    &0.68 (0.06)    &H &14.09          &$-$3.69        & 11.0 (0.4) & 90.6 (  3.7)  & 9.42          &1&               0009+501  &          \\
\enddata
\tablecomments{
(1) Double degenerate binary (or candidate). 
(2) \citet{gianninas2011}. 
(3) $D$ from \citet{vennes2012}. 
(4) DXP+DA: $\Te$, $\logg$ and $D$ come from \citet{liebert1993} and belong to the DA component; the values given for the companion are $\Te=(16,000\pm2000)$ K and $\logg=8.5\pm0.2$. 
(5) Trigonometric parallax from \citet{dahn82}. }
\tablenotetext{a}{Fit: (1) spectroscopic, (2) photometric.}
\tablenotetext{*}{New WD identified in this survey.}
\label{tab_40pcresults}
\end{deluxetable}

\clearpage

\figcaption[f1.ps]{Absolute visual magnitude as a
  function of photometric distance for spectroscopically confirmed
  white dwarfs (top) and our remaining candidates (bottom). In the
  upper panel, the filled circles represent the 325 new white dwarfs
  identified in our survey, while
  the open circles correspond to the 416 white dwarfs already known in
  the literature. Dashed lines in the figure are lines of constant
  apparent $V$ magnitudes. The 592 remaining white dwarf candidates in
  our survey, still awaiting spectroscopic confirmation, are shown in
  the lower panel. Lower-priority candidates (those identified on the
  basis USNO photographic magnitudes and those with $V>18$) are shown
  with cross symbols.\label{Mv_d}}

\figcaption[f2.ps]{Optical spectra for our new,
  spectroscopically confirmed DA white dwarfs, displayed in order of
  decreasing effective temperature (upper left to bottom right) and
  shifted vertically for clarity; 05431+3637 is a DAZ star. The
  \ha\ line is also shown when available, and normalized to a
  continuum set to unity. The last 3 spectra at the bottom of the
  right panel are new observations of stars already presented in Paper
  I that are double degenerate binary candidates.\label{DA_hot}}

\figcaption[f3.ps]{\ha\ line for our new, spectroscopically confirmed
  DA stars, too cool to exhibit the rest of the Balmer series. Spectra
  are displayed in order of right ascension, normalized to a
  continuum set to unity, and shifted vertically for
  clarity.\label{DA_cool}}

\figcaption[f4.ps]{Spectra of our newly identified, magnetic DA
  white dwarfs, shifted vertically for clarity. $04523+2519$ was
  classified as non-magnetic in Paper I. All others are new white
  dwarf discoveries.\label{DAH}}

\figcaption[f5.ps]{Spectra of our newly identified
  binary systems composed of a DA white dwarf and an M dwarf
  companion. Spectra are shifted vertically for clarity. The 
  hydrogen line cores are often contaminated by
  line emission from the chromospherically active M
  dwarf.\label{DA+dM}}

\figcaption[f6.ps]{New spectra of DZ (DZA), DB (DBZ),
  and DQ stars. Spectra for the DZ stars 01216+3440, 03196+3630,
  16477+2636, 21420+2252, and 23003+2204 represent new higher S/N
  observations of stars reported in Paper I, used to better constrain
  the metal abundances.\label{DZDQDB}}

\figcaption[f7.ps]{Optical spectra for our new,
  spectroscopically confirmed DC stars. All spectra are normalized to
  a continuum set to unity and are offset from each other by a factor
  of 0.9.\label{DC}}

\figcaption[f8.ps]{($u-g$, $g-r$) color-color diagram
  showing all stars in our survey for which $ugriz$ photometry is
  available. The 151 spectroscopically confirmed white dwarfs are
  shown with various black symbols explained in the legend, while the
  125 white dwarf candidates still lacking spectroscopic data are
  shown with red dots. The solid curves represent pure hydrogen model
  atmospheres at $\log g=7.0$, 8.0, and 9.0 (from bottom to top);
  effective temperatures are indicated in units of $10^3$ K. The
  dashed curve corresponds to pure helium atmospheres at $\log g=8.0$,
  and the dotted lines
  represent DQ models for 5 different compositions, from $\log
  \rm{C/He}=-9.0$ to $-5.0$.\label{color_gmr}}

\figcaption[f9.ps]{Upper panel: Equal cylindrical projection
  of the equatorial coordinates for the sample of white dwarfs
  identified from SUPERBLINK. The 416 previously known stars from the
  WD Catalog recovered by our selection criteria (open circles) are
  compared to the 325 new WD identifications (solid circles). Also
  shown by the bold solid line is the region of the galactic
  plane. Lower panel: Sky density as a function of right ascension,
  normalized to the total number of white dwarf stars (all lines are
  thus on a comparable scale). The dotted line represents the 416
  stars from the WD Catalog recovered by our selection criteria, while
  the dashed line corresponds to the 325 new identifications
  only. Finally, the solid line represents the sum of the
  contributions of the new and known white dwarfs.\label{plot_equal}}

\figcaption[f10.ps]{Sample fits to the photometric energy
  distributions (represented by error bars) of 5 new white dwarf
  identifications with pure hydrogen models (filled circles) and with
  pure helium models (open circles). The adopted solution is indicated
  in red. In the right panels are shown the observed normalized
  spectra together with the synthetic line profiles calculated with
  the atmospheric parameters corresponding to the pure hydrogen
  solutions.\label{ex_photo}}

\figcaption[f11.ps]{Same as Figure \ref{ex_photo} but for
  all DC and cool DA stars in our sample.\label{photoDADC}}

\figcaption[f12.ps]{Fits to the observed energy distributions
  for the new DQ identifications. The filled circles correspond to our
  best fit to the photometry, and the atmospheric parameters and
  carbon abundances are given in each panel. In the right panels are
  shown the observed spectra together with the predicted model fit (in
  red).\label{DQfits}}

\figcaption[f13.ps]{Fits to the energy distributions for the
  new DZ identifications. The filled circles correspond to our best
  fit with the atmospheric parameters given in each panel. In the right
  panels are shown the observed spectra together with the predicted
  model fit (in red); the insert shows our fit to \ha, when
  detected.\label{DZfits}}

\figcaption[f14.ps]{Fits to the optical spectra of the DA
  stars in our sample. The lines range from either \ha\ (when
  available) or \hb\ (bottom) to H8 (top), each offset vertically by a
  factor of 0.2. Theoretical line profiles shown in green are not used
  in the fitting procedure.\label{DAspectro2}}

\figcaption[f15.ps]{Fits to the optical spectra for the 4 new DB
  stars in our sample; the atmospheric parameters ($\Te$, $\logg$, and
  $\log \rm{H/He}$) of each object are given in the figure. When
  available, the region near \ha\ is used to measure the hydrogen
  abundance.\label{DBfit}}

\figcaption[f16.ps]{Same as Figure \ref{ex_photo} but for 4
  double degenerate binary candidates in our sample.\label{fitsDADC}}

\figcaption[f17.ps]{Spectroscopic fits to 2 of our double
  degenerate binary candidates (04263+4820 and 11598+0007) and to the
  DA + DA white dwarf binary SDSS 1257+5428 (see text).\label{DA+DA}}

\figcaption[f18.ps]{Our best spectroscopic fit to the
  helium-rich DAZB white dwarf 01489+1902 (GD 16); the insert shows the region
  covering the \ha\ line profile.\label{fit_01489}}

\figcaption[f19.ps]{Cumulative number of stars as a function of
  distance, for our northern hemisphere census. The solid curve shows
  the expected number of white dwarfs in one hemisphere, assuming an
  average space density of $4.8\times10^{-3}$ pc$^{-3}$, while the
  dotted curve represents the expected number of white dwarfs on the
  whole celestial sphere.\label{NvsD}}

\figcaption[f20.ps]{Velocity-space projections for the white
  dwarfs within 40 pc of the Sun. Velocities are calculated assuming
  $V_{\rm rad}=0$, and using only the projections having the smallest
  contribution to $V_{\rm rad}$. Each star is thus displayed in one
  panel only. The new white dwarf identifications are plotted with solid
  circles, while those already known in the literature are shown 
  by open circles; the red circles at $(0,0)$ indicate our
  limit of detection.\label{uvw}}

\figcaption[f21.ps]{Mass as a function of effective
  temperature for the 288 white dwarfs in the 40 pc sample with mass
  determinations; the 204 white dwarfs fitted with an assumed
  $\logg=8.0$ value are displayed at the bottom the figure. Also shown
  are theoretical isochrones for our C/O core evolutionary models with
  thick hydrogen layers labeled in Gyr; solid lines correspond to
  white dwarf cooling ages only, while the dotted lines also include
  the main sequence lifetime. The dashed line represents a 0.661
  $M_{\odot}$ sequence, which corresponds to the median mass of our
  sample.\label{correltm}}

\figcaption[f22.ps]{Mass distribution for the white
  dwarfs in our sample that have mass determinations. The thick solid
  line histogram shows the distribution for the 288 stars with $D<40$ pc,
  while the red and blue shaded histograms correspond, respectively,
  to the subsamples of 248 hydrogen- and 40 helium-atmosphere white
  dwarfs.  The corresponding mean values and standard deviations are
  given in the figure.\label{histo_mass_spectro}}

\figcaption[f23.ps]{Mass distribution for the white
  dwarfs in the 40 pc sample with mass determinations, compared with
  the mass distribution for the 20 pc sample \citep{noemi2012}, for
  the DA stars in the SDSS DR4 \citep{tremblay11a}, and for the DA
  stars in the WD Catalog \citep{gianninas2011}.  Note that the peaks
  of the WD Catalog and 40 pc sample distributions are
  superposed.\label{histo_compare}}

\figcaption[f24.ps]{Distribution of the various spectral types
  for the white dwarfs in the 40 pc sample as a function of effective
  temperature. The number of stars of each spectral type is indicated
  in the panels.\label{spec_evol}}

\figcaption[f25.ps]{Left panel: total number of white dwarfs
  (solid-line histogram) and hydrogen-atmosphere white dwarfs (hatched
  histogram) as a function of effective temperature. Right panel:
  ratio of helium-atmosphere white dwarfs to the total number of stars
  as a function of effective temperature.\label{ratio_H}}

\figcaption[f26.ps]{Left panel: total number of white dwarfs
  in the $7000-9000$~K temperature bin (solid-line histogram) and
  hydrogen-atmosphere white dwarfs (hatched histogram) as a function
  distance. Right panel: ratio of helium-atmosphere white dwarfs to
  the total number of stars in the same temperature bin as a function
  of distance.\label{ratio_D}}

\figcaption[f27.ps]{Luminosity function for our sample of white
  dwarfs within 40 pc of the Sun as a function of $\mbol$ (red line),
  compared to the luminosity functions obtained by \citet{noemi2012}
  for the 20 pc sample, by \citet{harris06} for white dwarfs in the
  SDSS, and by \citet{bergeron2011} for the DA and DB stars in the PG
  survey; the number of stars in each magnitude bin is given for the
  40 pc sample only. The approximate temperature scale for a $M = 0.6$
  \msun\ sequence is shown at the top of the figure.\label{lf}}

\figcaption[f28.ps]{Luminosity function for our sample of white
  dwarfs within 40 pc of the Sun as a function of $\mbol$ (red line),
  given in half-magnitude bins, compared with theoretical luminosity
  functions from \citet{fontaine01} for a total age of 10, 11, and 12
  Gyr. The approximate temperature scale for a $M = 0.6$
  \msun\ sequence is shown at the top of the figure.\label{lf2}}

\clearpage

\begin{figure}[p]
\plotone{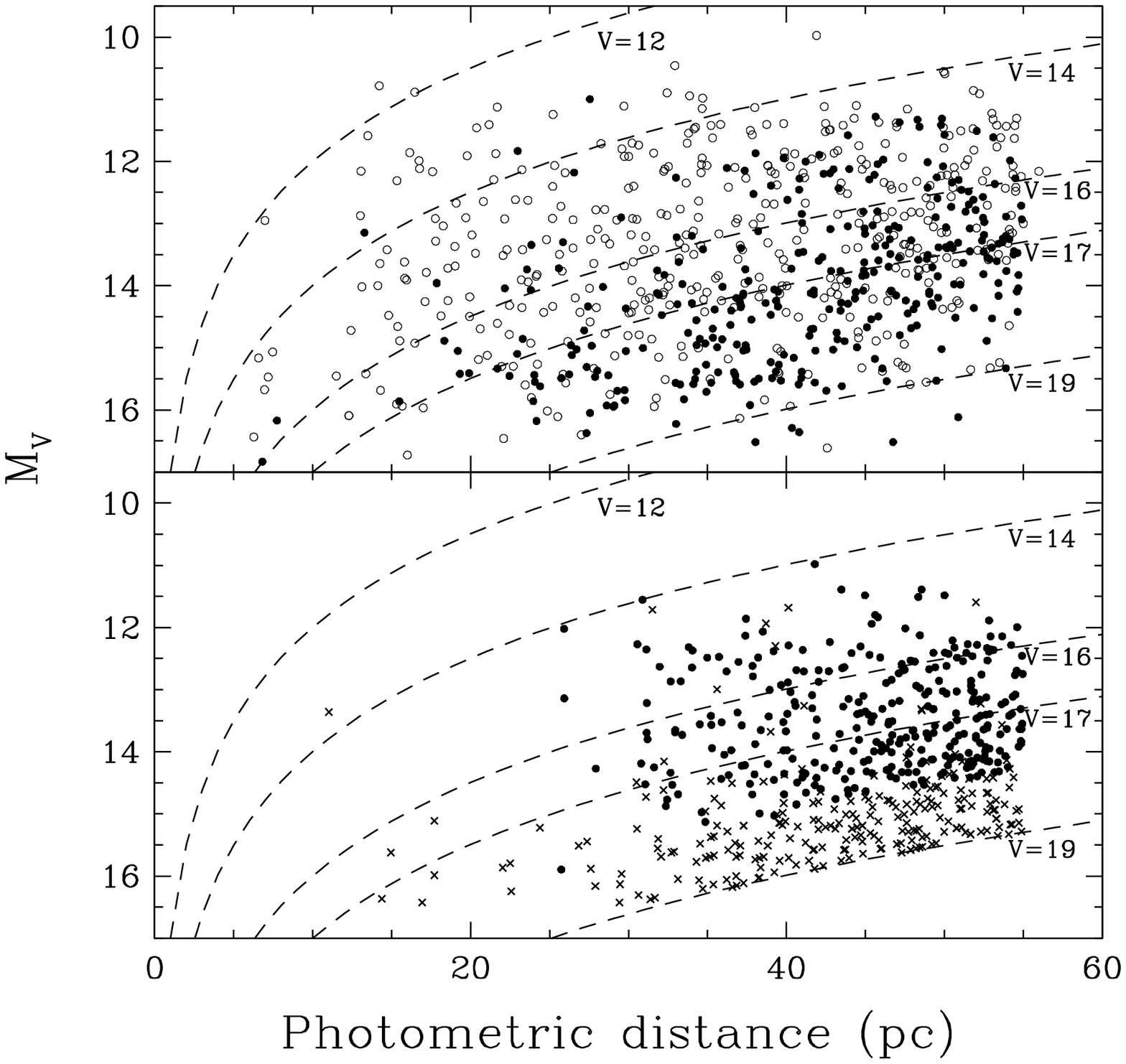}
\begin{flushright}
Figure \ref{Mv_d}
\end{flushright}
\end{figure}

\clearpage

\begin{figure}[p]
\plotone{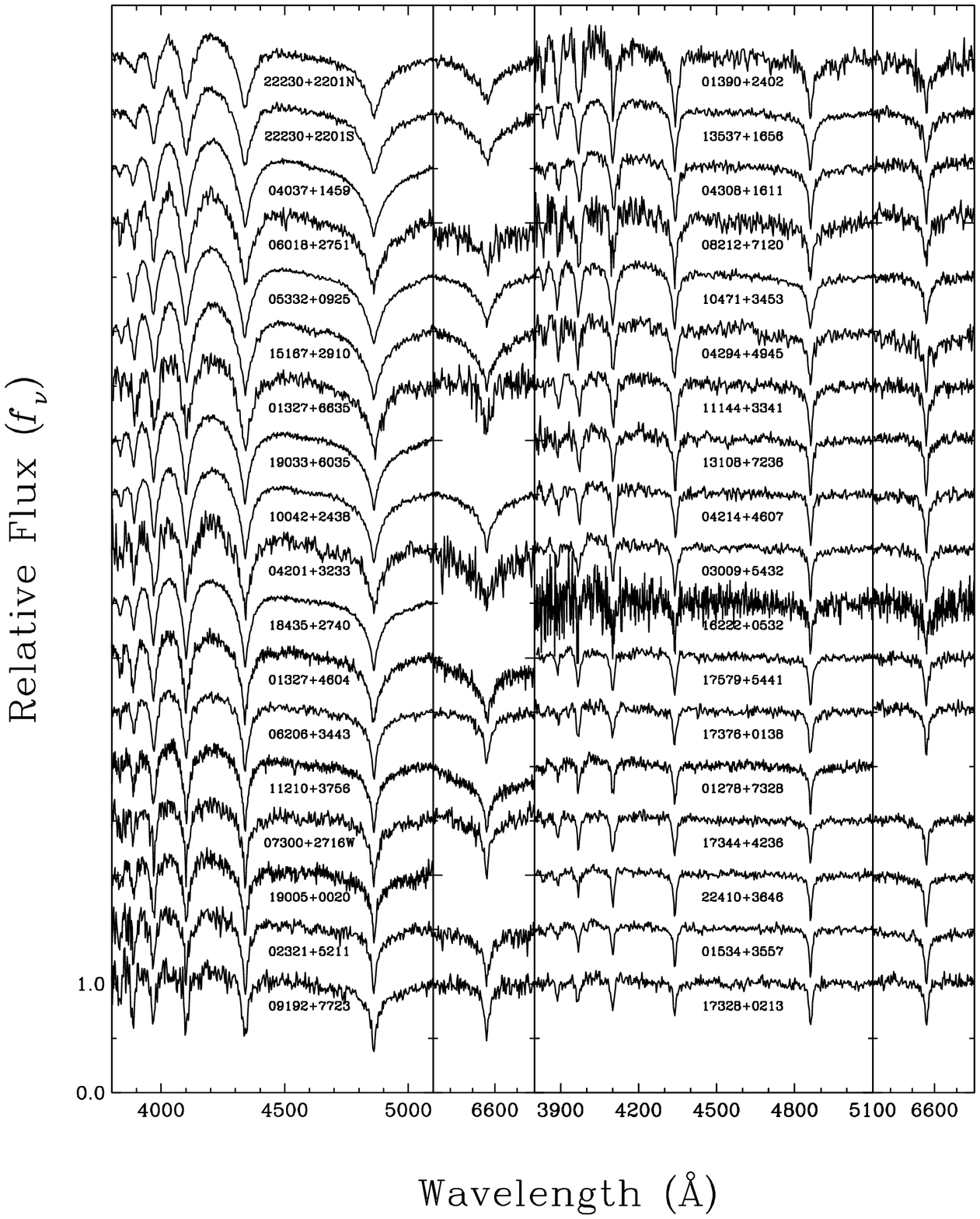}
\begin{flushright}
Figure \ref{DA_hot}a
\end{flushright}
\end{figure}

\clearpage

\begin{figure}[p]
\plotone{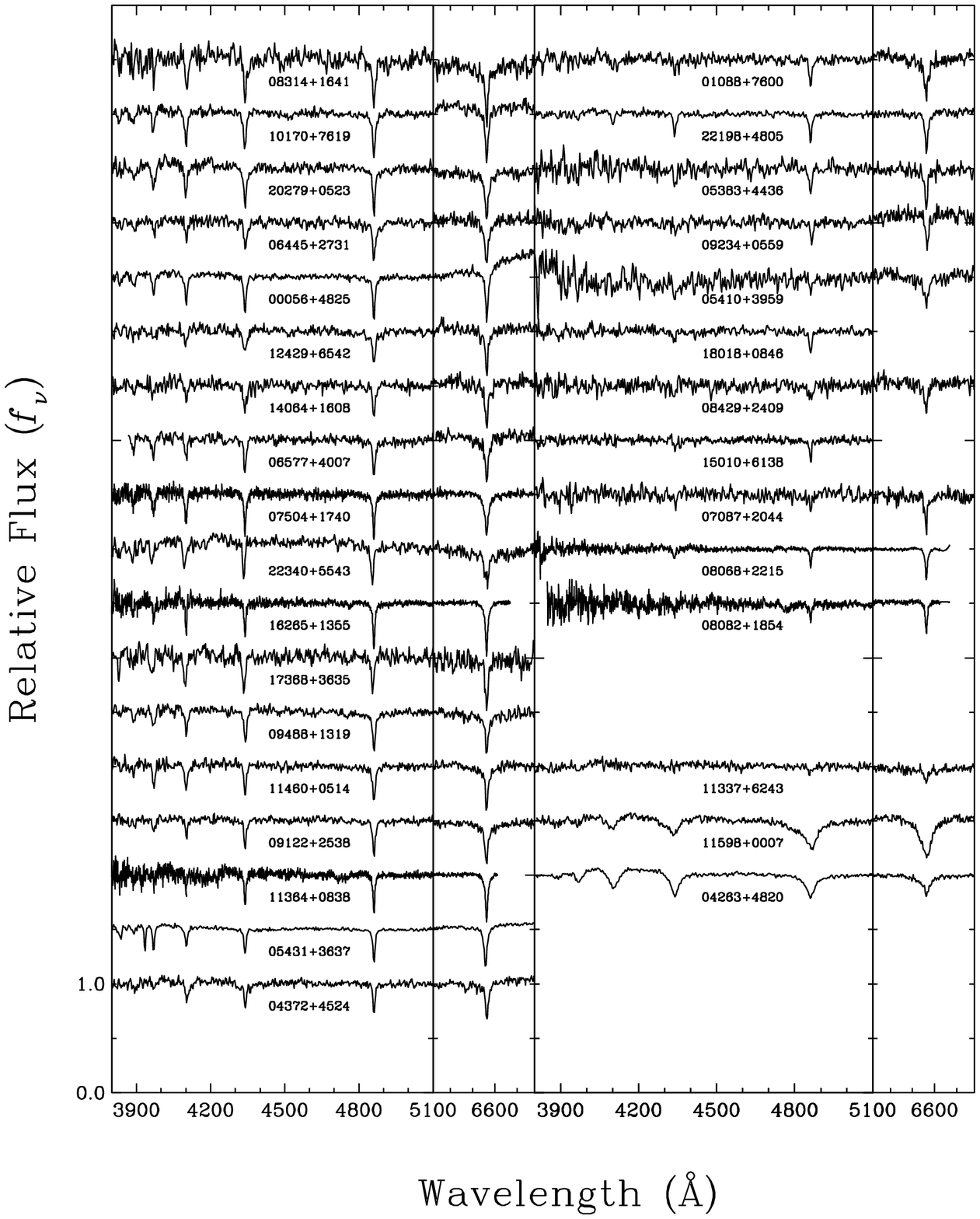}
\begin{flushright}
Figure \ref{DA_hot}b
\end{flushright}
\end{figure}

\clearpage

\begin{figure}[p]
\plotone{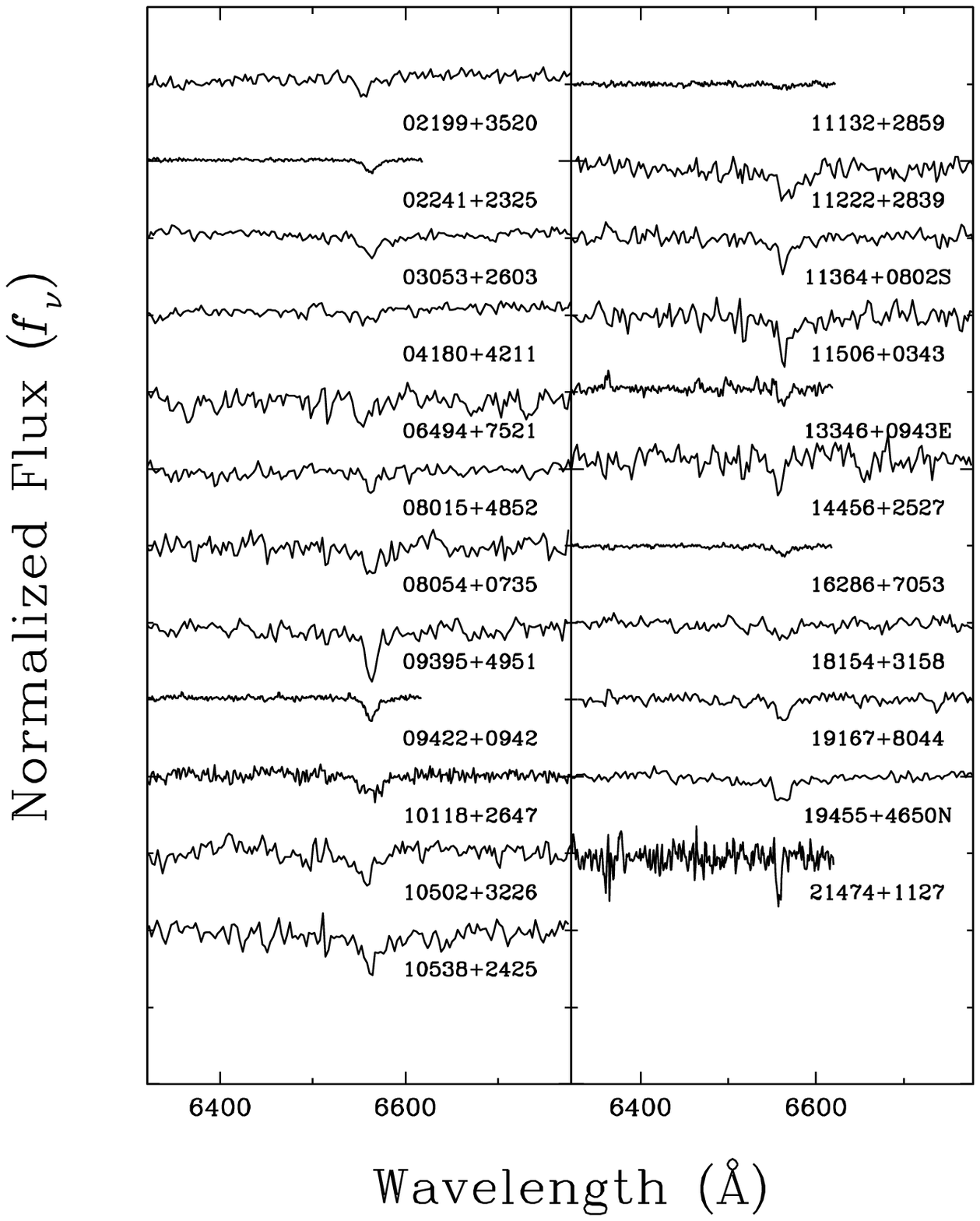}
\begin{flushright}
Figure \ref{DA_cool}
\end{flushright}
\end{figure}

\clearpage

\begin{figure}[p]
\plotone{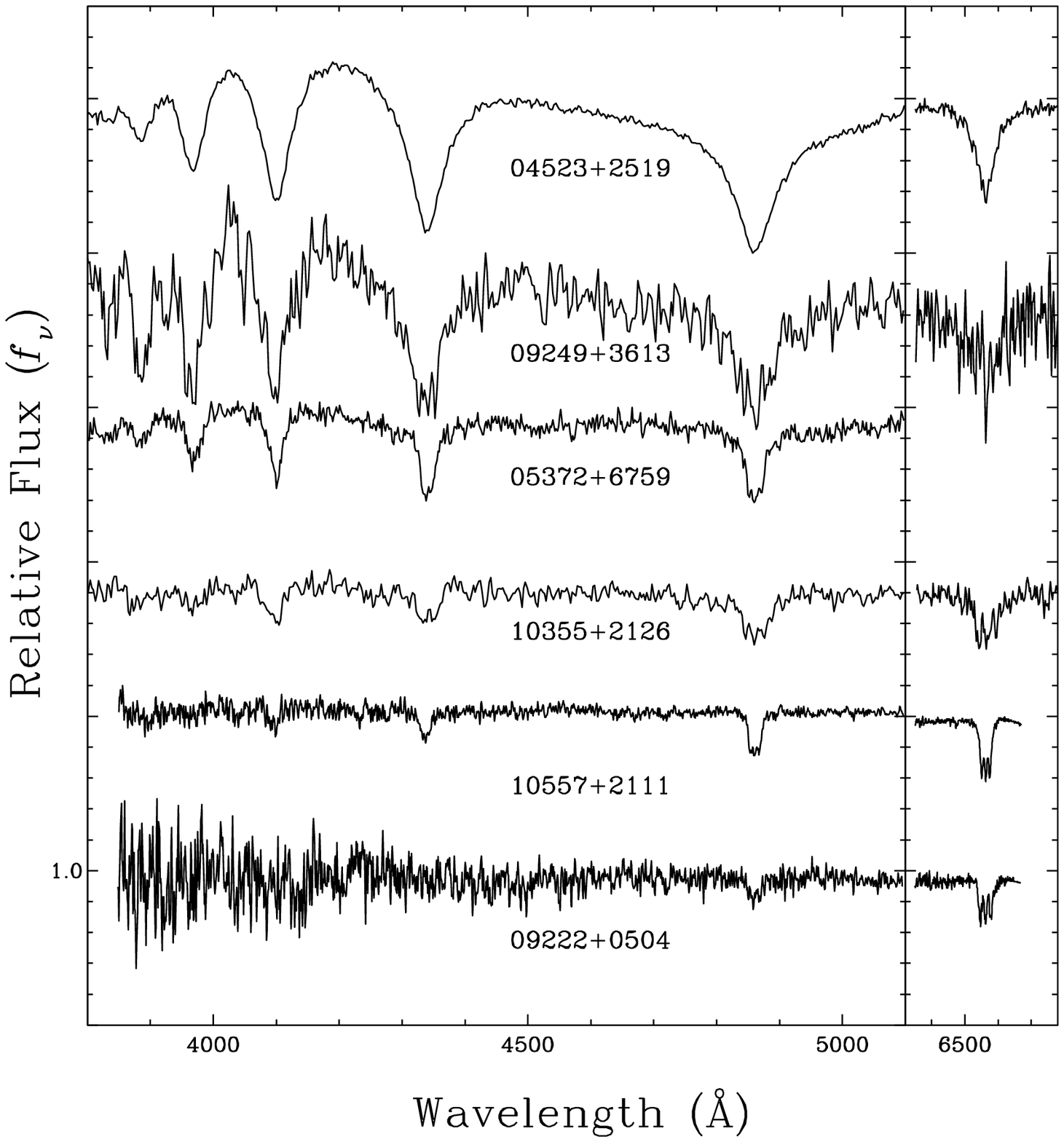}
\begin{flushright}
Figure \ref{DAH}
\end{flushright}
\end{figure}

\clearpage

\begin{figure}[p]
\plotone{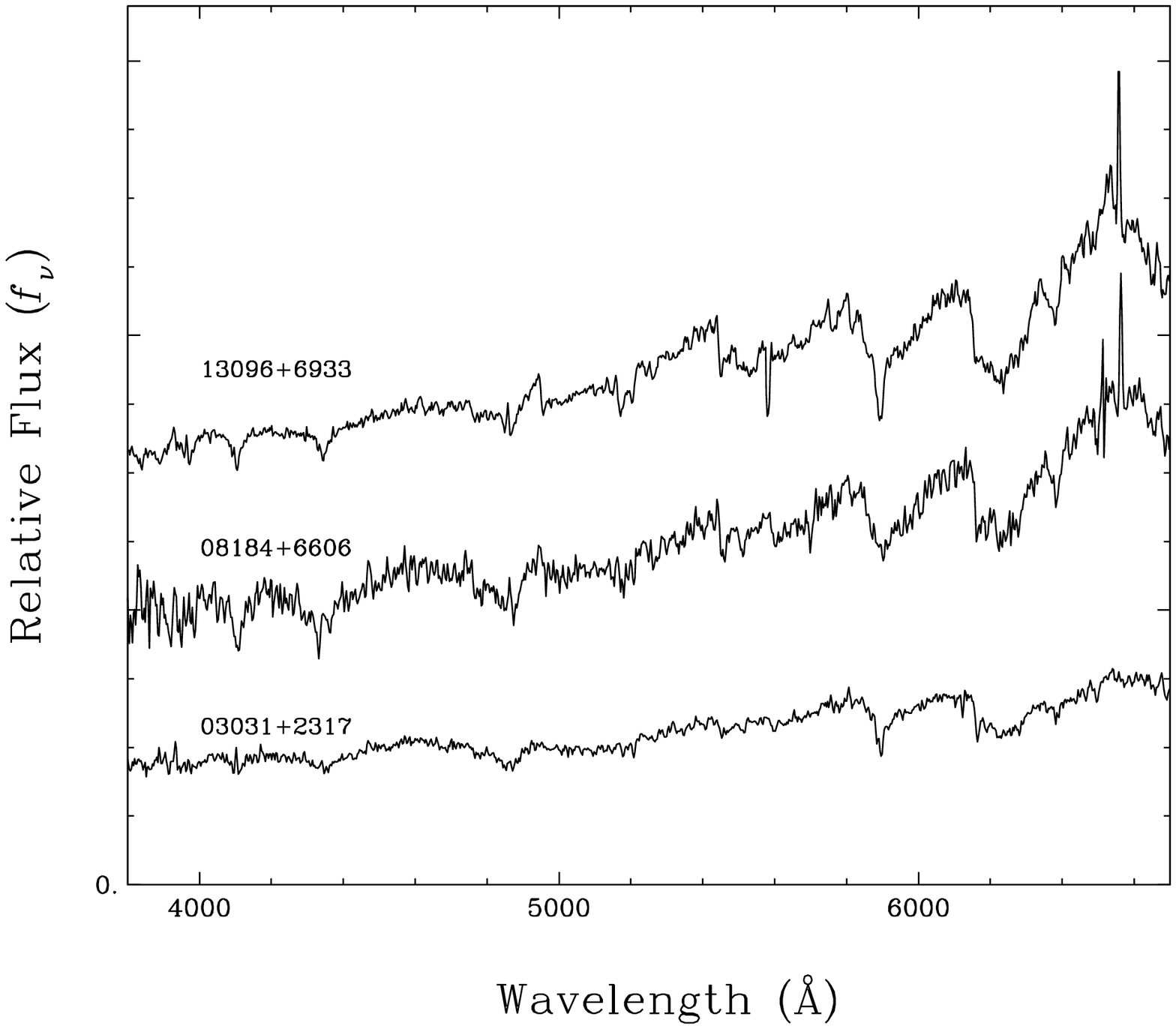}
\begin{flushright}
Figure \ref{DA+dM}
\end{flushright}
\end{figure}

\clearpage

\begin{figure}[p]
\plotone{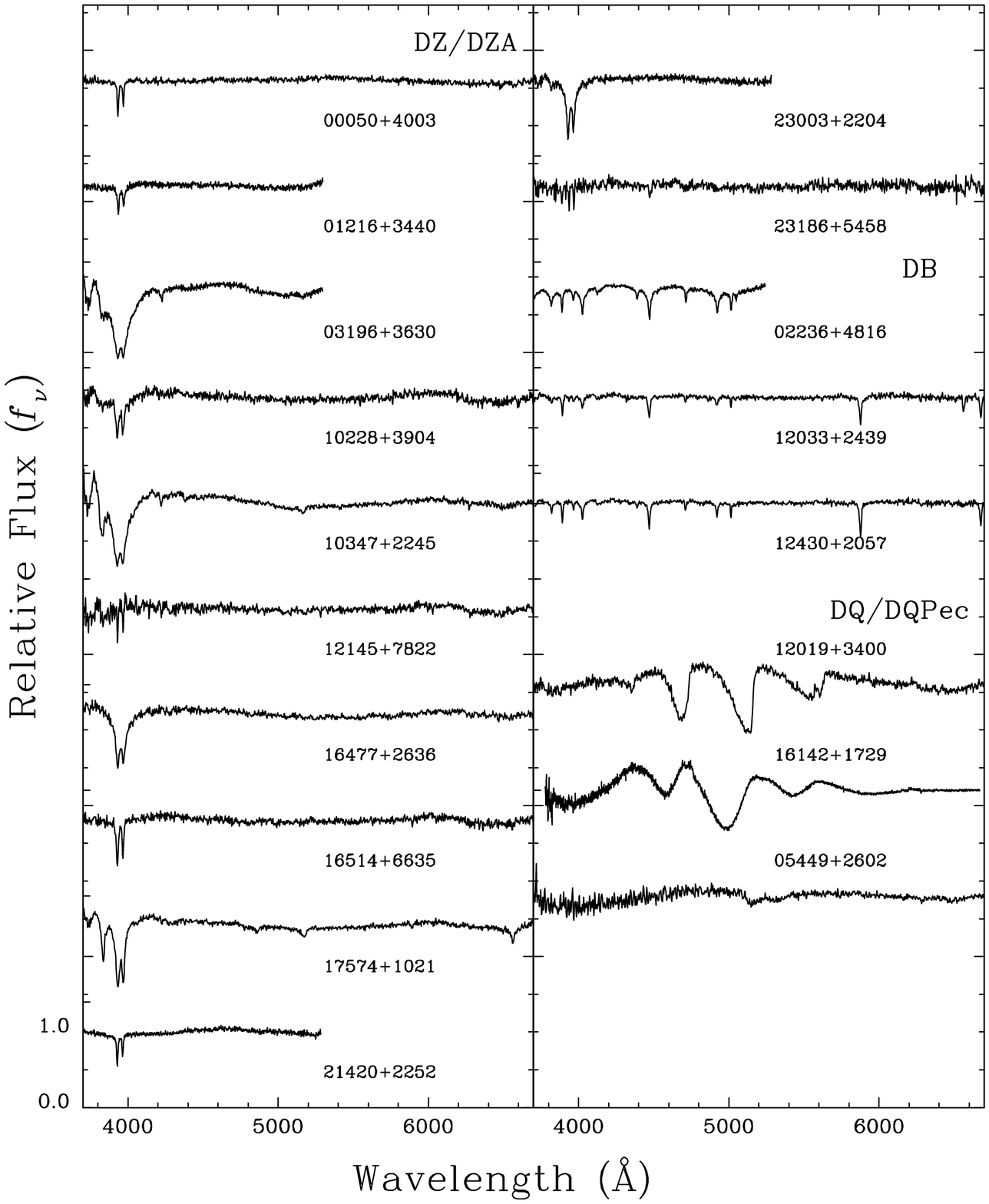}
\begin{flushright}
Figure \ref{DZDQDB}
\end{flushright}
\end{figure}

\clearpage

\begin{figure}[p]
\plotone{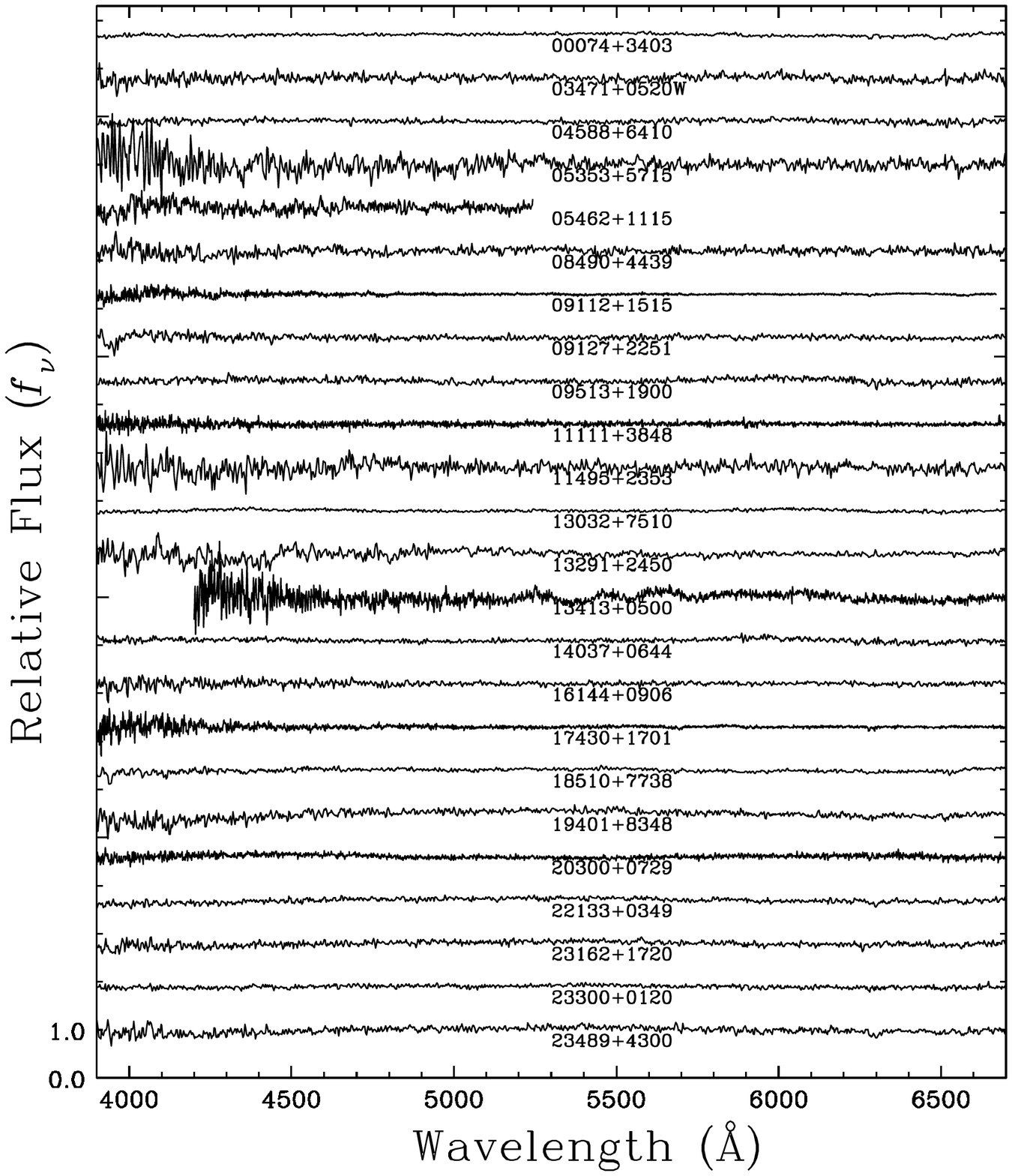}
\begin{flushright}
Figure \ref{DC}
\end{flushright}
\end{figure}

\clearpage

\begin{figure}[p]
\plotone{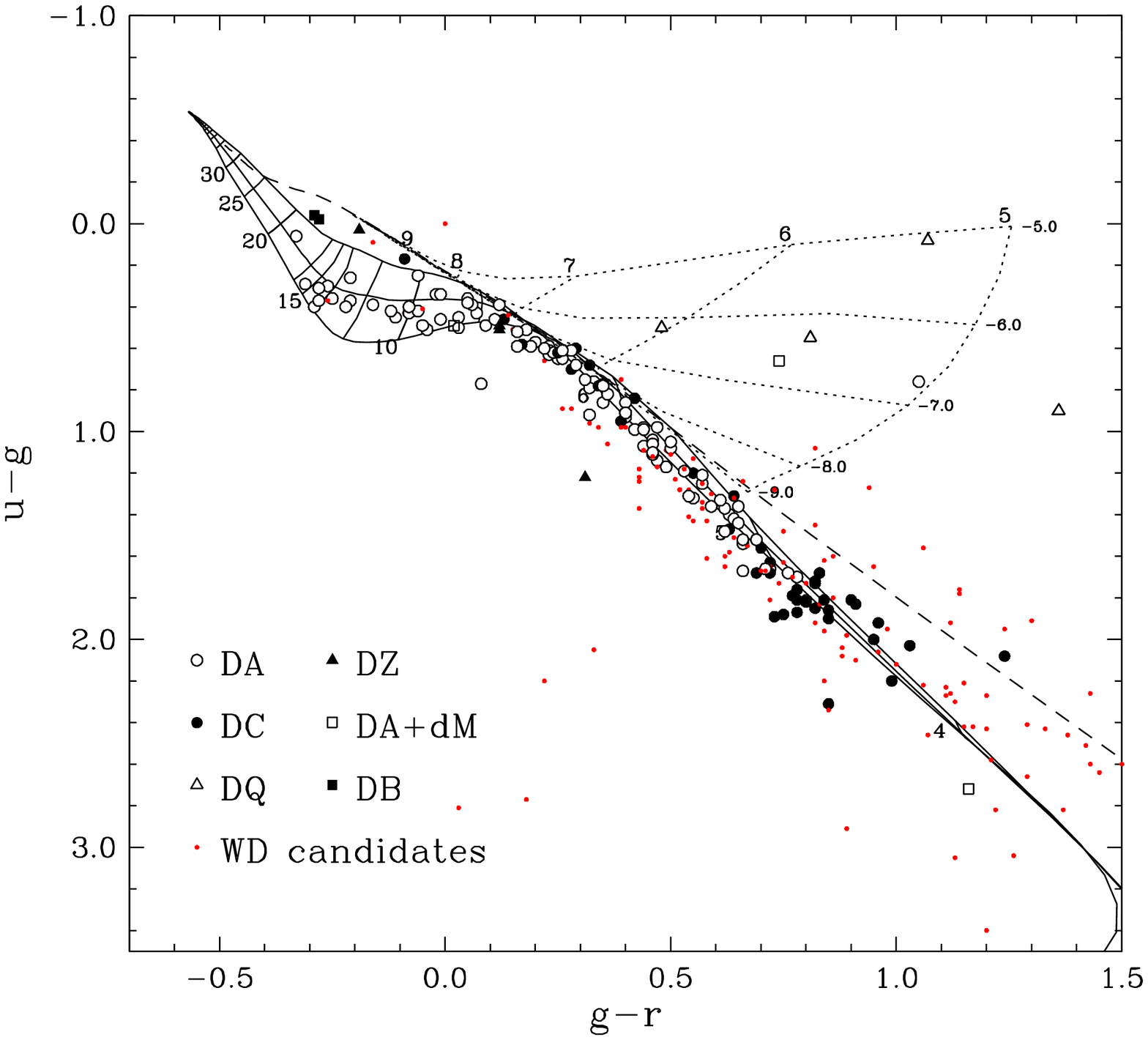}
\begin{flushright}
Figure \ref{color_gmr}
\end{flushright}
\end{figure}

\clearpage

\begin{figure}[p]
\plotone{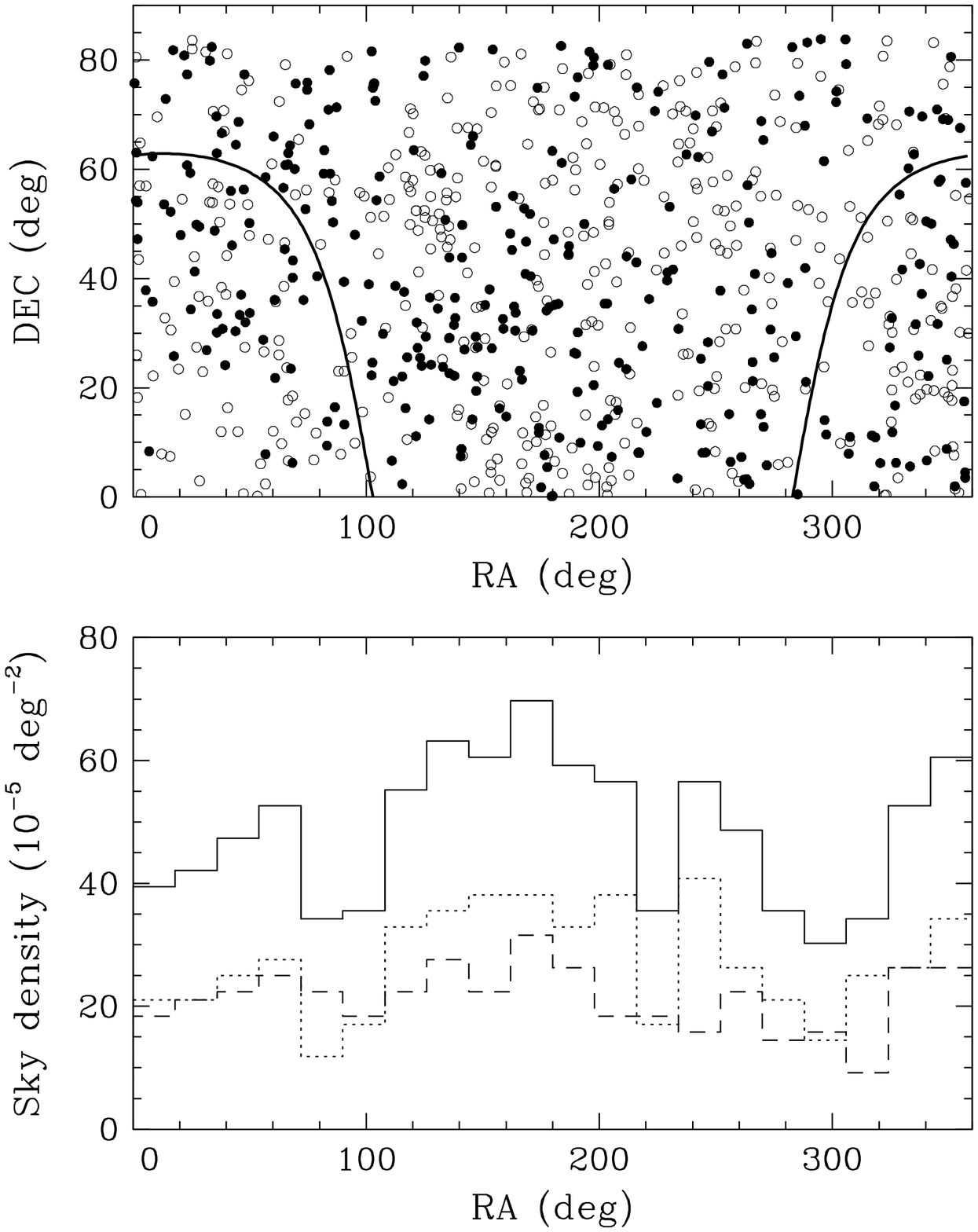}
\begin{flushright}
Figure \ref{plot_equal}
\end{flushright}
\end{figure}

\clearpage

\begin{figure}[p]
\plotone{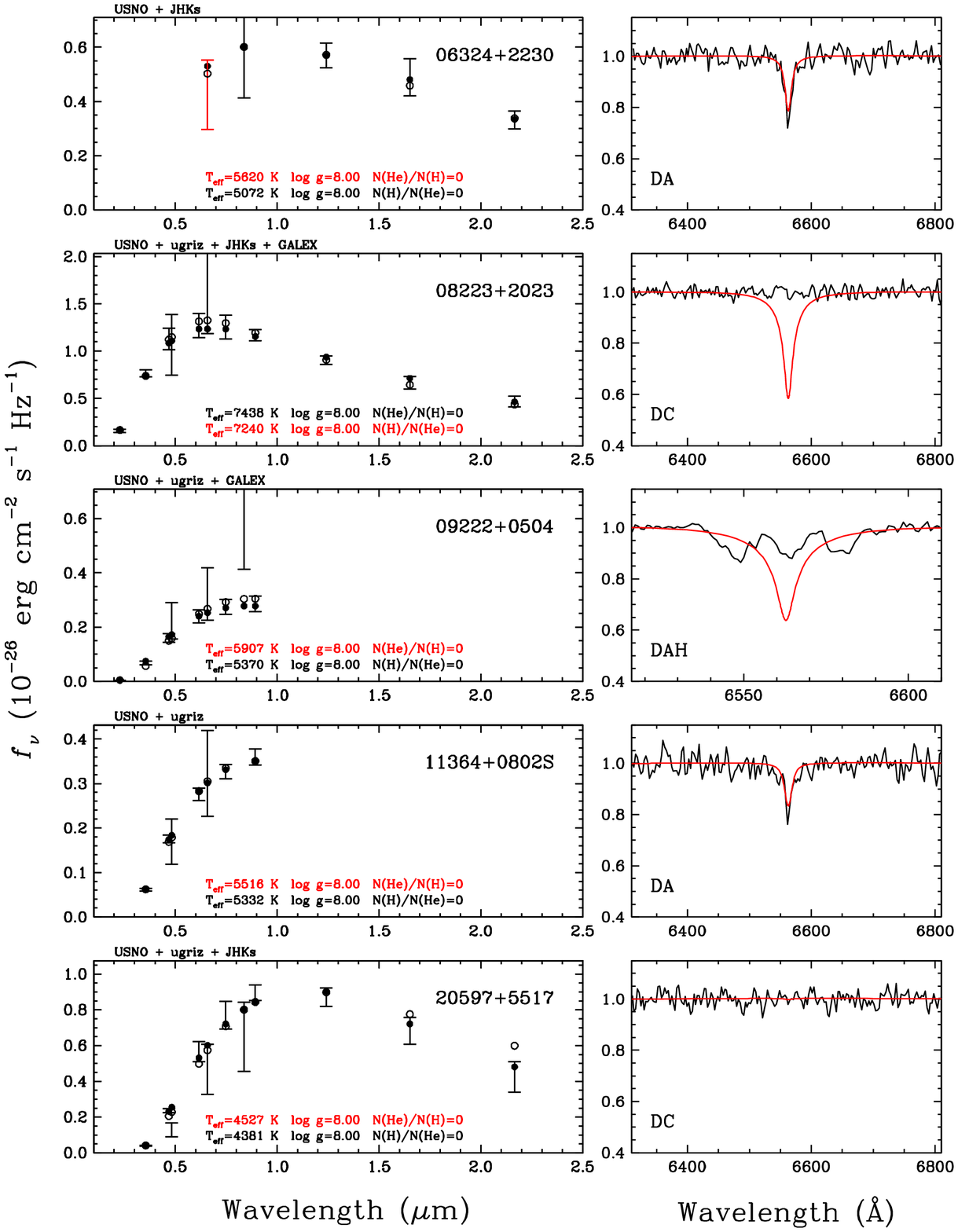}
\begin{flushright}
Figure \ref{ex_photo}
\end{flushright}
\end{figure}

\clearpage

\begin{figure}[p]
\plotone{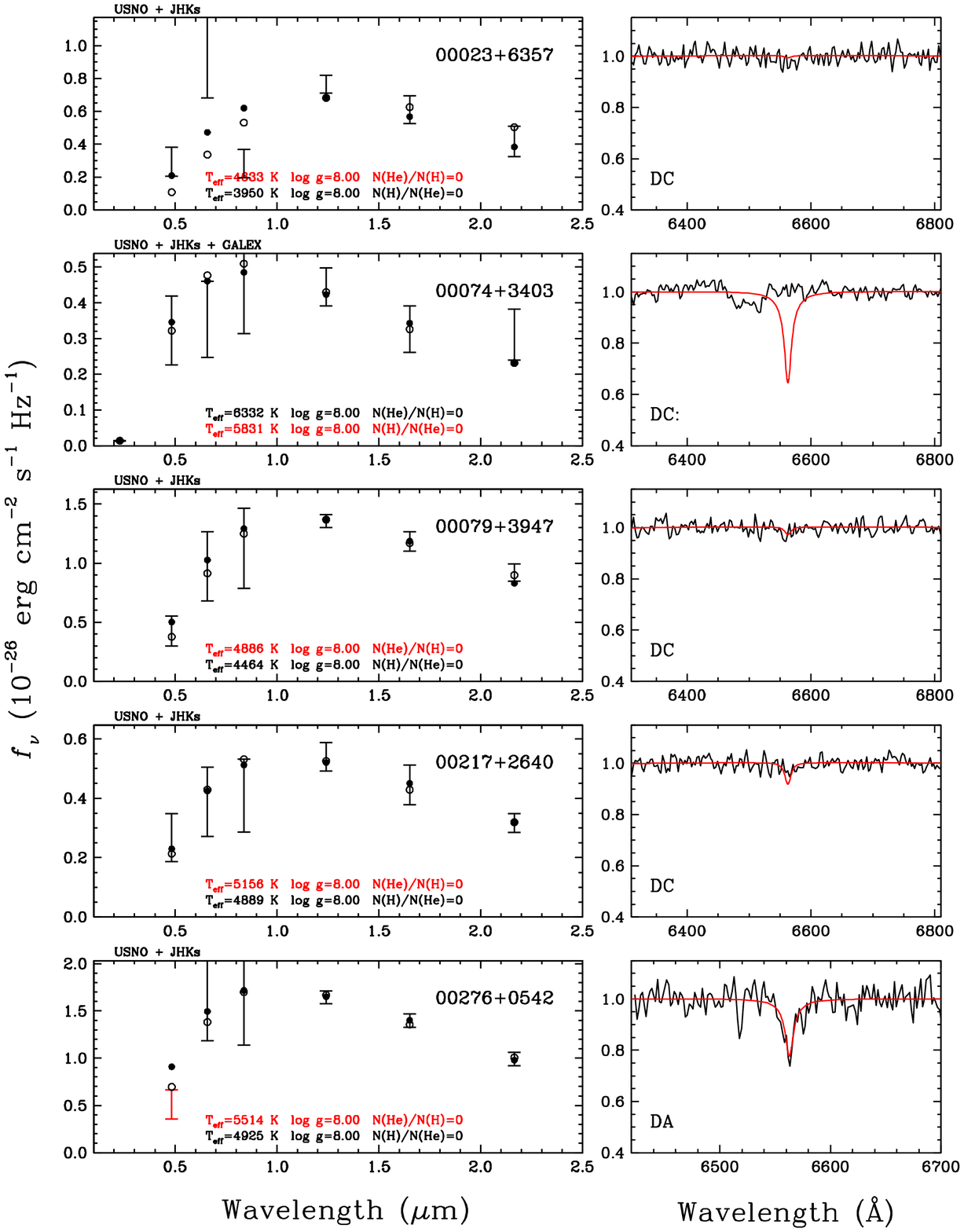}
\begin{flushright}
Figure \ref{photoDADC}a
\end{flushright}
\end{figure}

\clearpage

\begin{figure}[p]
\plotone{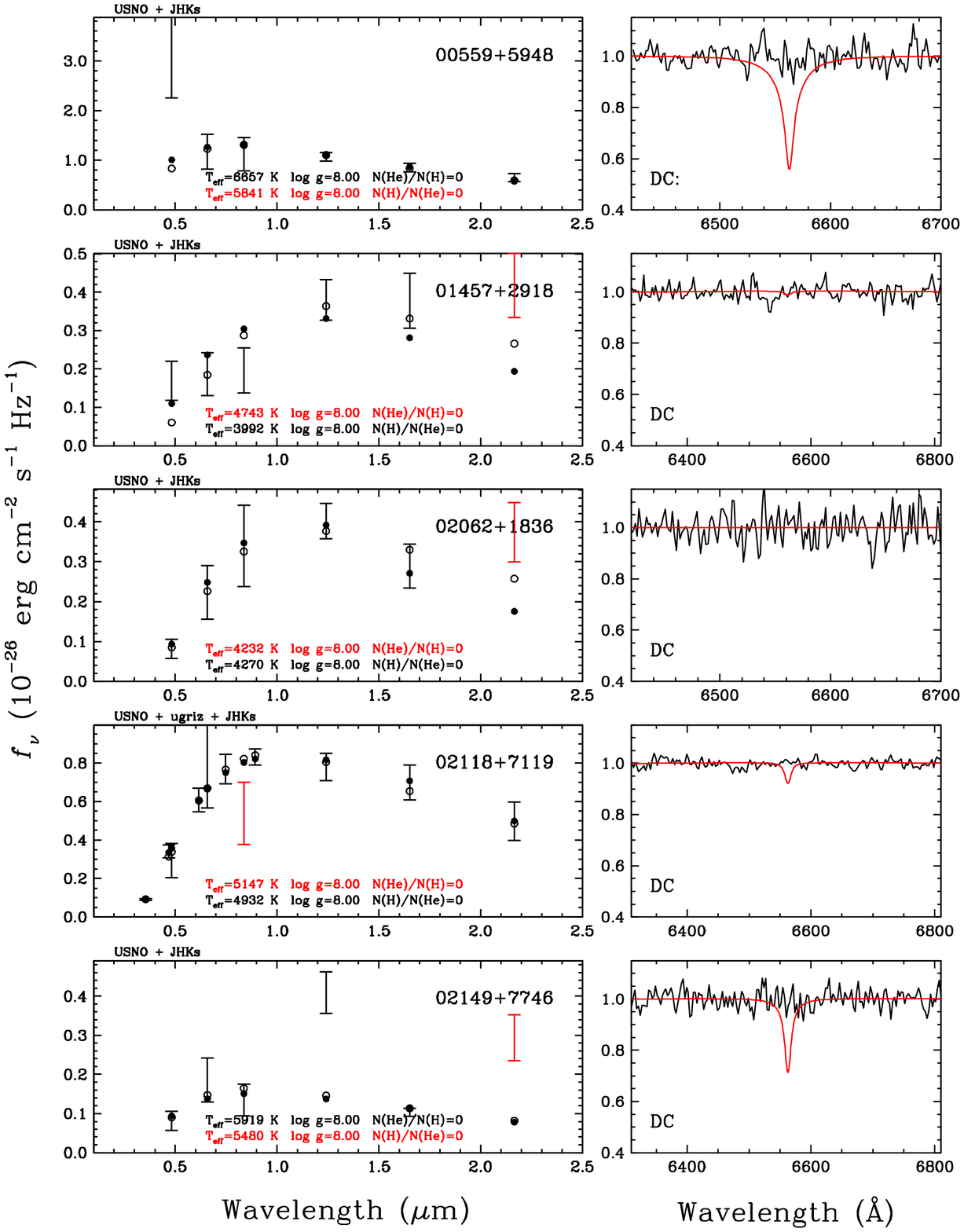}
\begin{flushright}
Figure \ref{photoDADC}b
\end{flushright}
\end{figure}

\clearpage

\begin{figure}[p]
\plotone{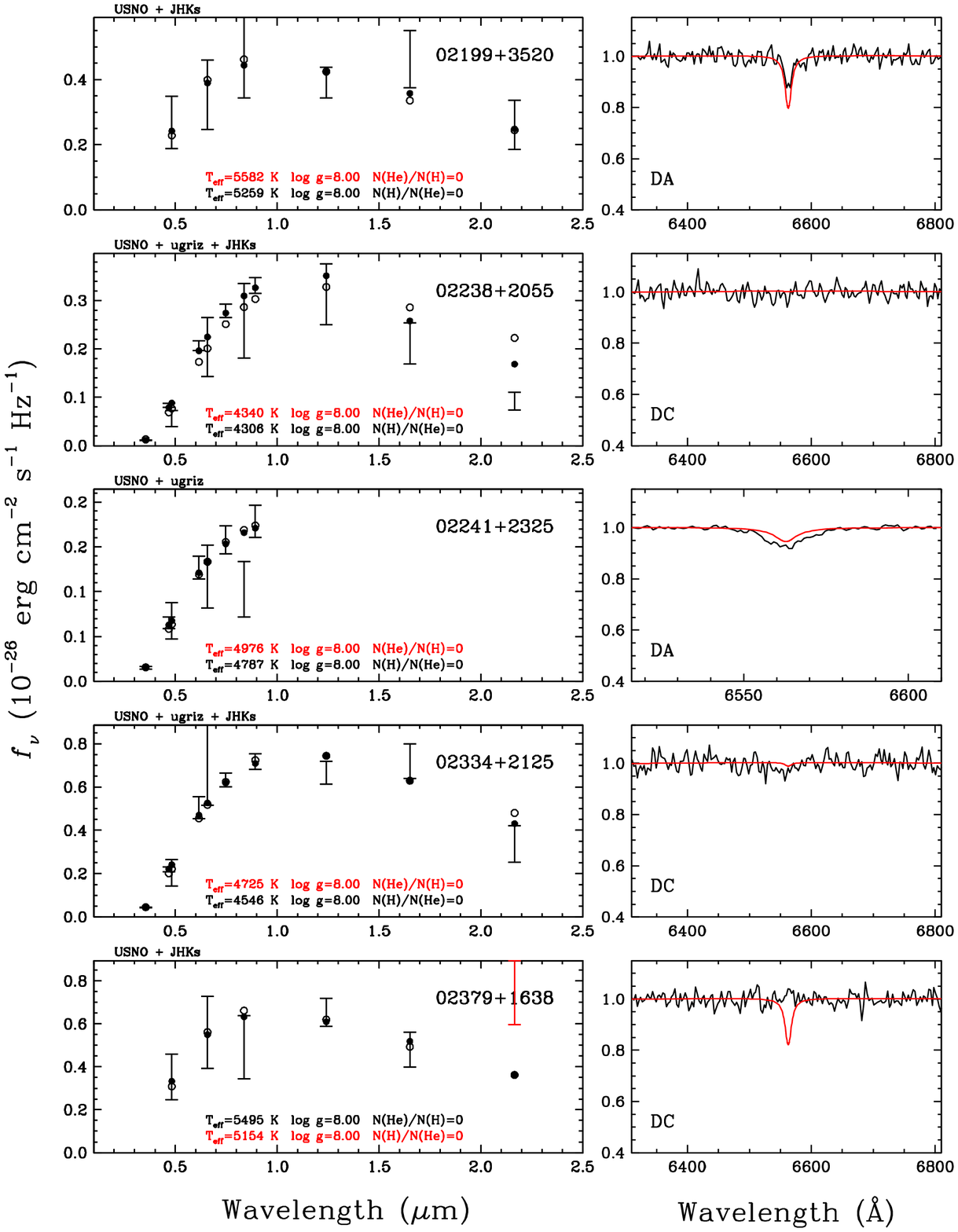}
\begin{flushright}
Figure \ref{photoDADC}c
\end{flushright}
\end{figure}

\clearpage

\begin{figure}[p]
\plotone{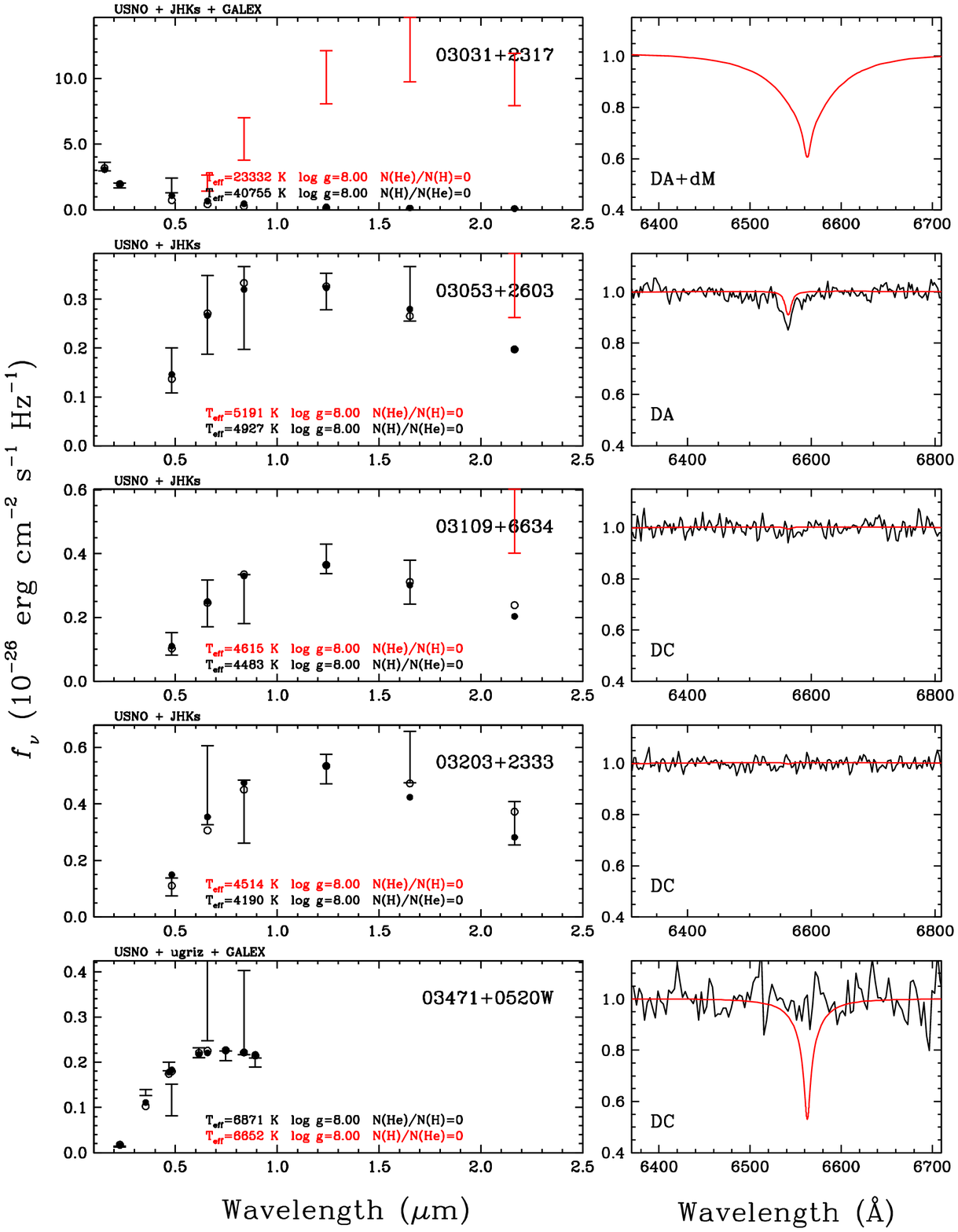}
\begin{flushright}
Figure \ref{photoDADC}d
\end{flushright}
\end{figure}

\clearpage

\begin{figure}[p]
\plotone{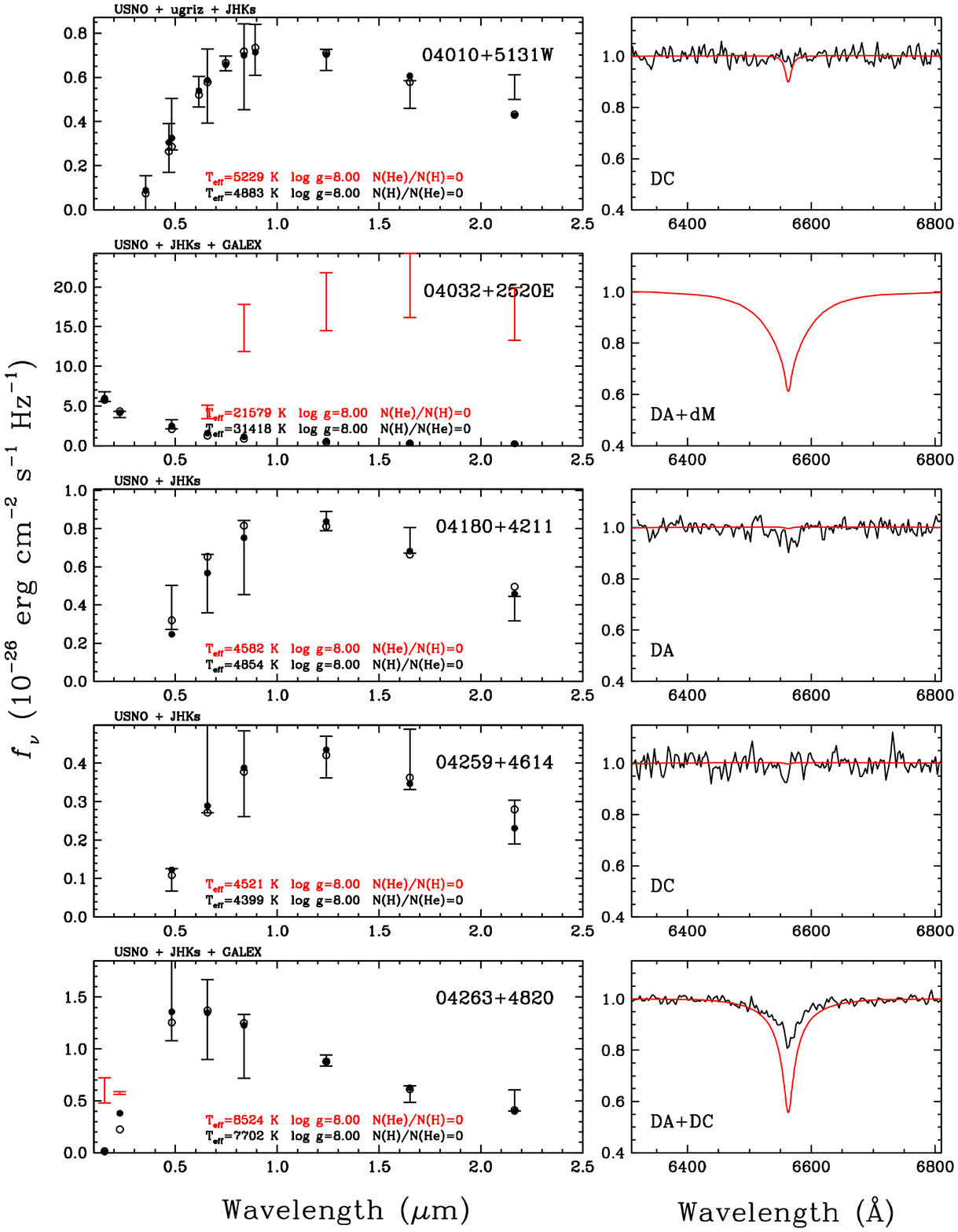}
\begin{flushright}
Figure \ref{photoDADC}e
\end{flushright}
\end{figure}

\clearpage

\begin{figure}[p]
\plotone{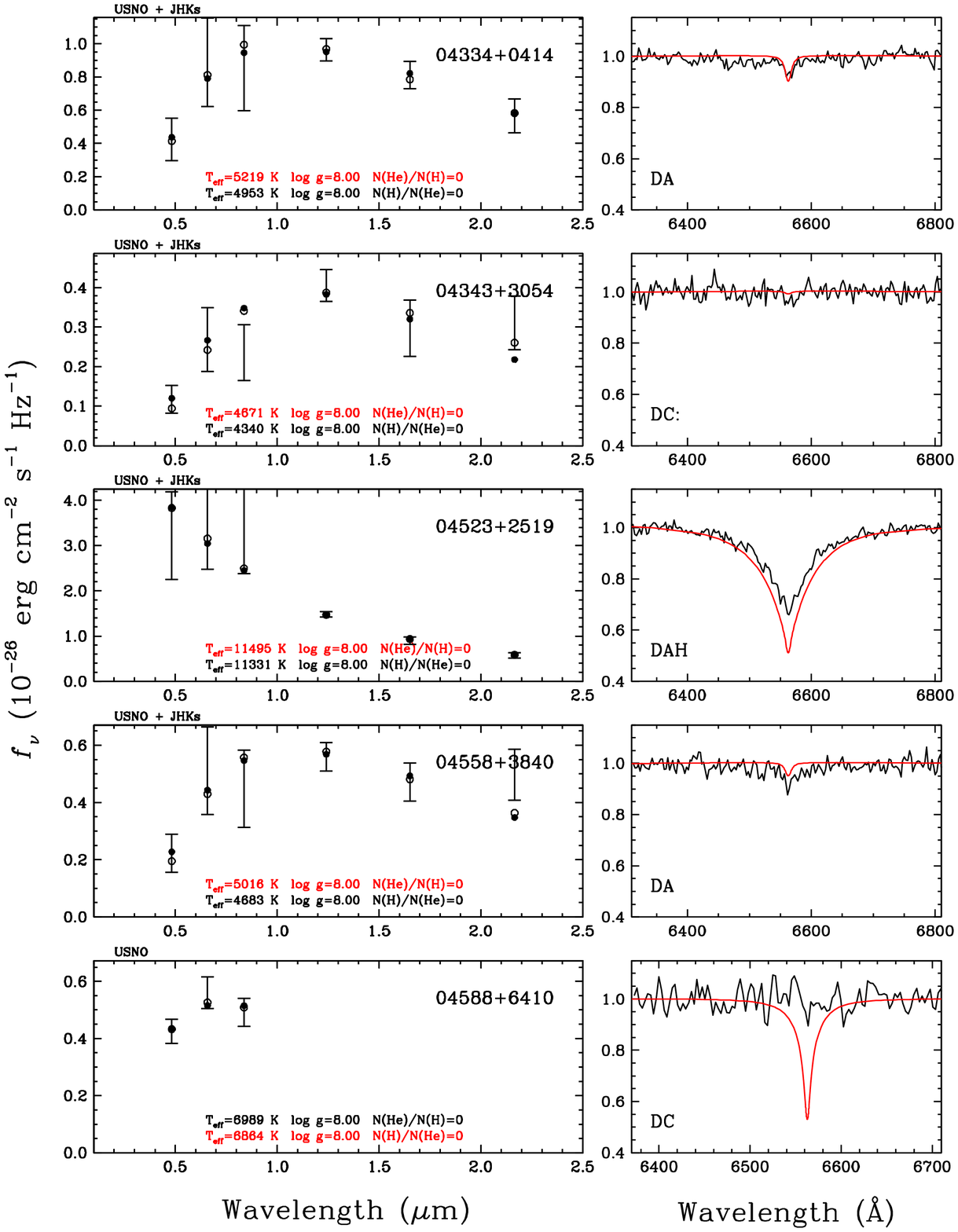}
\begin{flushright}
Figure \ref{photoDADC}f
\end{flushright}
\end{figure}

\clearpage

\begin{figure}[p]
\plotone{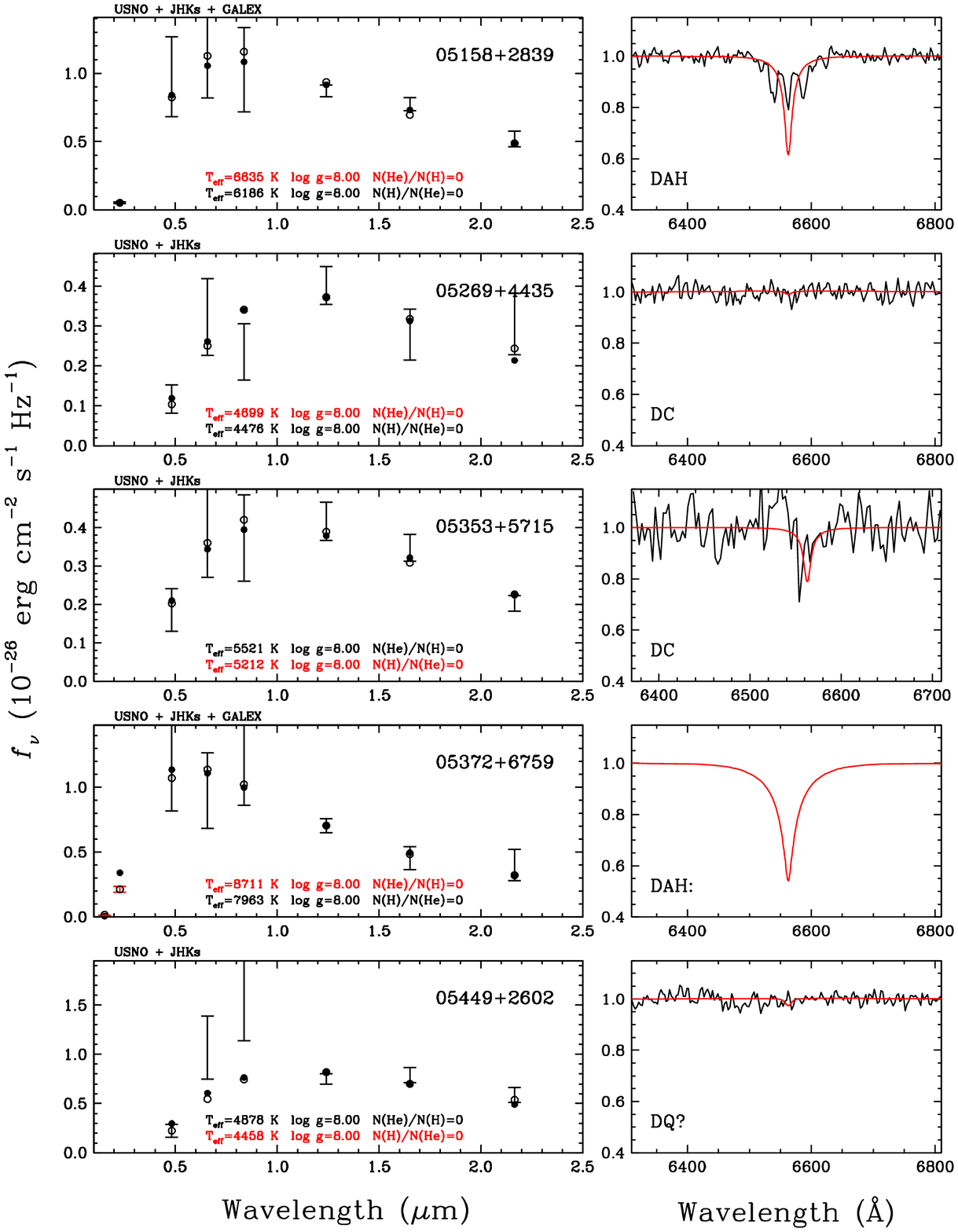}
\begin{flushright}
Figure \ref{photoDADC}g
\end{flushright}
\end{figure}

\clearpage

\begin{figure}[p]
\plotone{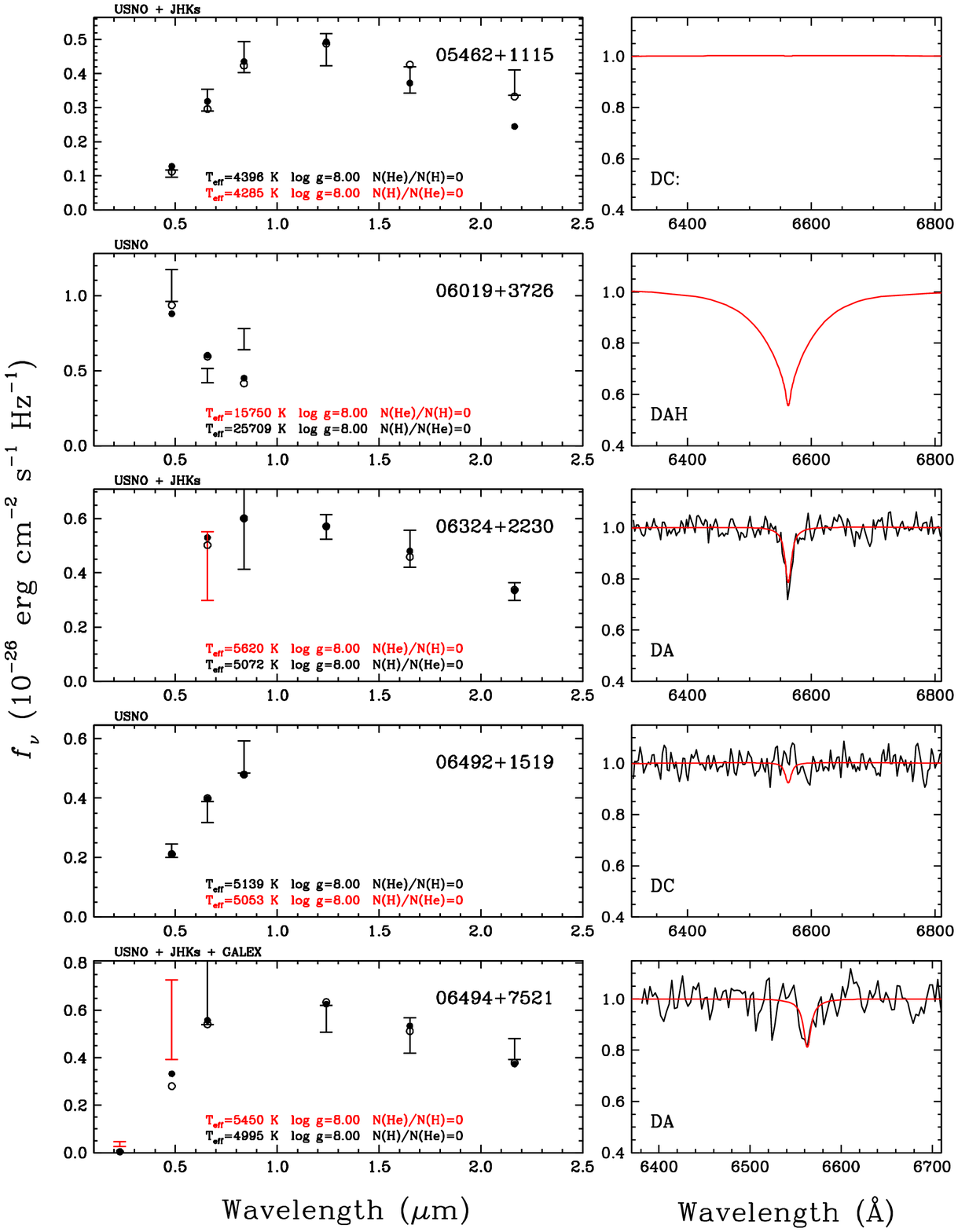}
\begin{flushright}
Figure \ref{photoDADC}h
\end{flushright}
\end{figure}

\clearpage

\begin{figure}[p]
\plotone{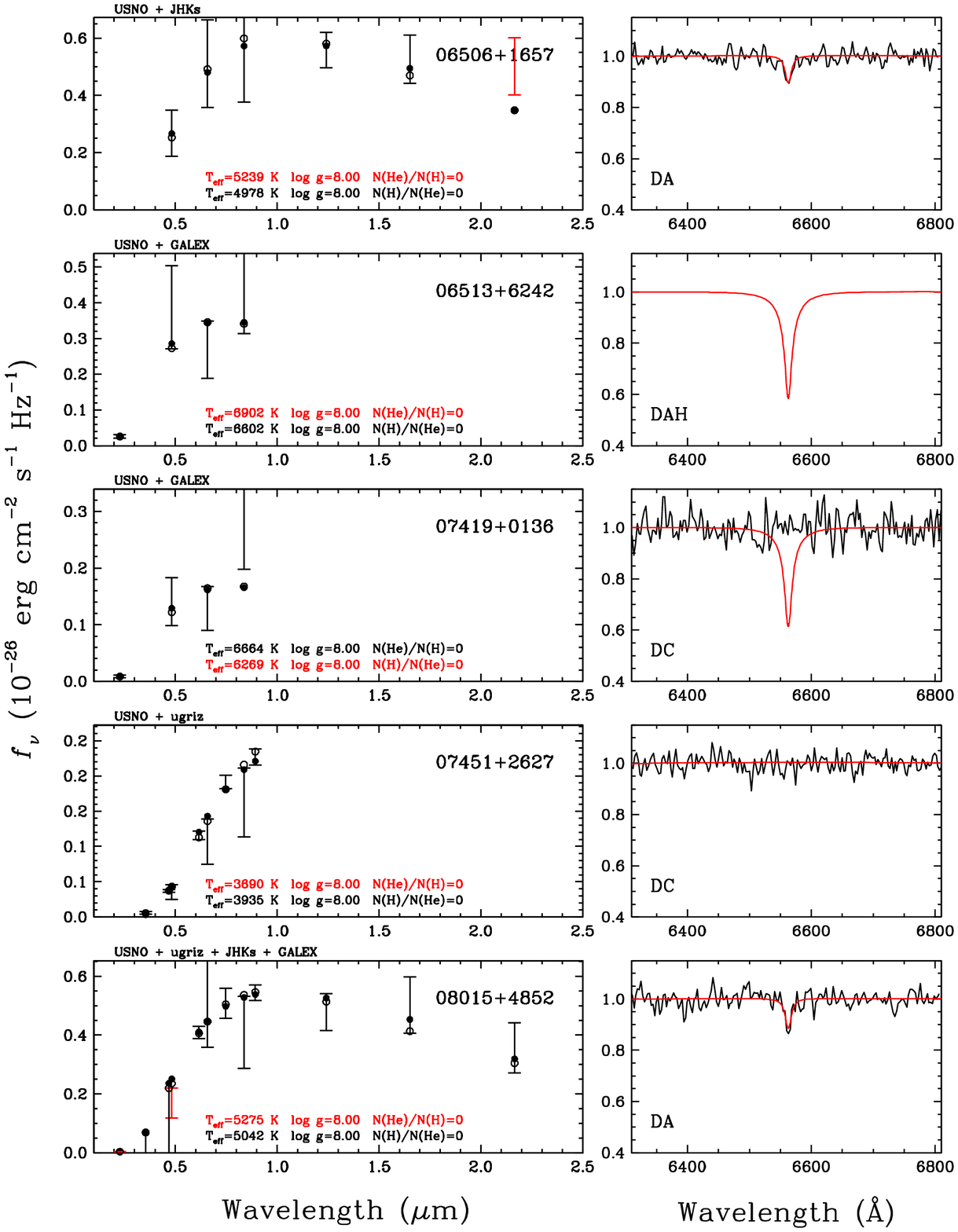}
\begin{flushright}
Figure \ref{photoDADC}i
\end{flushright}
\end{figure}

\clearpage

\begin{figure}[p]
\plotone{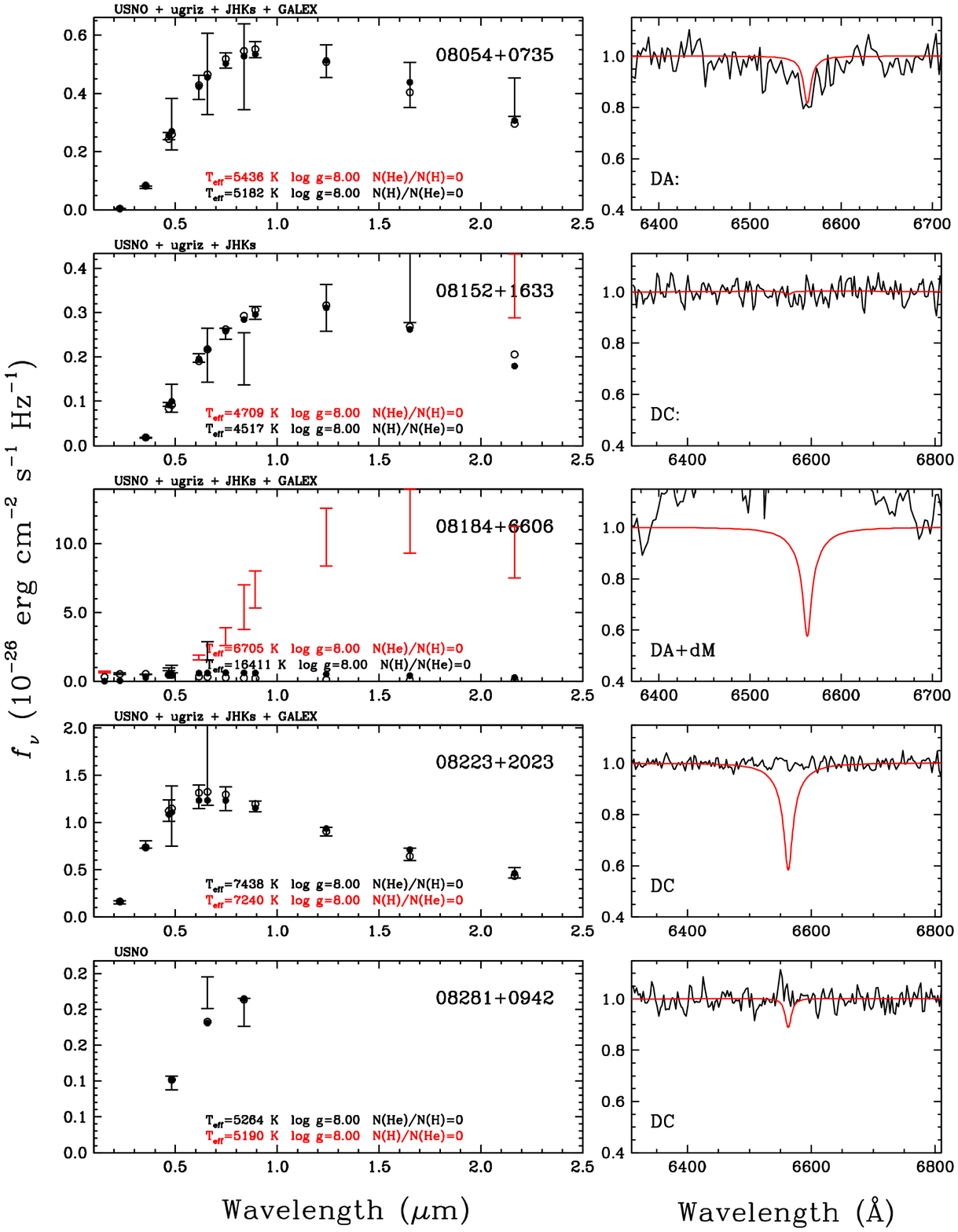}
\begin{flushright}
Figure \ref{photoDADC}j
\end{flushright}
\end{figure}

\clearpage

\begin{figure}[p]
\plotone{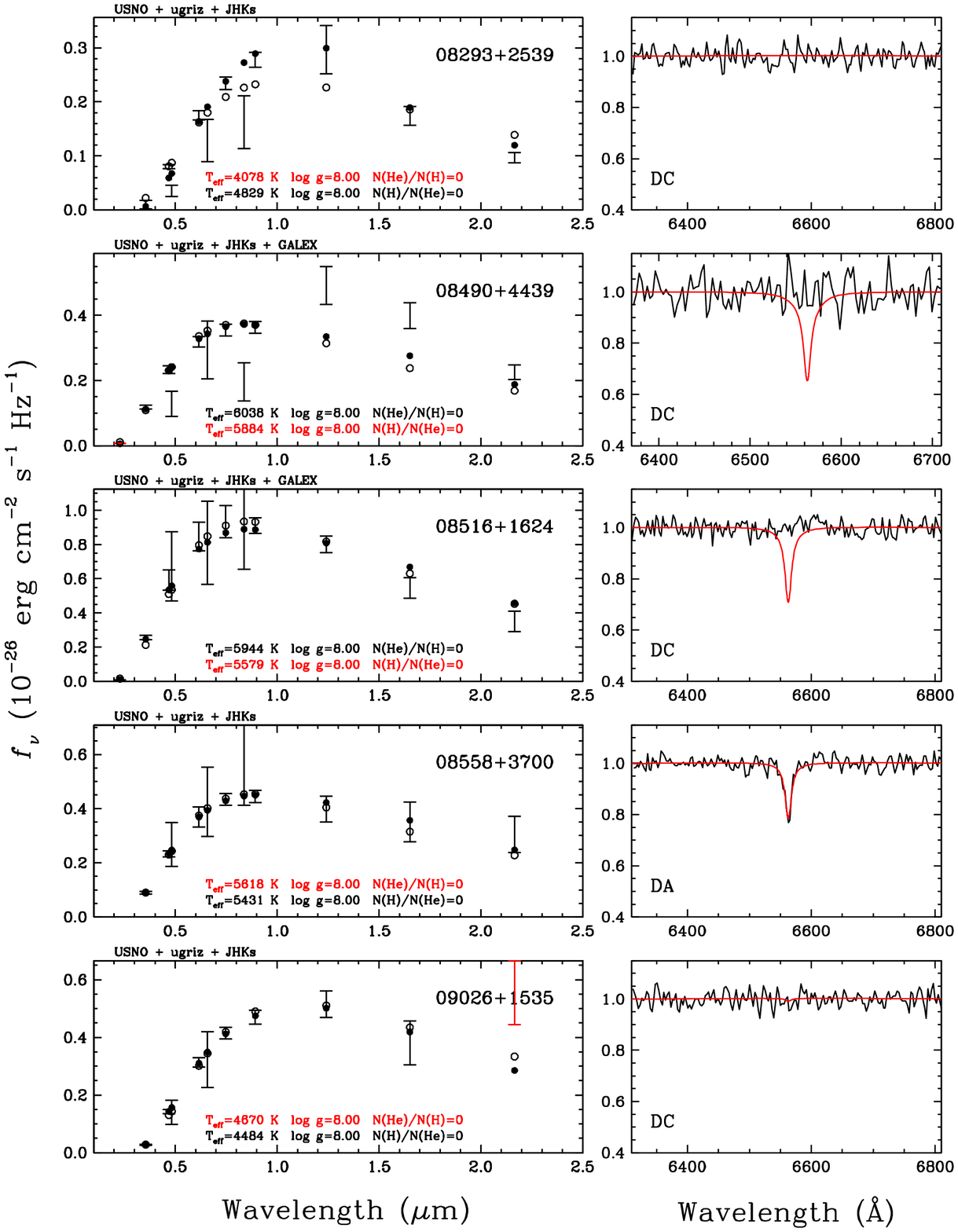}
\begin{flushright}
Figure \ref{photoDADC}k
\end{flushright}
\end{figure}

\clearpage

\begin{figure}[p]
\plotone{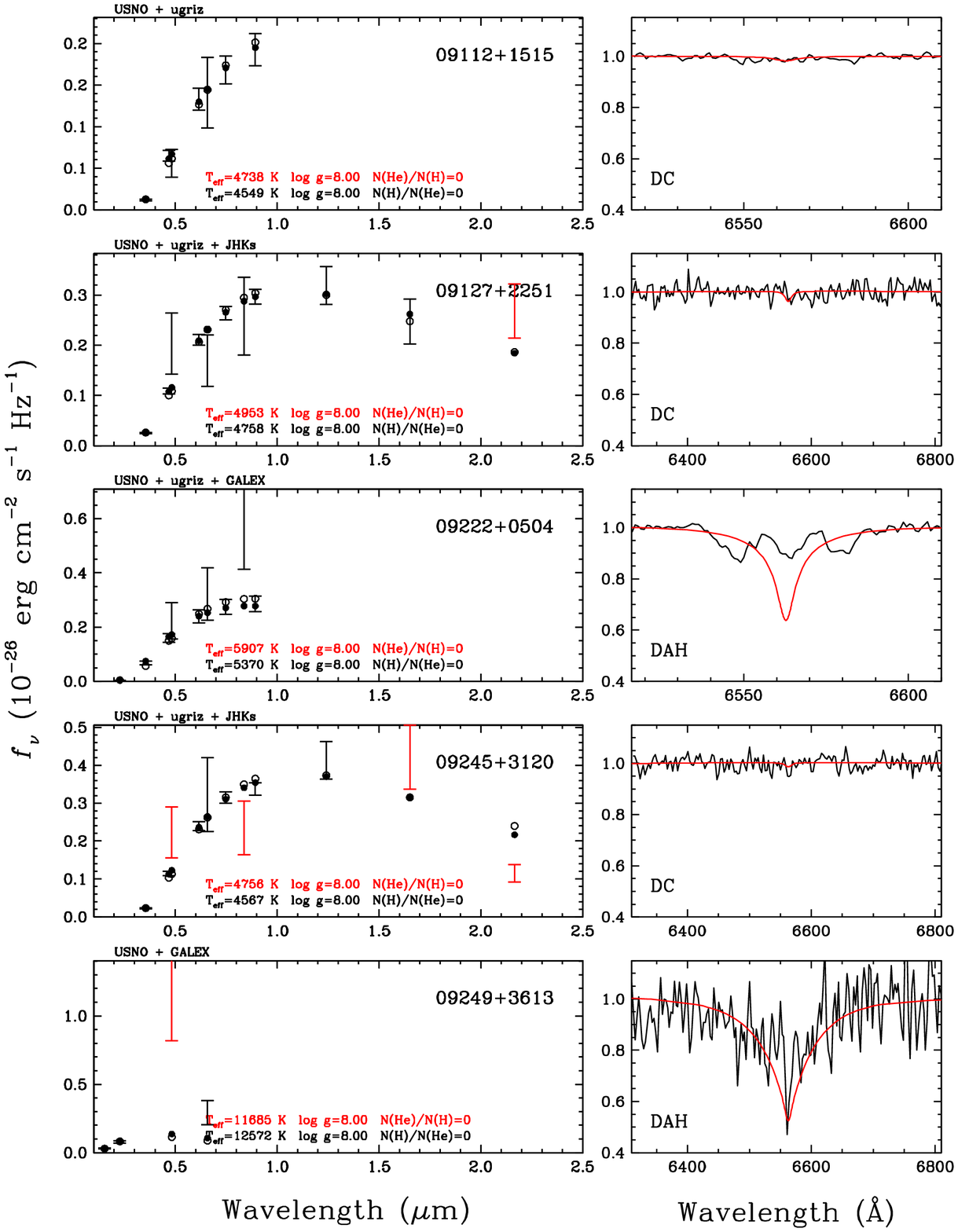}
\begin{flushright}
Figure \ref{photoDADC}l
\end{flushright}
\end{figure}

\clearpage

\begin{figure}[p]
\plotone{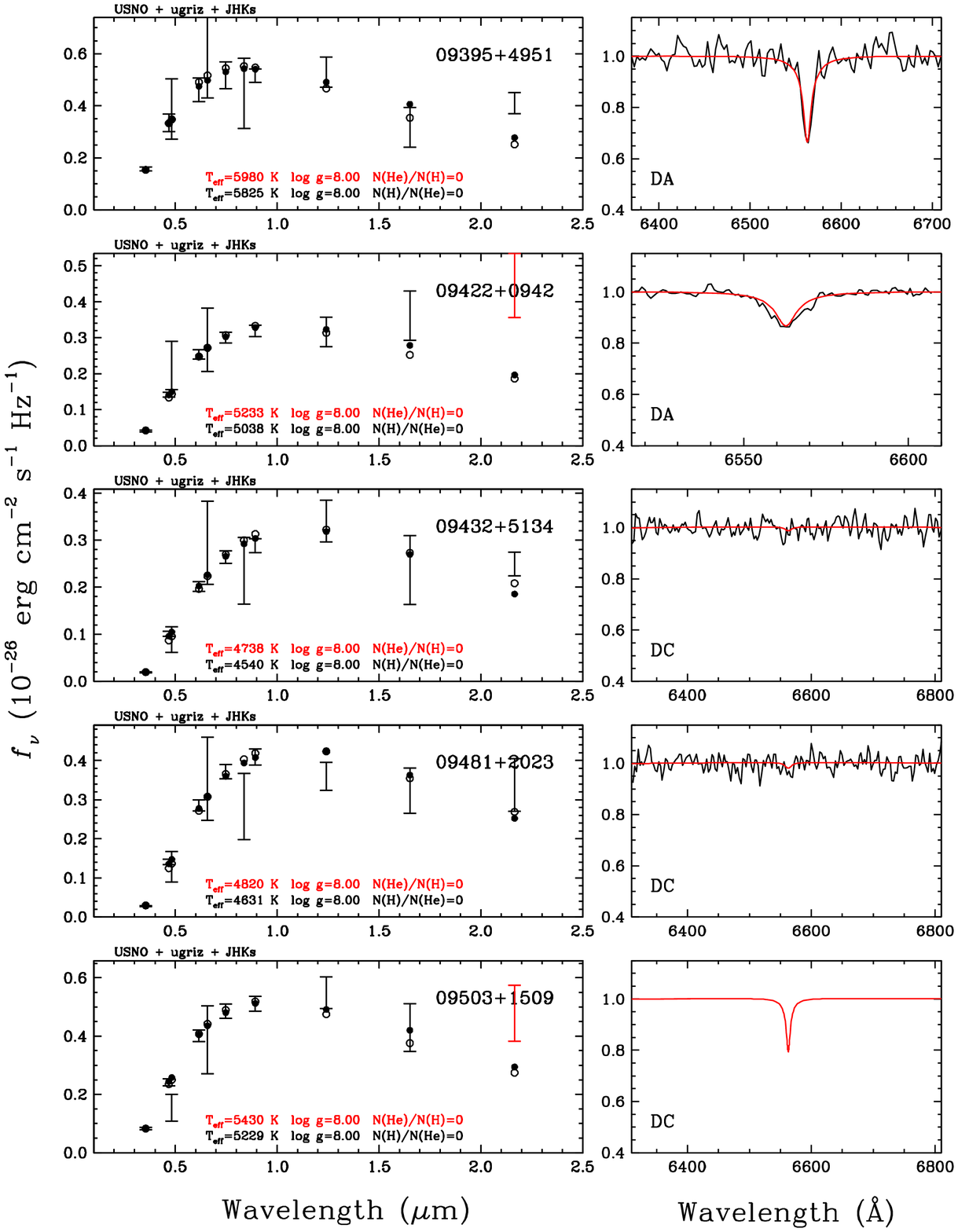}
\begin{flushright}
Figure \ref{photoDADC}m
\end{flushright}
\end{figure}

\clearpage

\begin{figure}[p]
\plotone{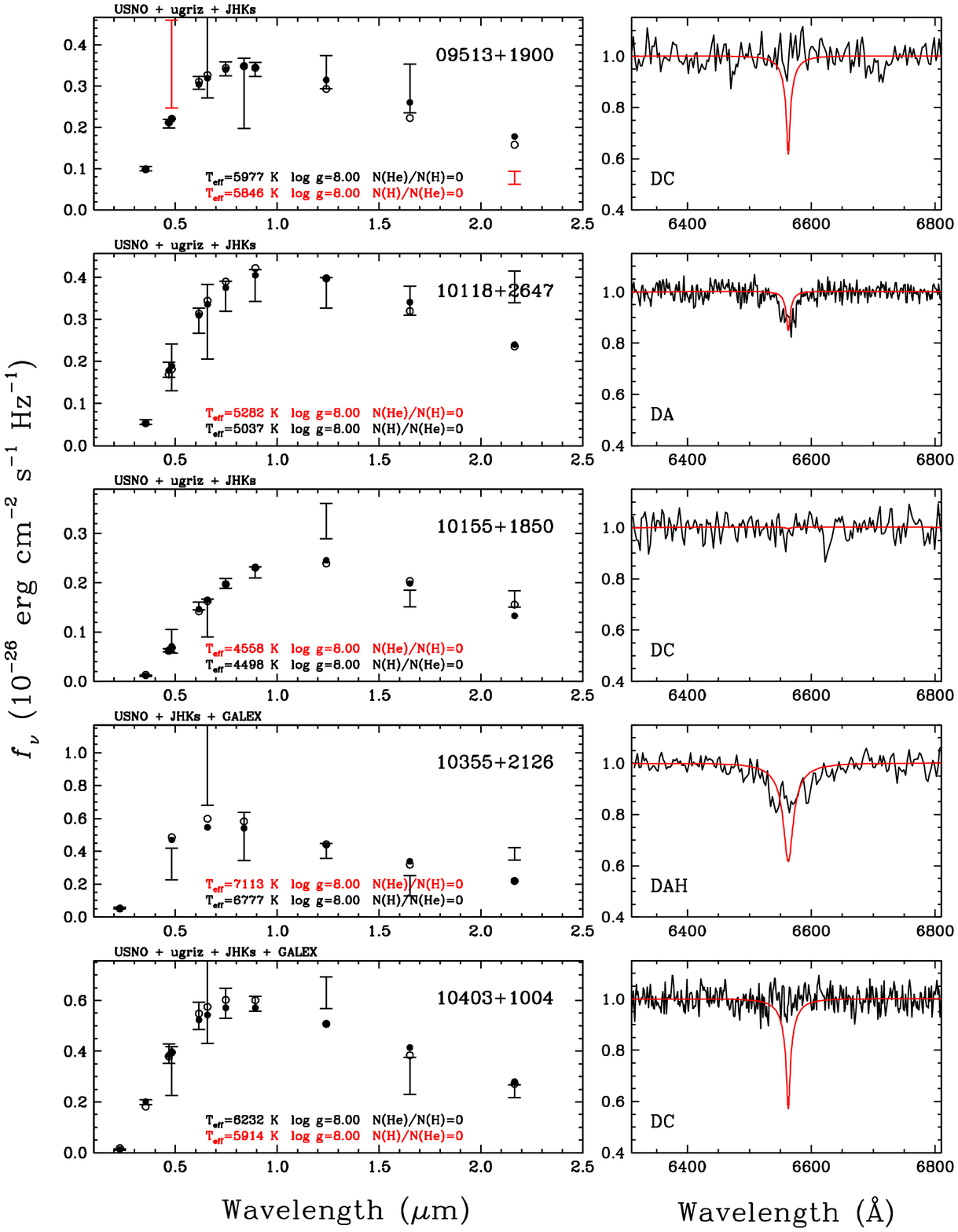}
\begin{flushright}
Figure \ref{photoDADC}n
\end{flushright}
\end{figure}

\clearpage

\begin{figure}[p]
\plotone{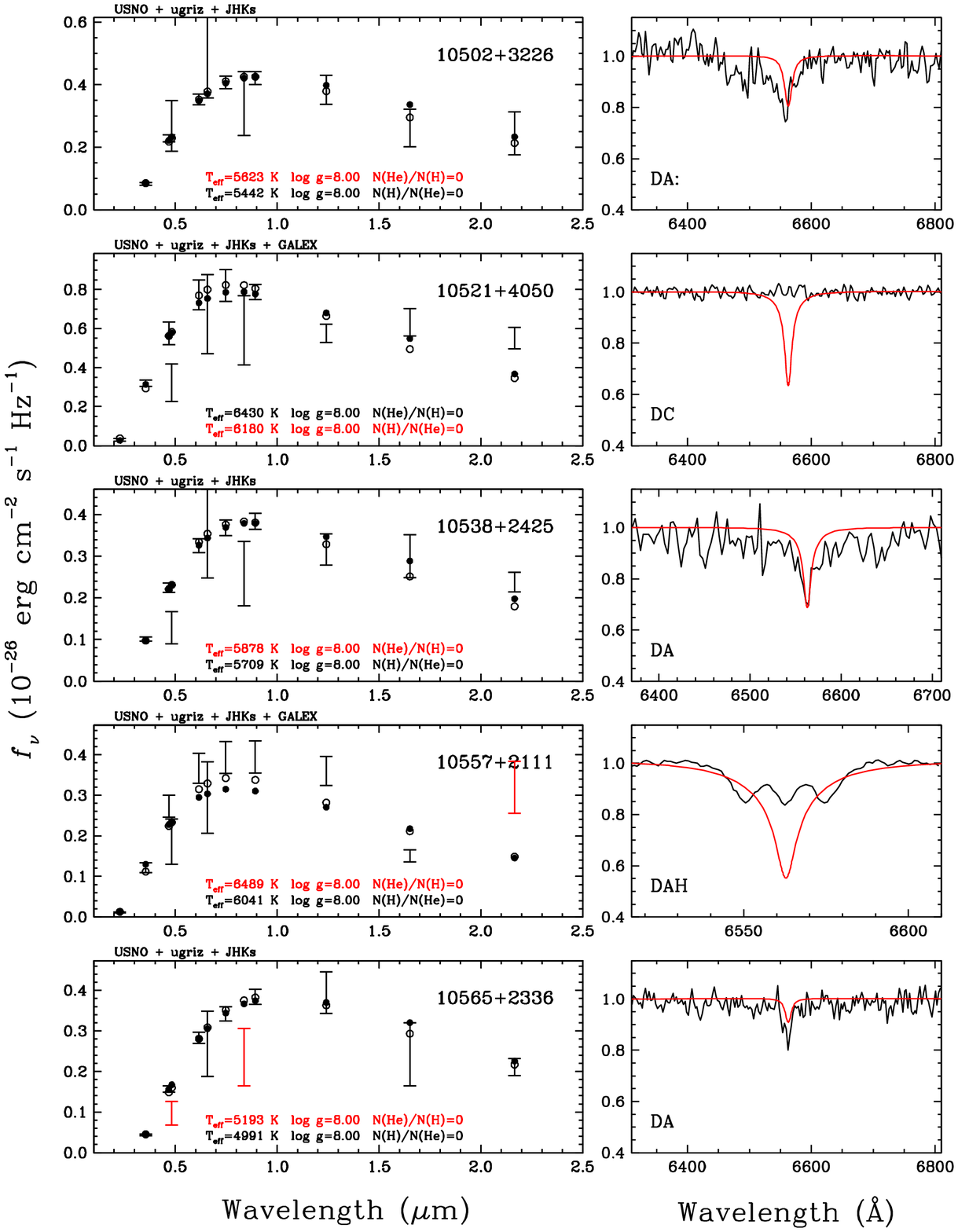}
\begin{flushright}
Figure \ref{photoDADC}o
\end{flushright}
\end{figure}

\clearpage

\begin{figure}[p]
\plotone{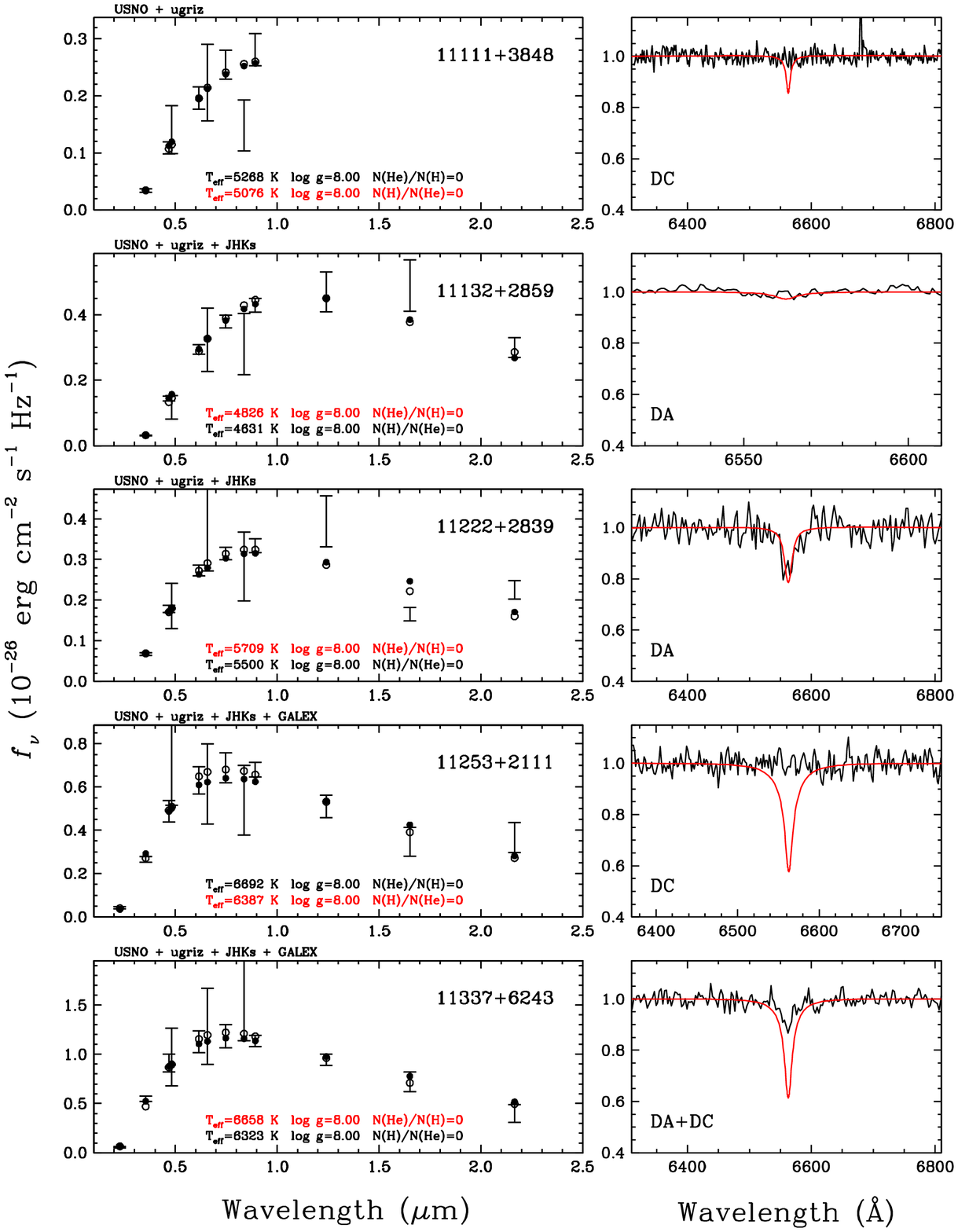}
\begin{flushright}
Figure \ref{photoDADC}p
\end{flushright}
\end{figure}

\clearpage

\begin{figure}[p]
\plotone{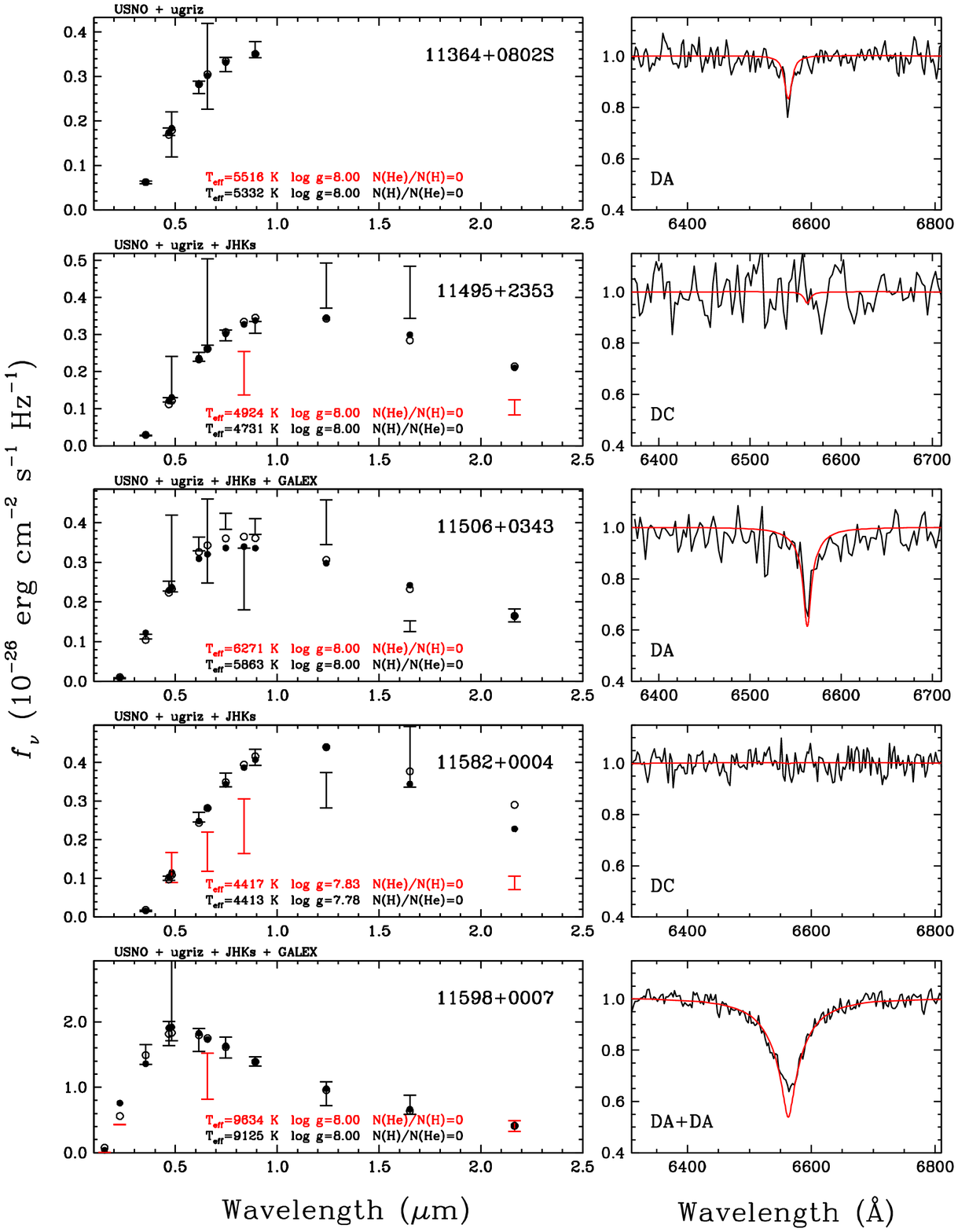}
\begin{flushright}
Figure \ref{photoDADC}q
\end{flushright}
\end{figure}

\clearpage

\begin{figure}[p]
\plotone{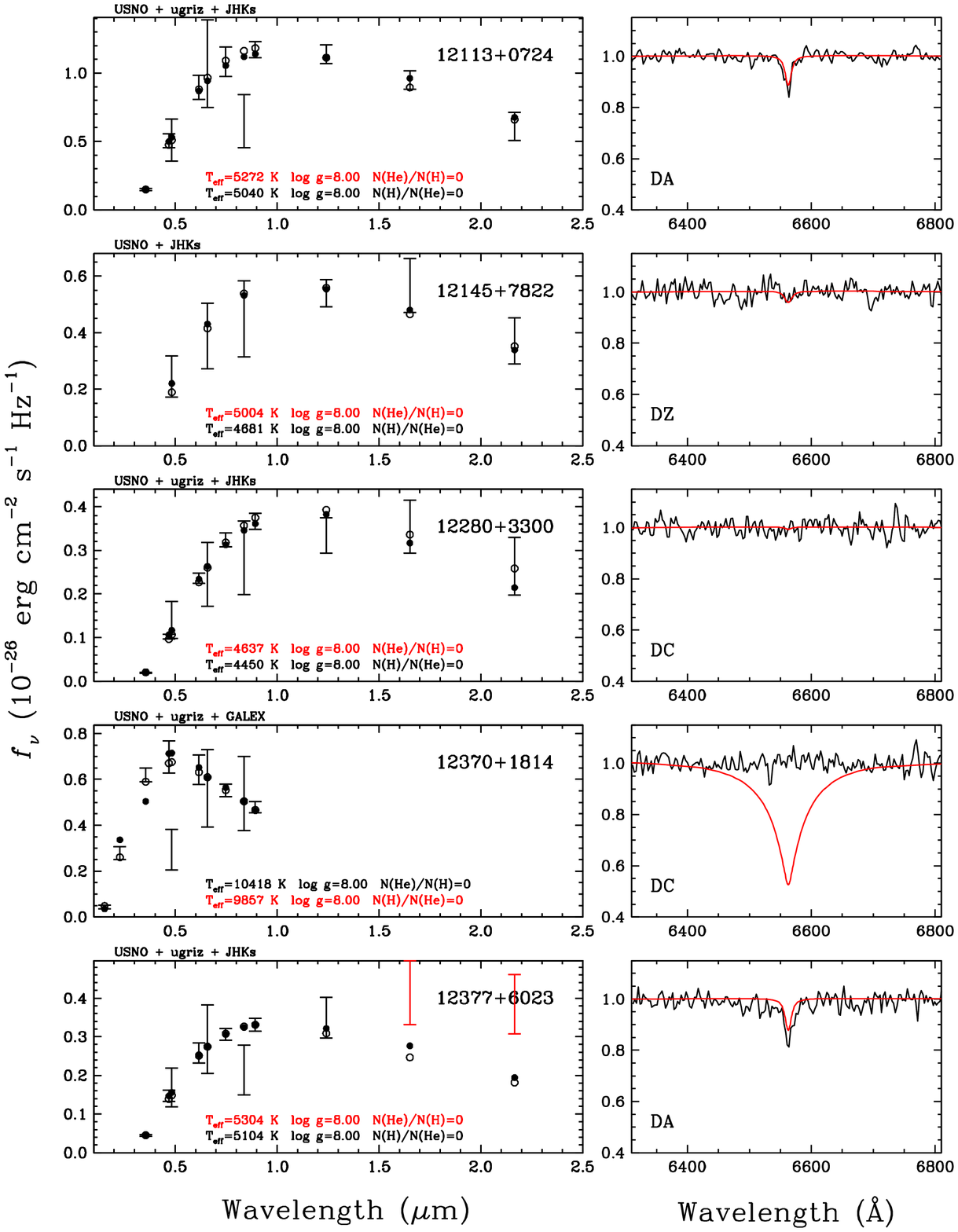}
\begin{flushright}
Figure \ref{photoDADC}r
\end{flushright}
\end{figure}

\clearpage

\begin{figure}[p]
\plotone{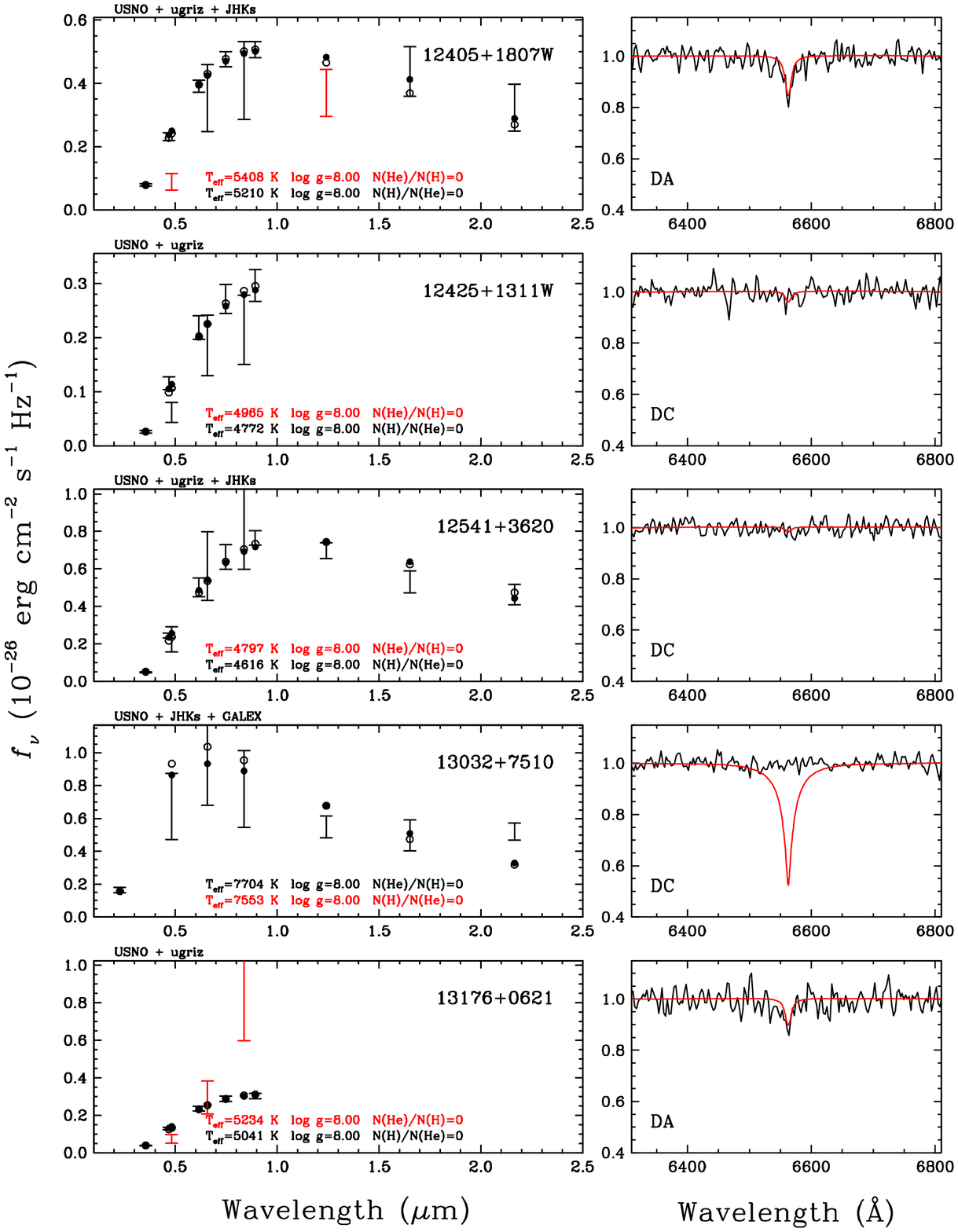}
\begin{flushright}
Figure \ref{photoDADC}s
\end{flushright}
\end{figure}

\clearpage

\begin{figure}[p]
\plotone{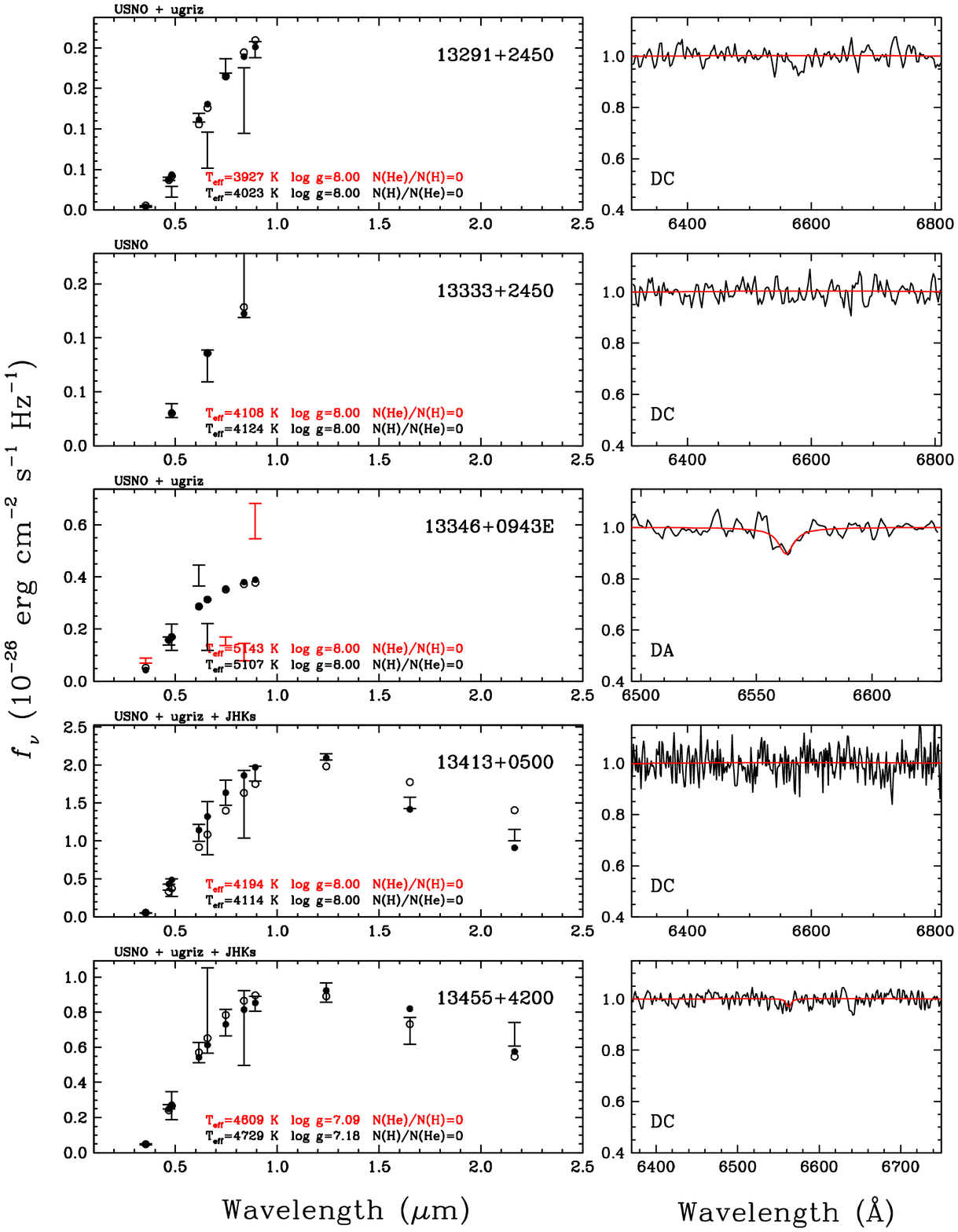}
\begin{flushright}
Figure \ref{photoDADC}t
\end{flushright}
\end{figure}

\clearpage

\begin{figure}[p]
\plotone{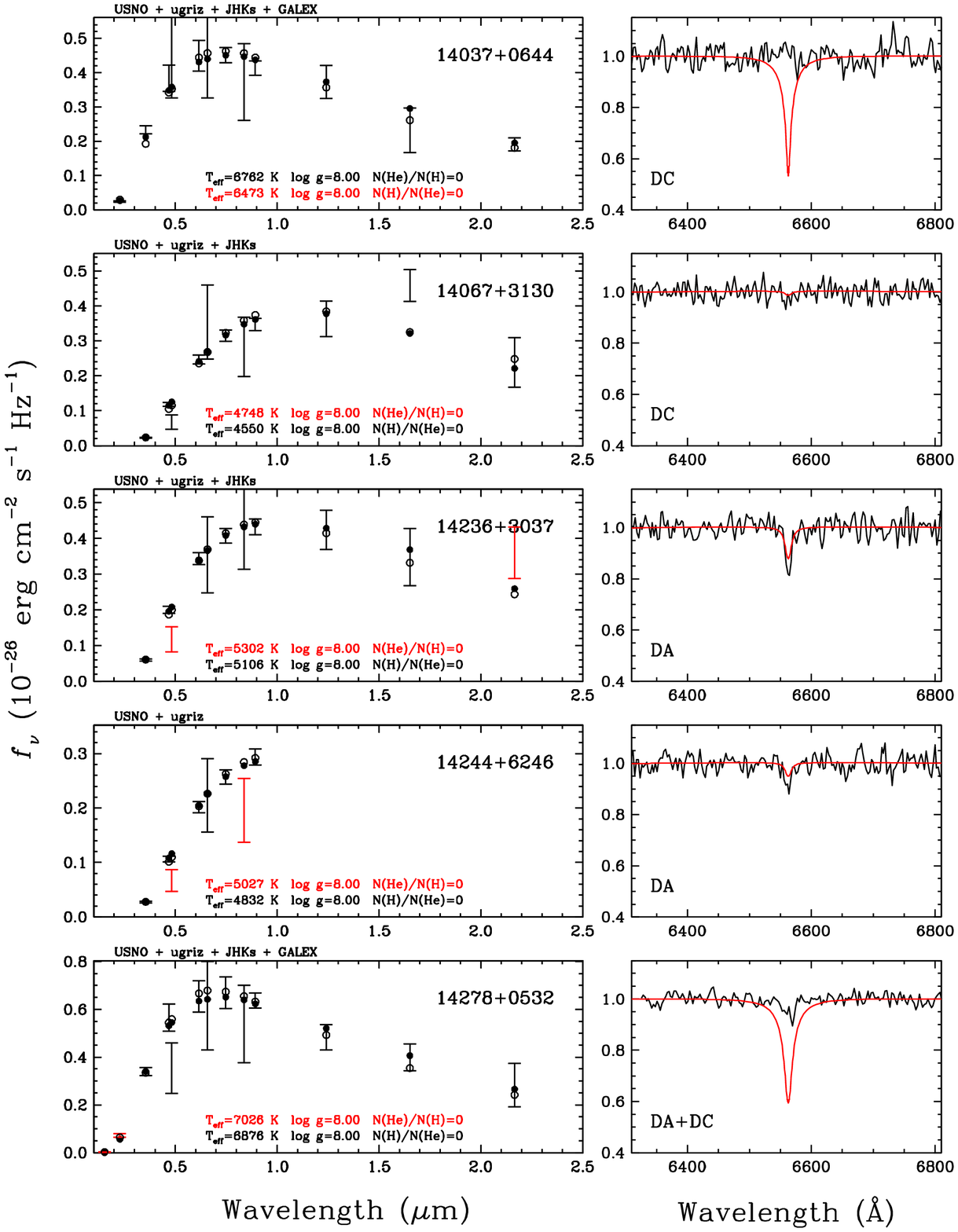}
\begin{flushright}
Figure \ref{photoDADC}u
\end{flushright}
\end{figure}

\clearpage

\begin{figure}[p]
\plotone{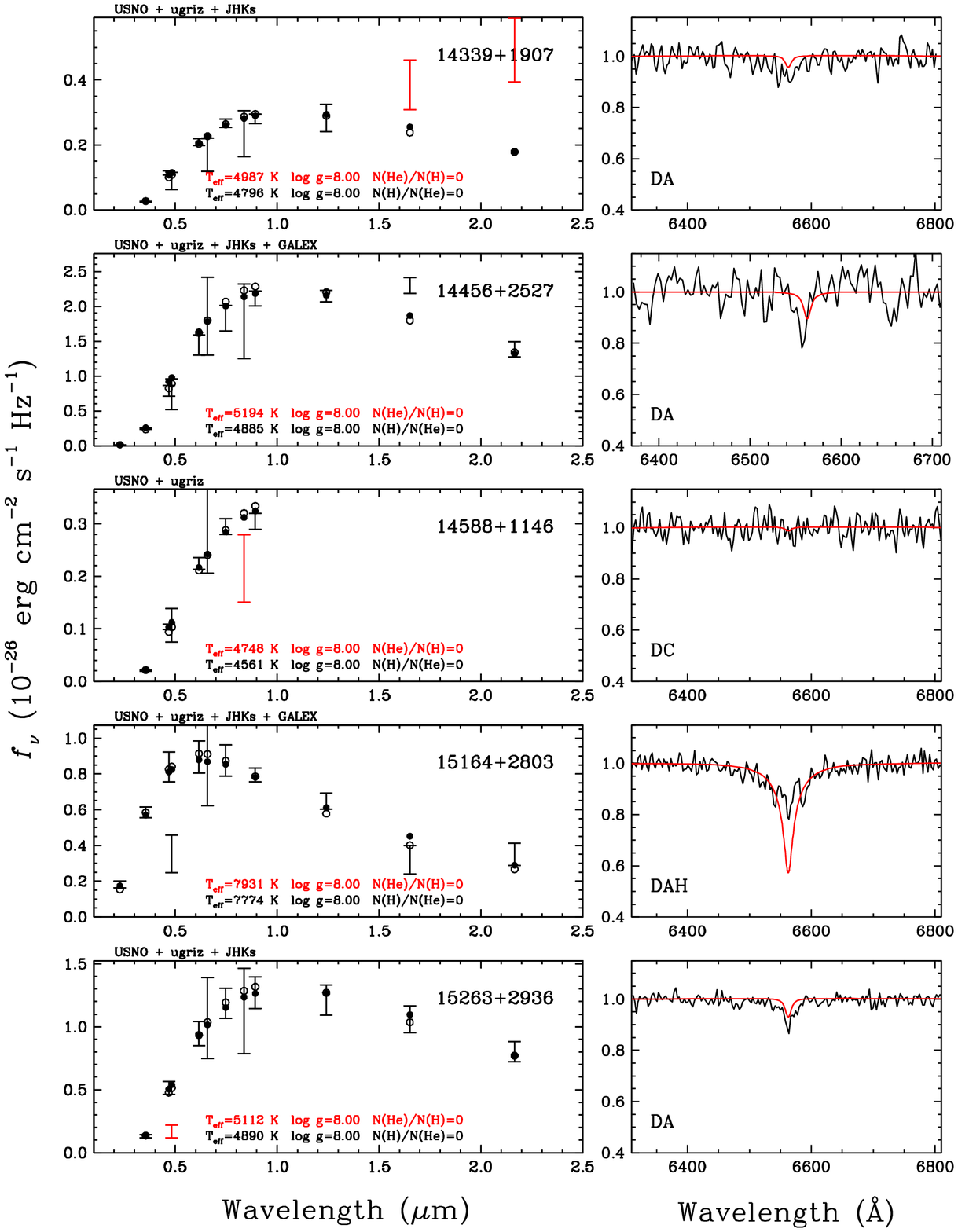}
\begin{flushright}
Figure \ref{photoDADC}v
\end{flushright}
\end{figure}

\clearpage

\begin{figure}[p]
\plotone{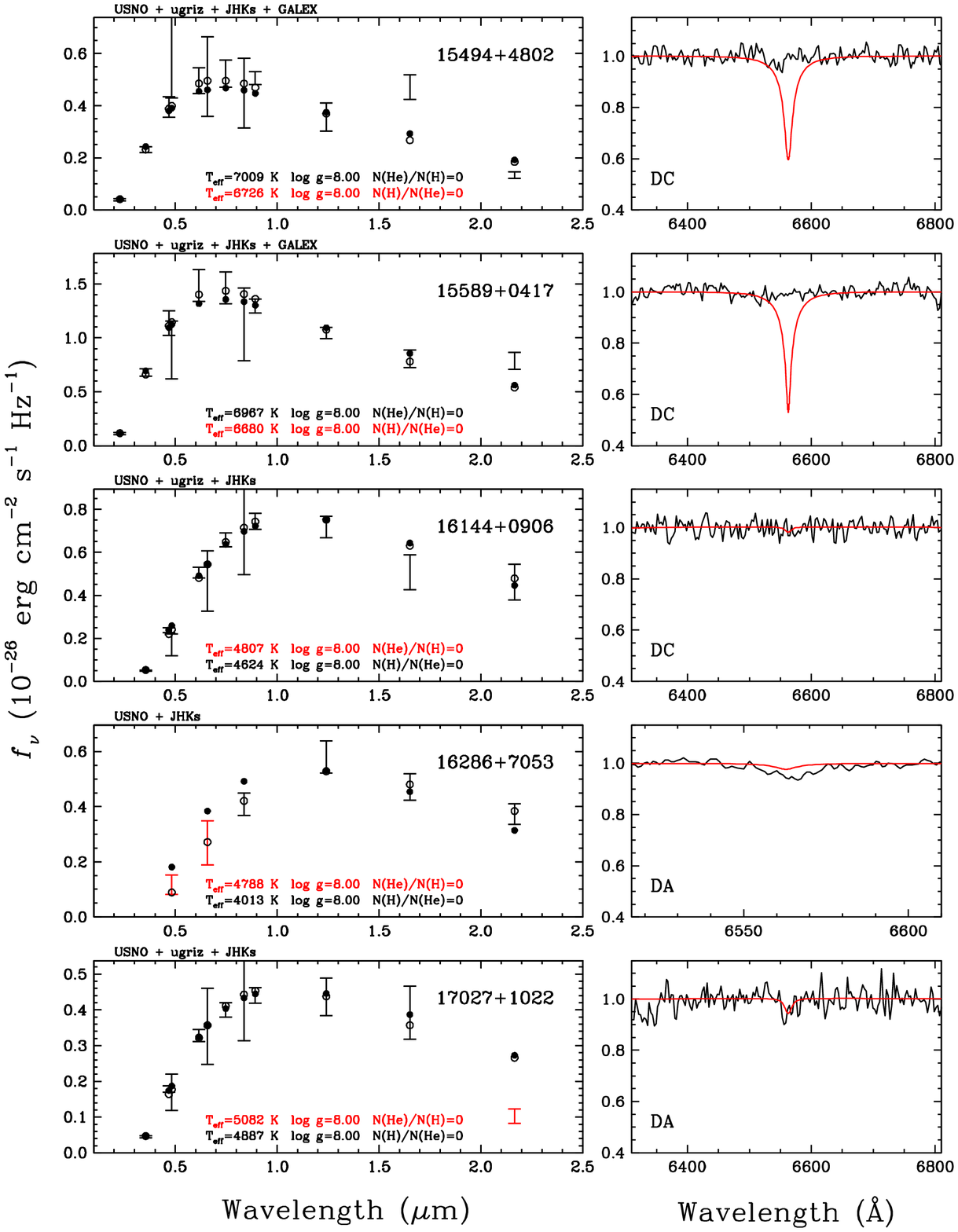}
\begin{flushright}
Figure \ref{photoDADC}w
\end{flushright}
\end{figure}

\clearpage

\begin{figure}[p]
\plotone{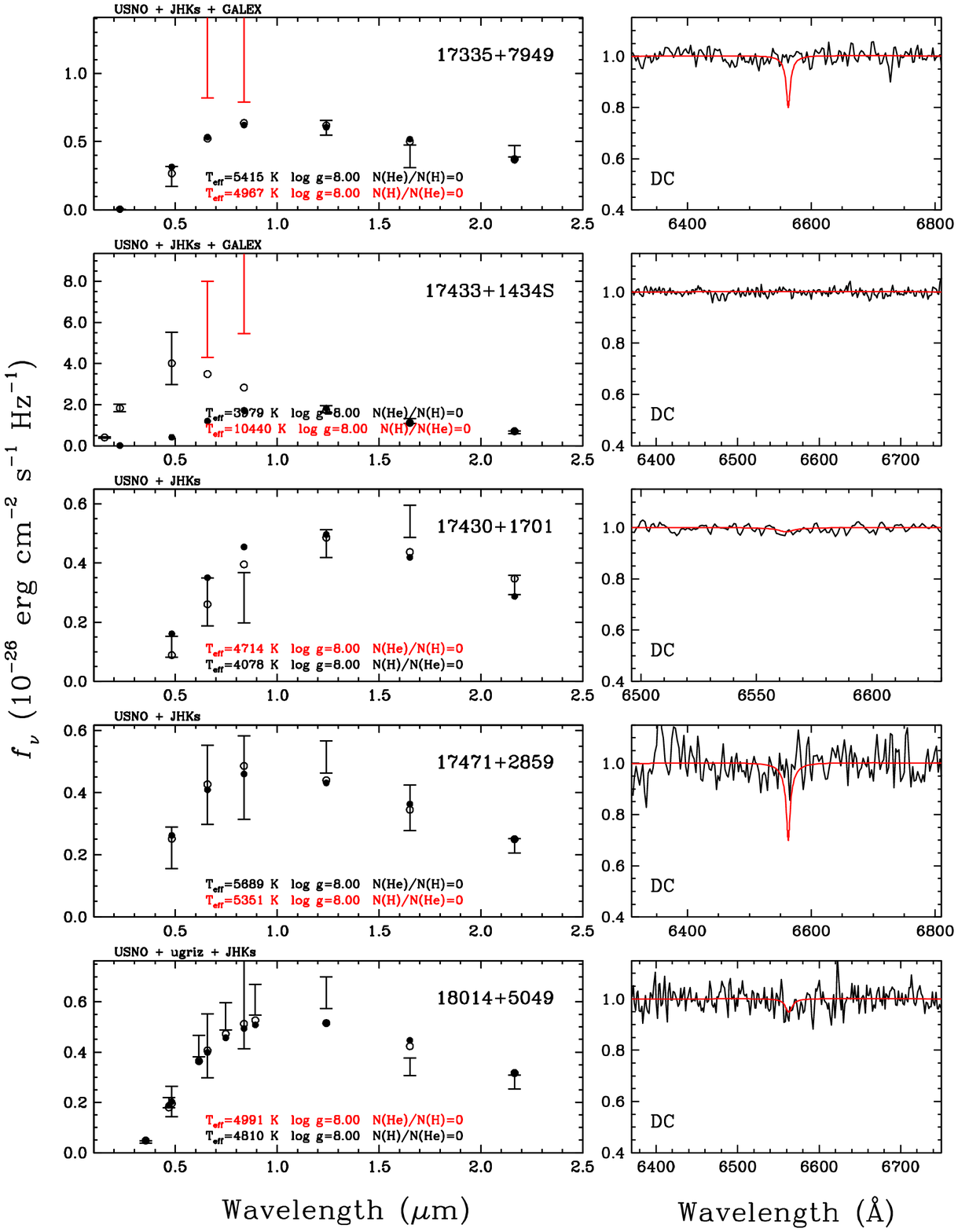}
\begin{flushright}
Figure \ref{photoDADC}x
\end{flushright}
\end{figure}

\clearpage

\begin{figure}[p]
\plotone{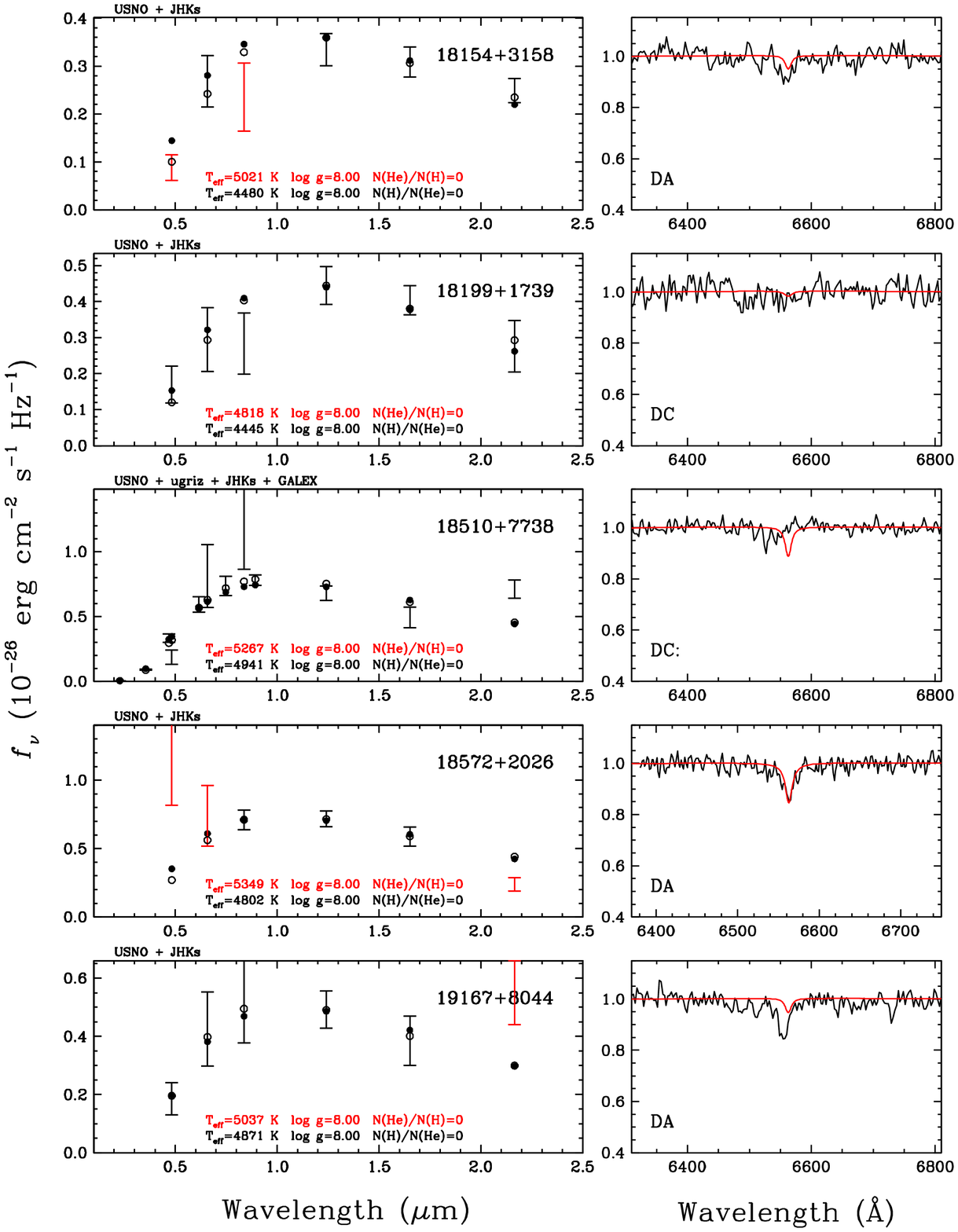}
\begin{flushright}
Figure \ref{photoDADC}y
\end{flushright}
\end{figure}

\clearpage

\begin{figure}[p]
\plotone{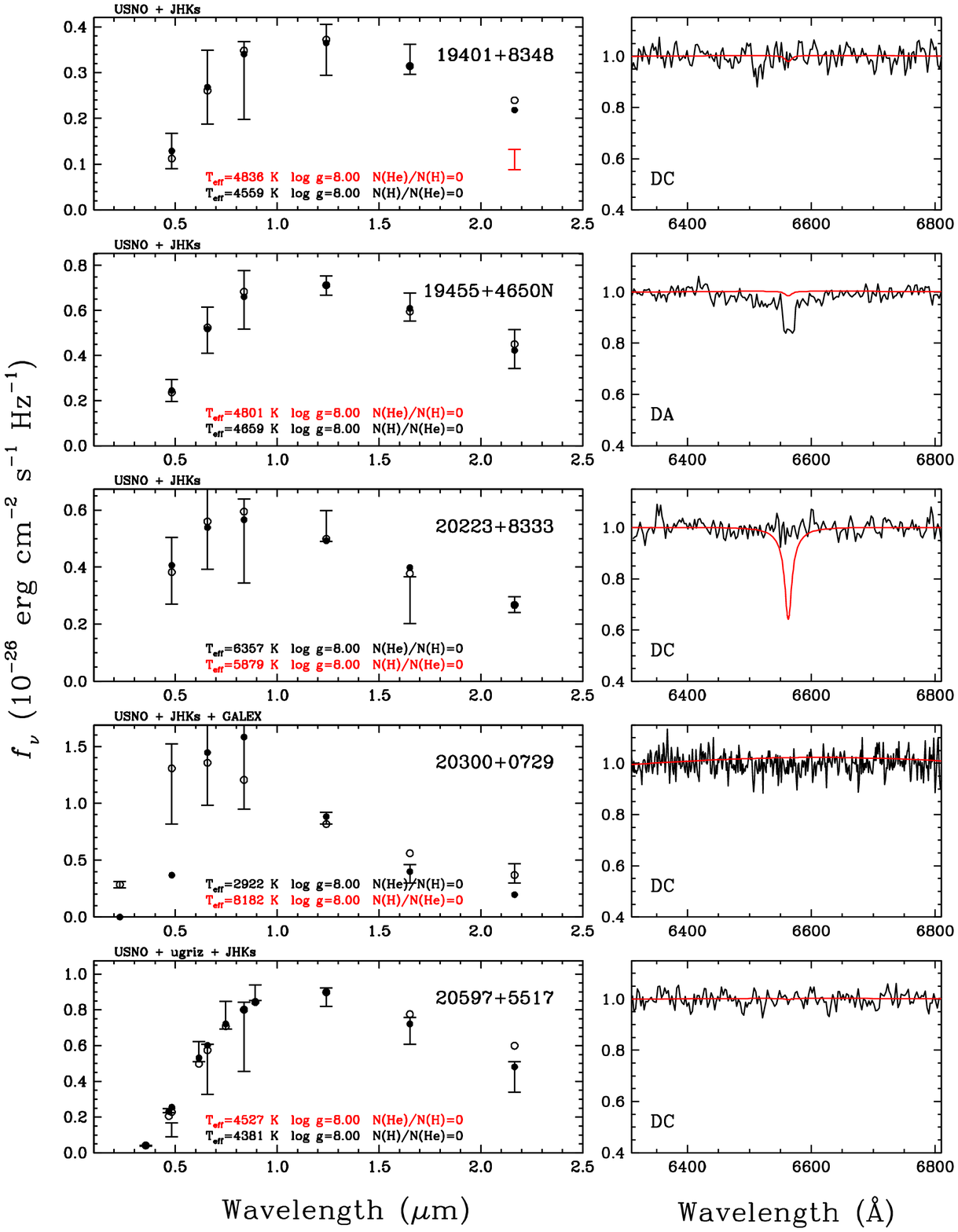}
\begin{flushright}
Figure \ref{photoDADC}z
\end{flushright}
\end{figure}

\clearpage

\begin{figure}[p]
\plotone{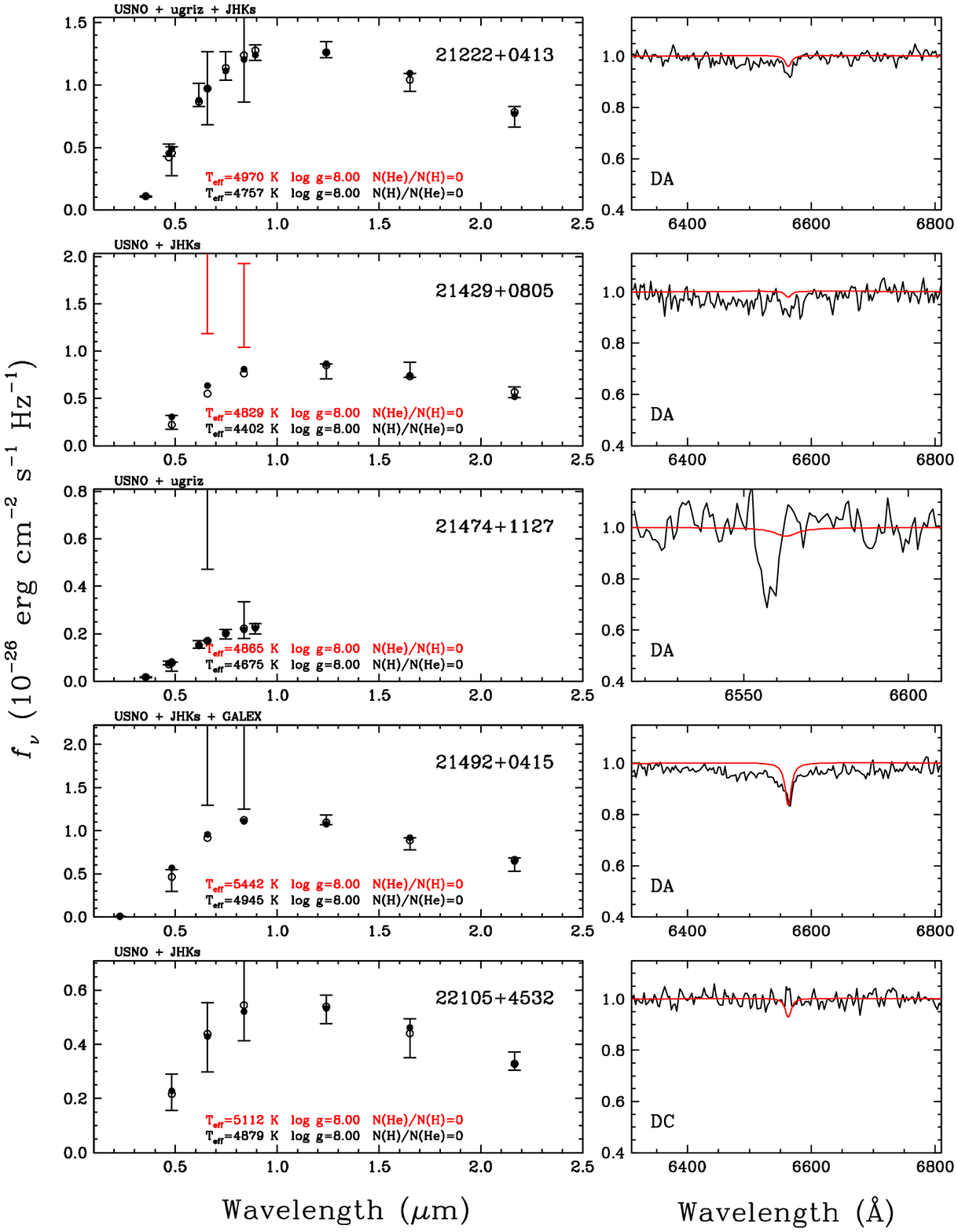}
\begin{flushright}
Figure \ref{photoDADC}A
\end{flushright}
\end{figure}

\clearpage

\begin{figure}[p]
\plotone{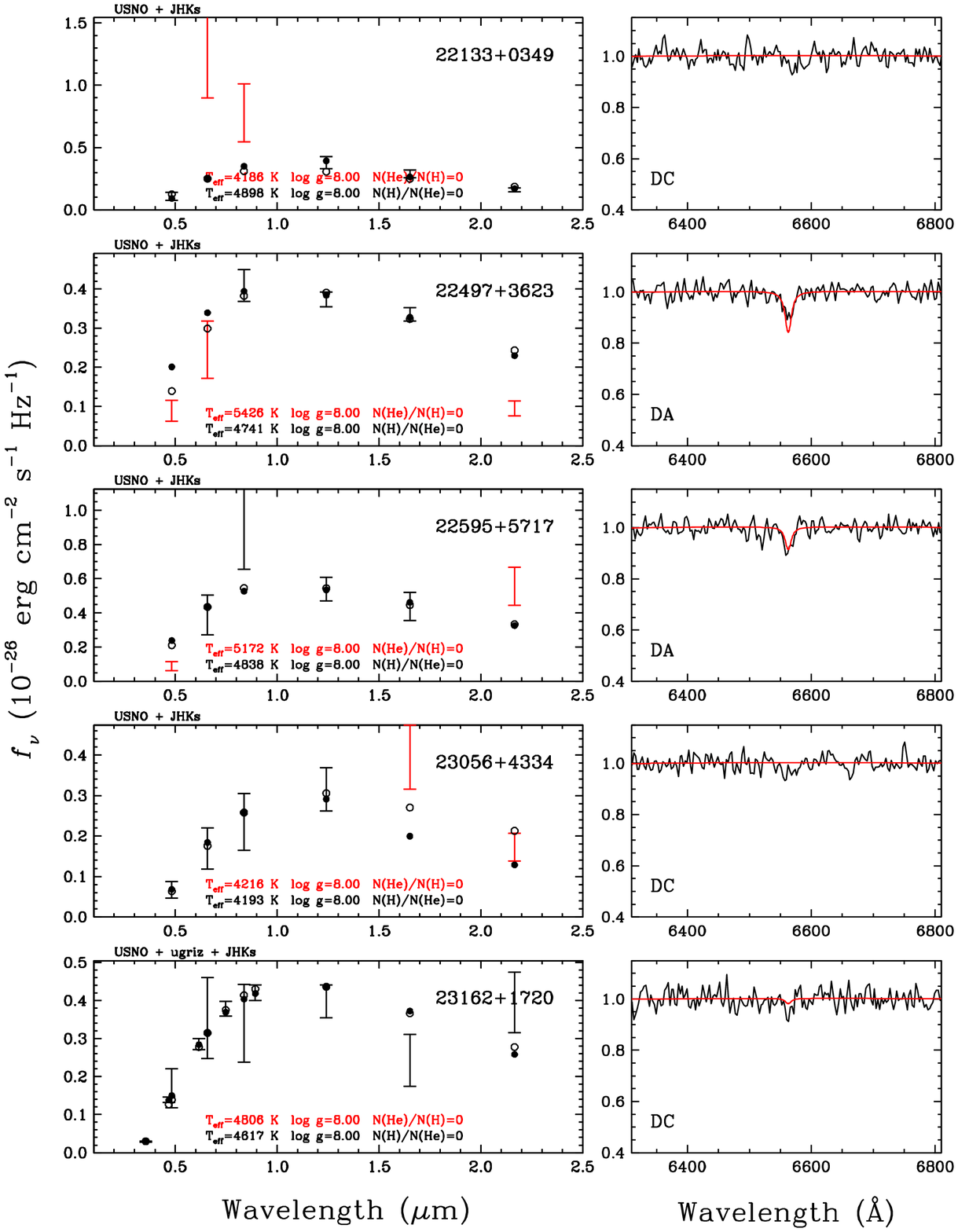}
\begin{flushright}
Figure \ref{photoDADC}B
\end{flushright}
\end{figure}

\clearpage

\begin{figure}[p]
\plotone{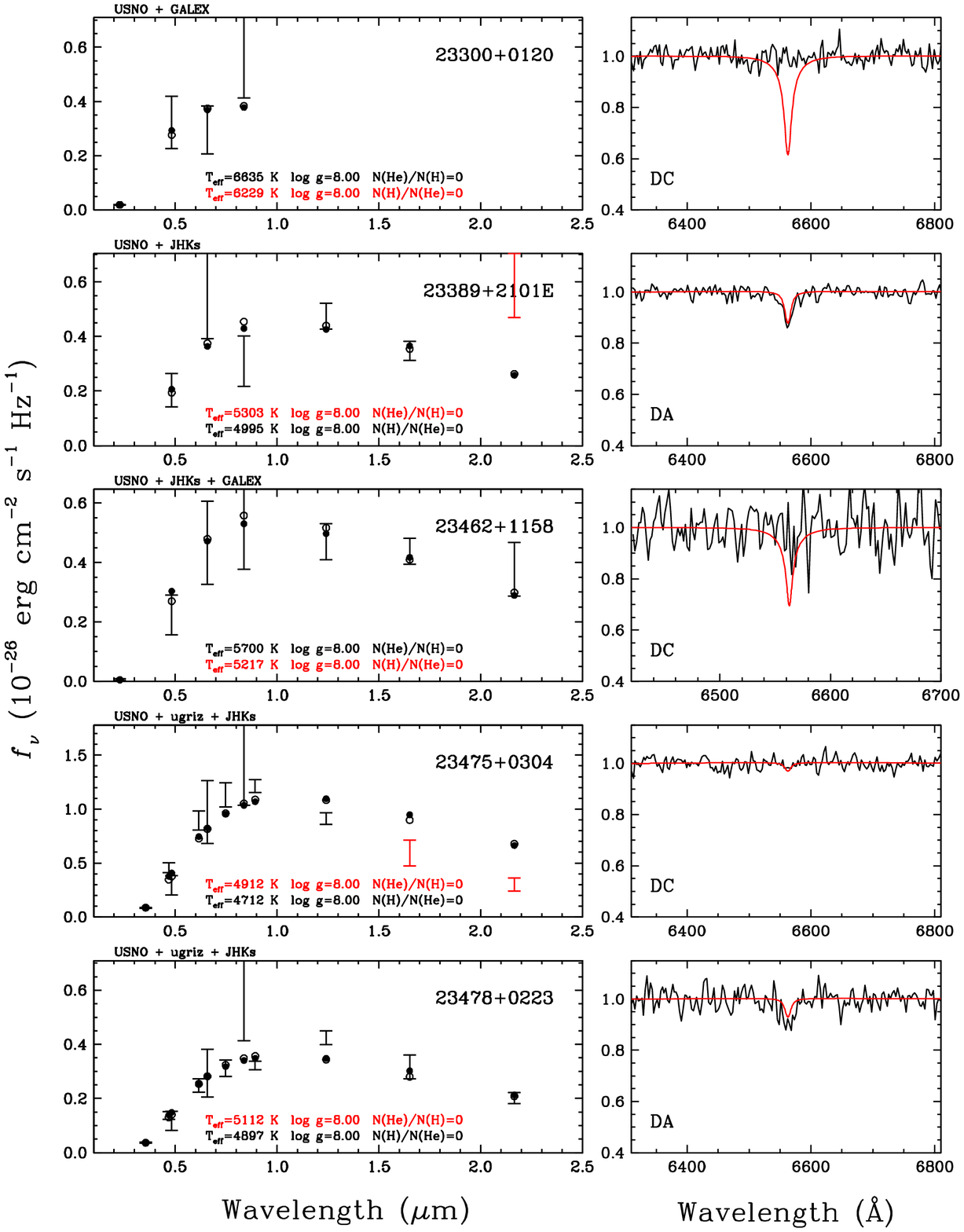}
\begin{flushright}
Figure \ref{photoDADC}C
\end{flushright}
\end{figure}

\clearpage

\begin{figure}[p]
\plotone{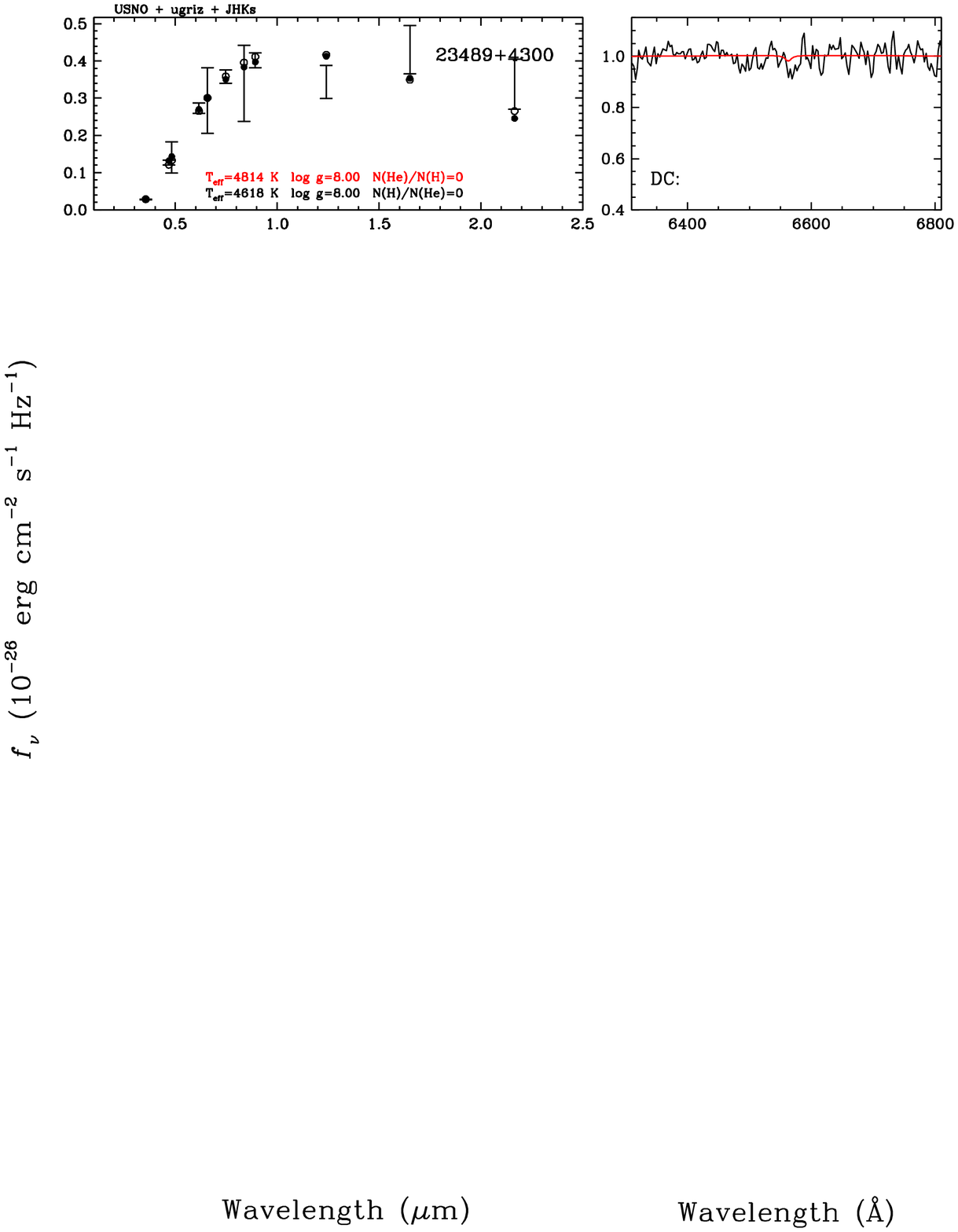}
\begin{flushright}
Figure \ref{photoDADC}D
\end{flushright}
\end{figure}

\clearpage

\begin{figure}[p]
\plotone{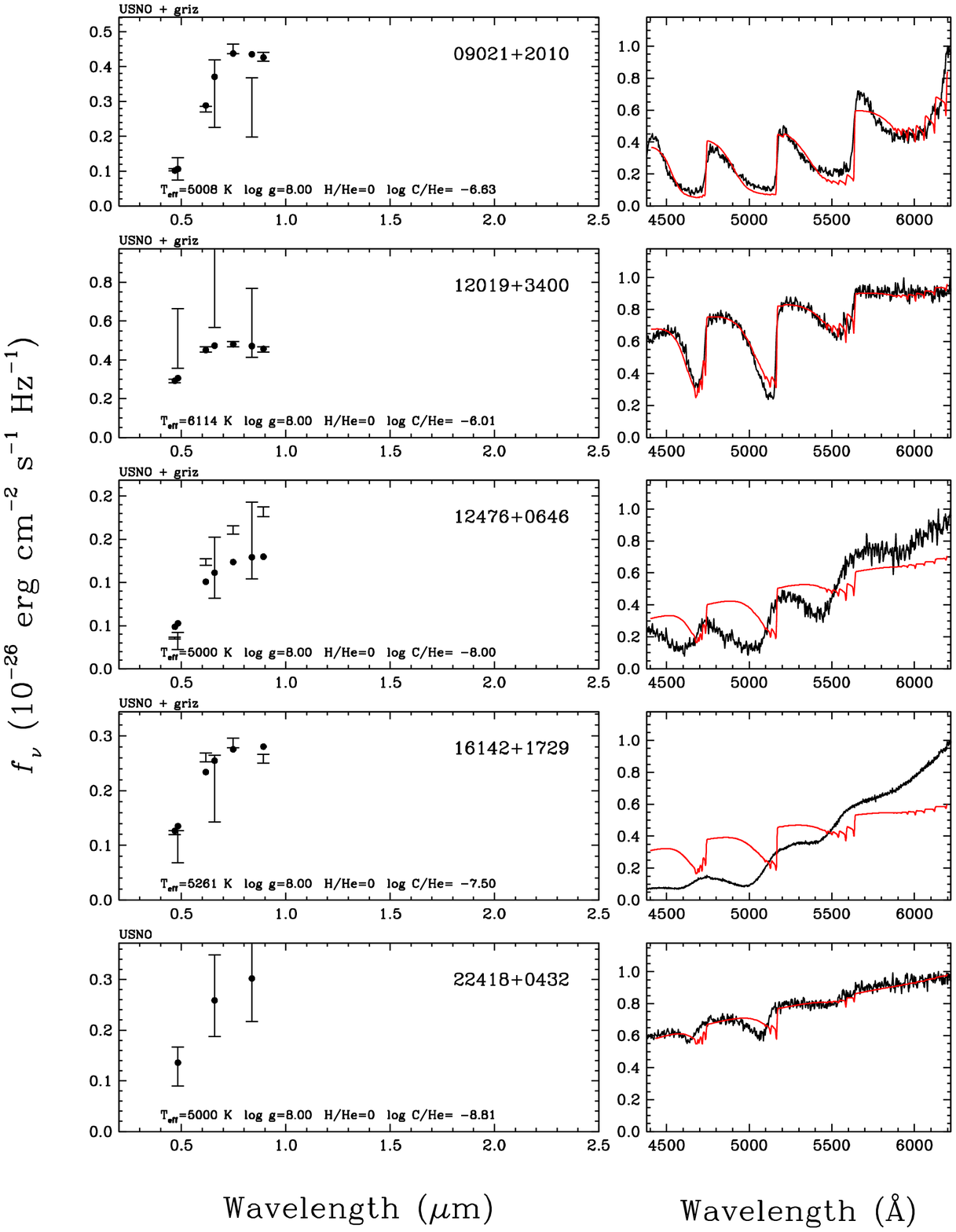}
\begin{flushright}
Figure \ref{DQfits}
\end{flushright}
\end{figure}

\clearpage

\begin{figure}[p]
\plotone{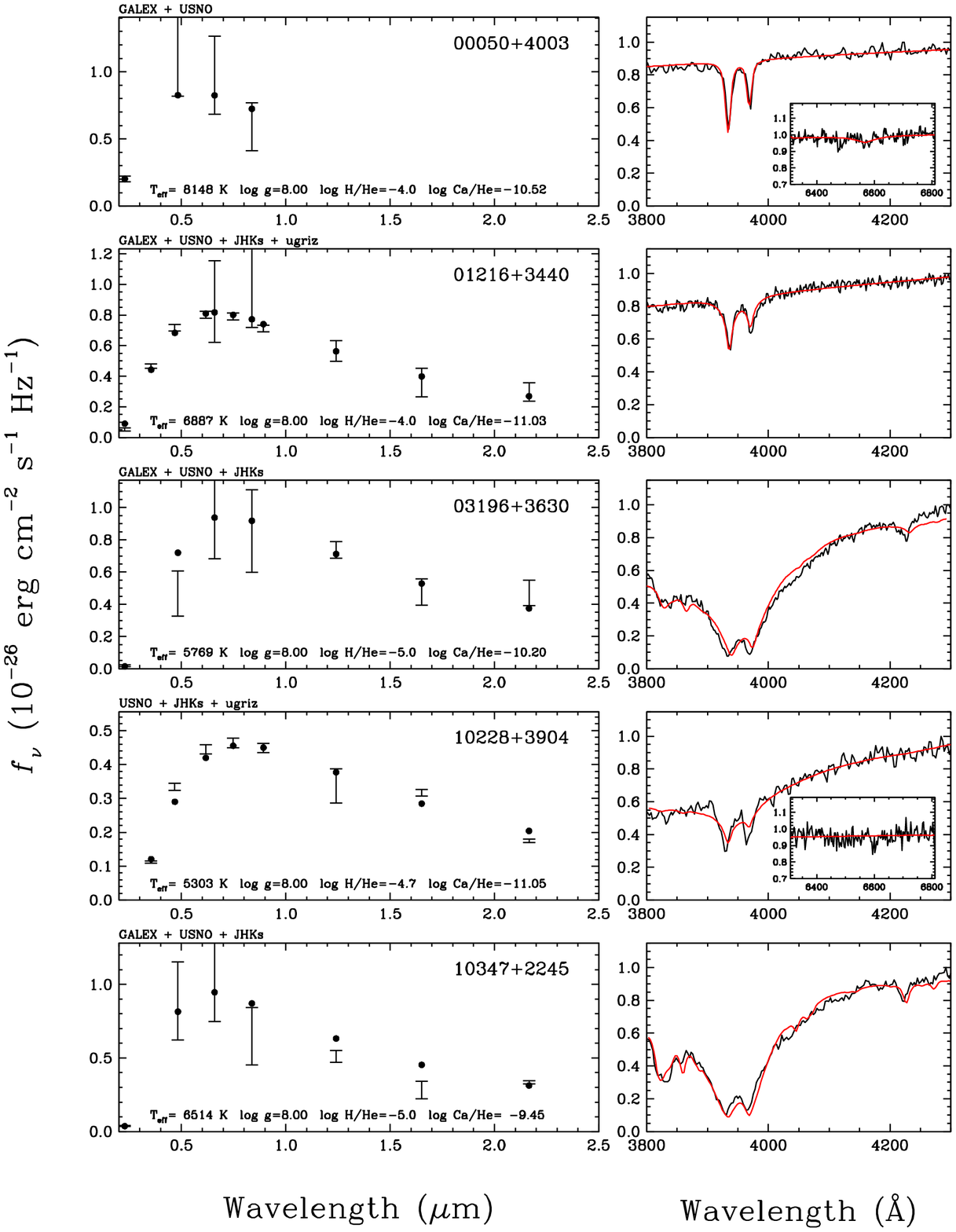}
\begin{flushright}
Figure \ref{DZfits}a
\end{flushright}
\end{figure}

\clearpage

\begin{figure}[p]
\plotone{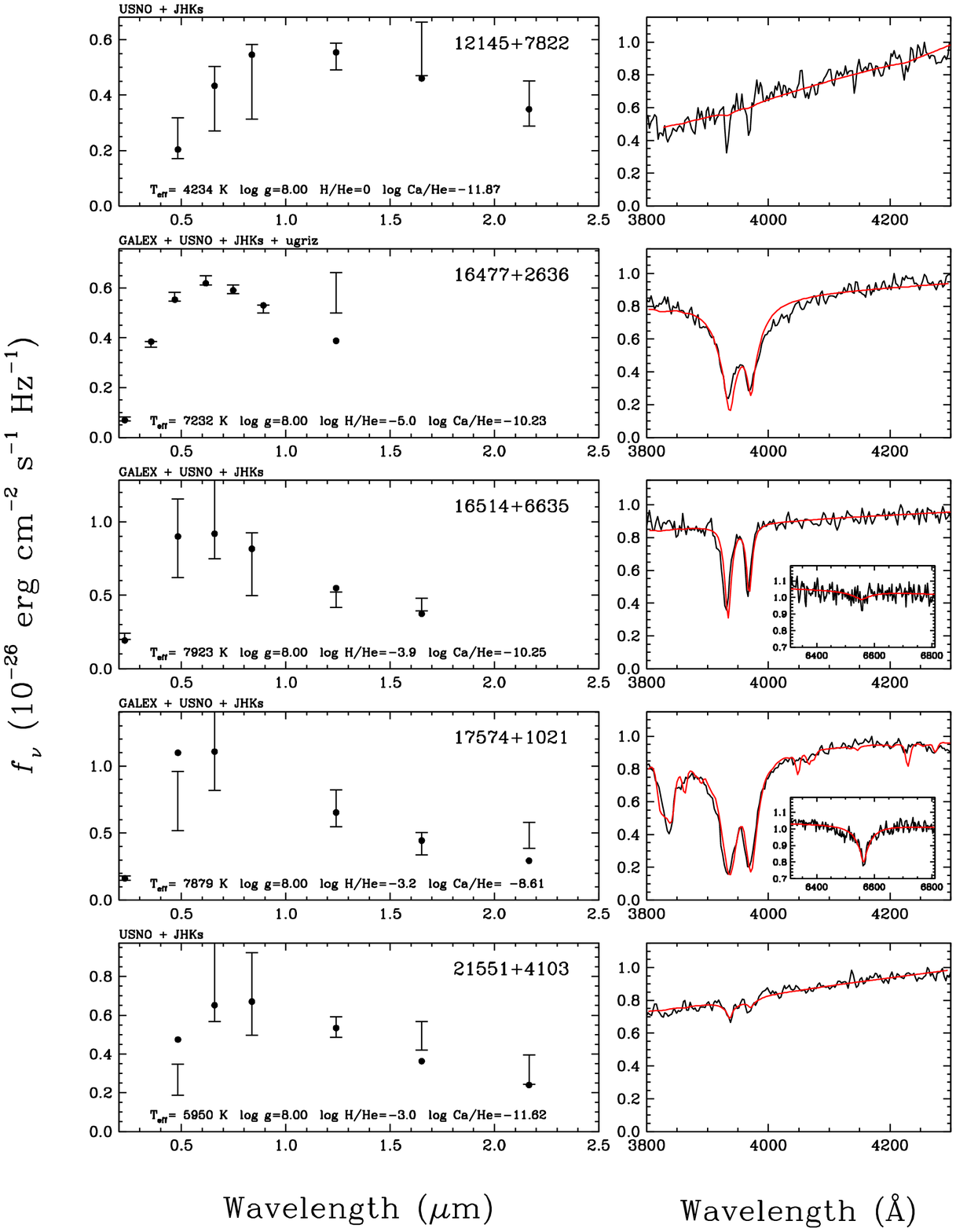}
\begin{flushright}
Figure \ref{DZfits}b
\end{flushright}
\end{figure}

\clearpage

\begin{figure}[p]
\plotone{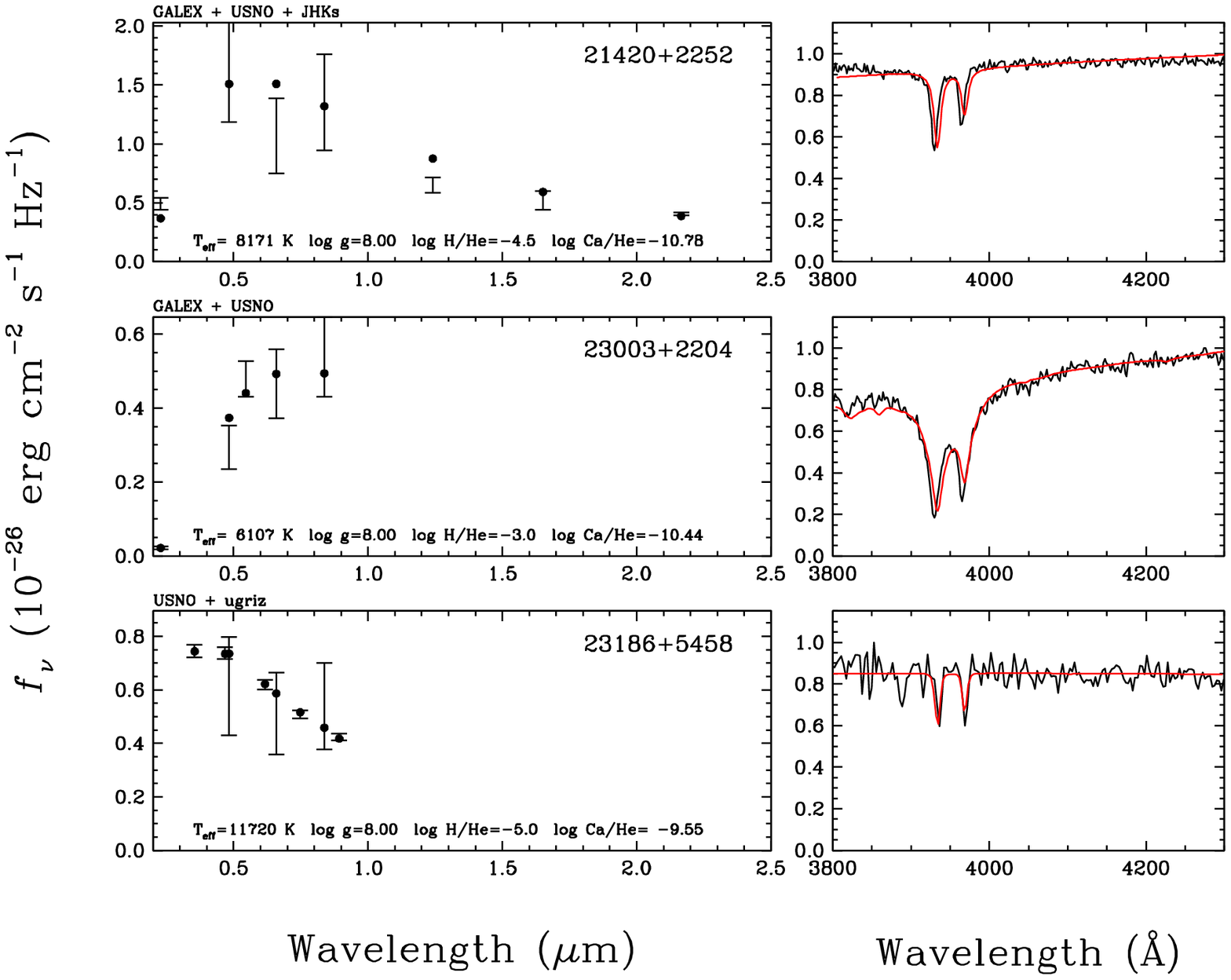}
\begin{flushright}
Figure \ref{DZfits}c
\end{flushright}
\end{figure}

\clearpage

\begin{figure}[p]
\plotone{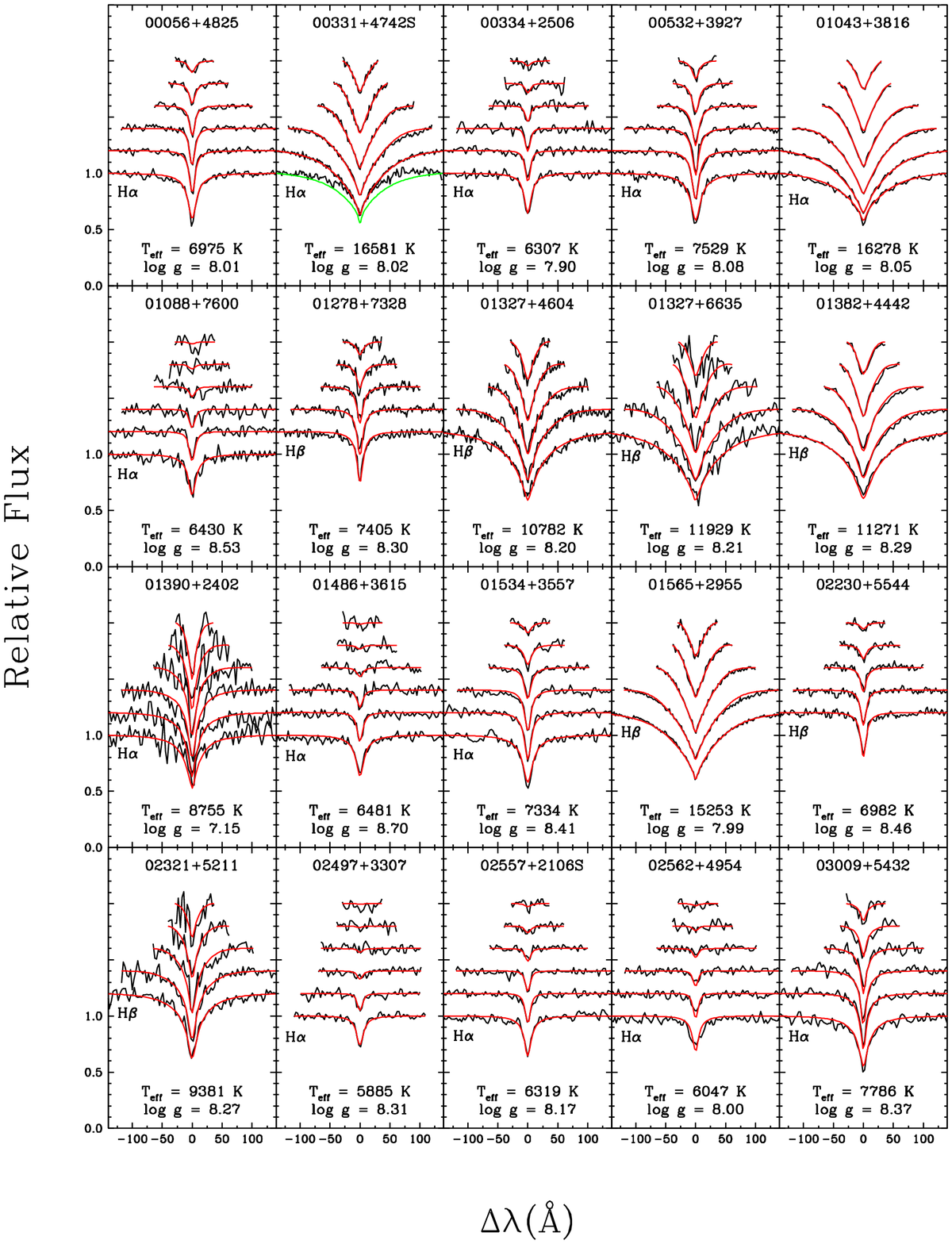}
\begin{flushright}
Figure \ref{DAspectro2}a
\end{flushright}
\end{figure}

\clearpage

\begin{figure}[p]
\plotone{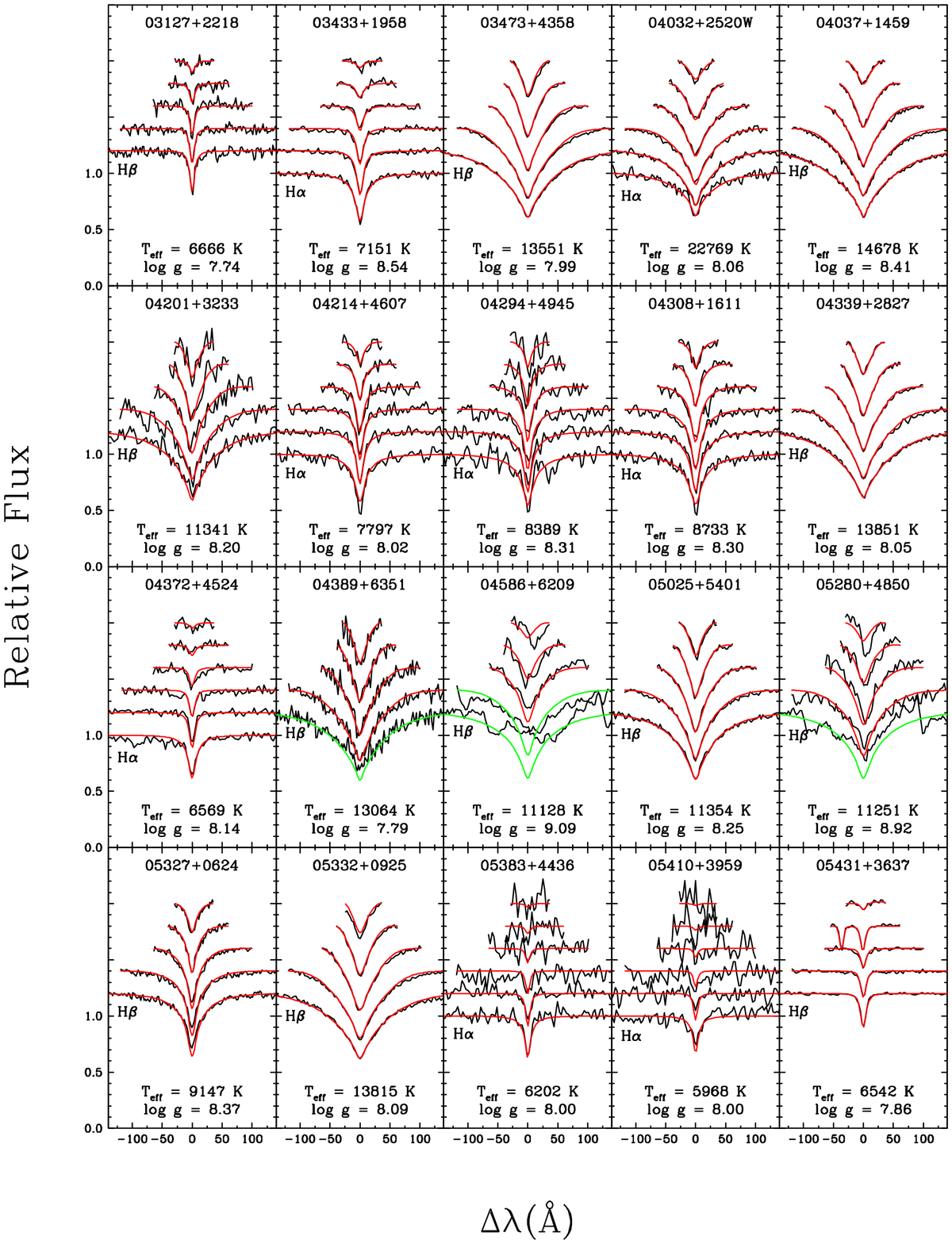}
\begin{flushright}
Figure \ref{DAspectro2}b
\end{flushright}
\end{figure}

\clearpage

\begin{figure}[p]
\plotone{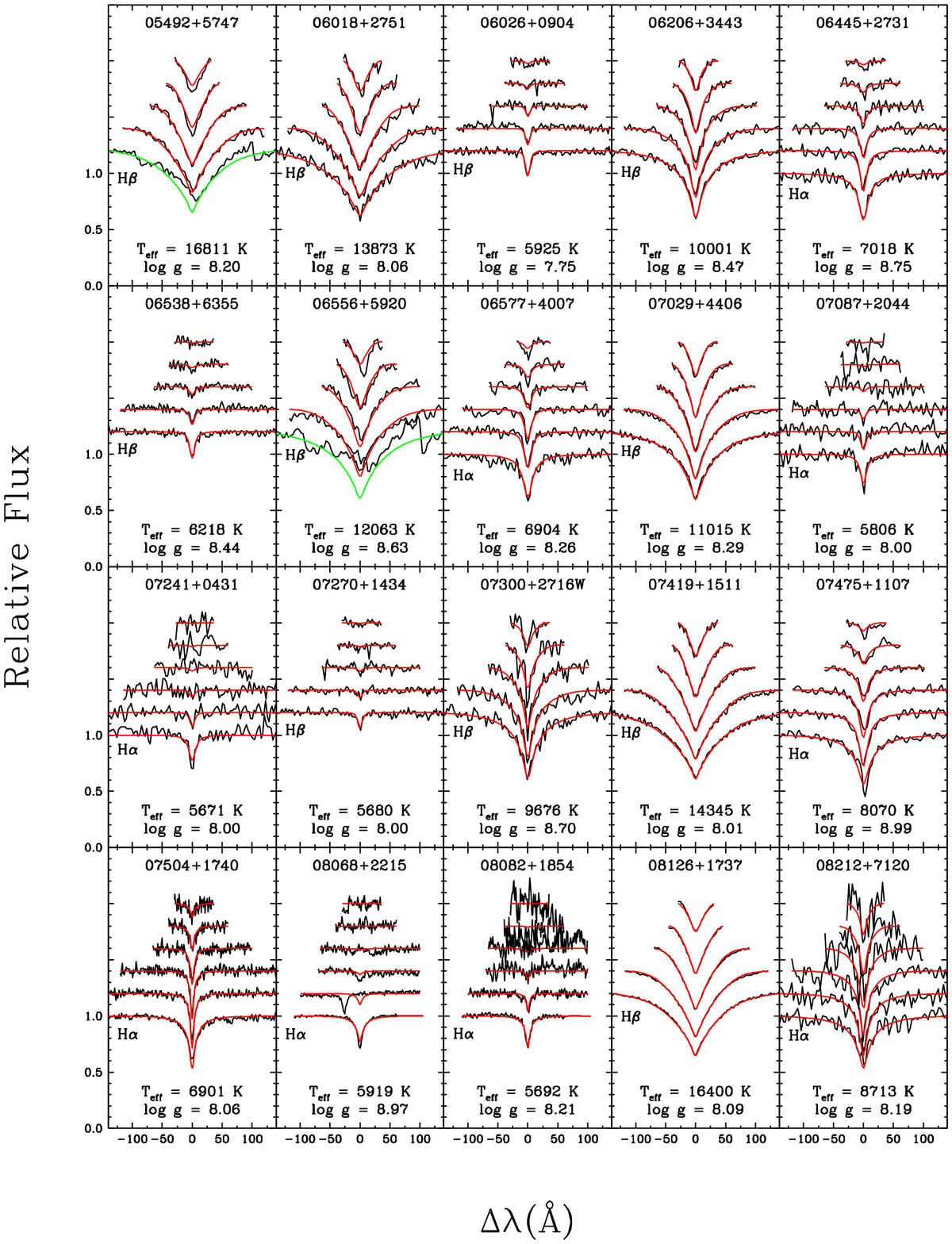}
\begin{flushright}
Figure \ref{DAspectro2}c
\end{flushright}
\end{figure}

\clearpage

\begin{figure}[p]
\plotone{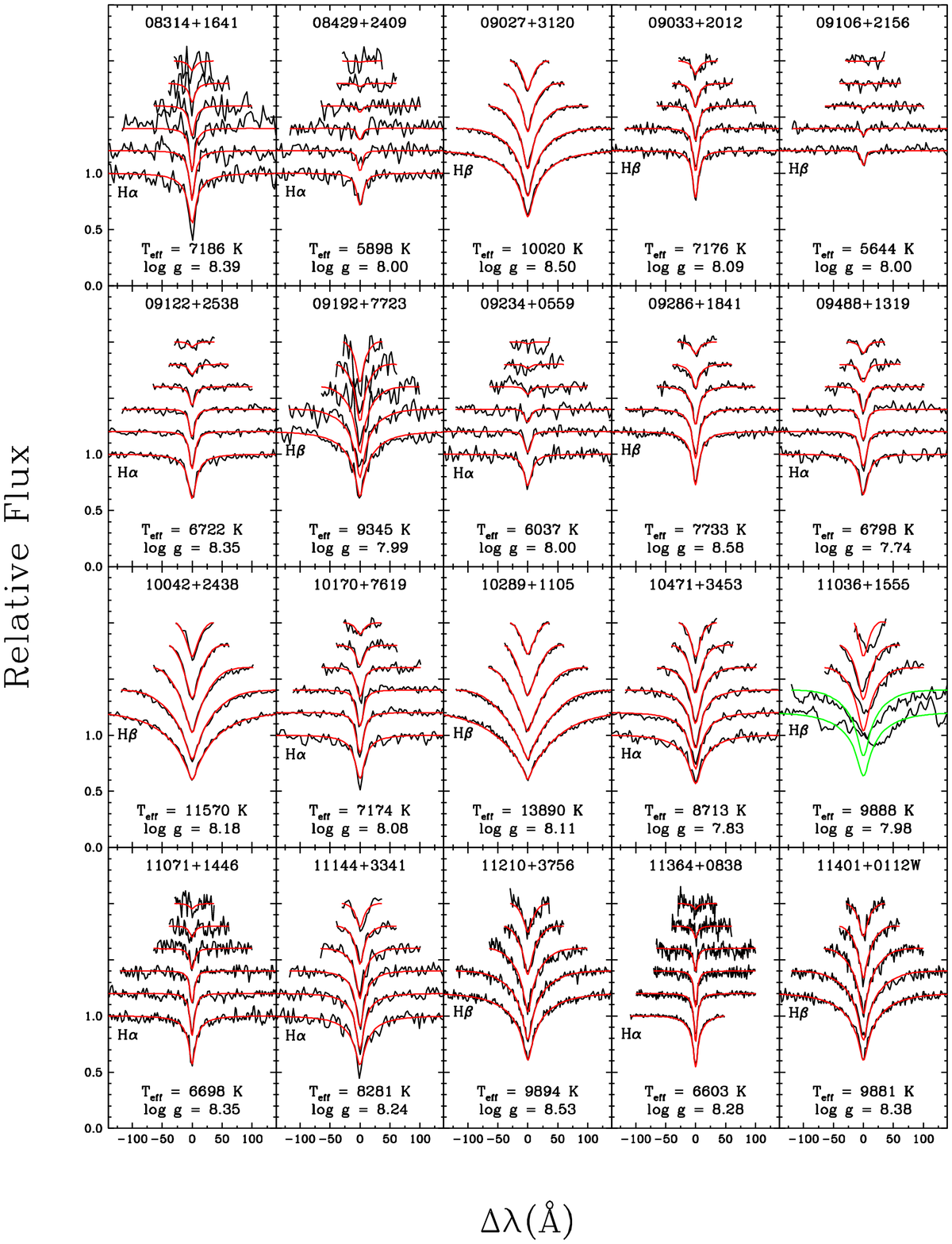}
\begin{flushright}
Figure \ref{DAspectro2}d
\end{flushright}
\end{figure}

\clearpage

\begin{figure}[p]
\plotone{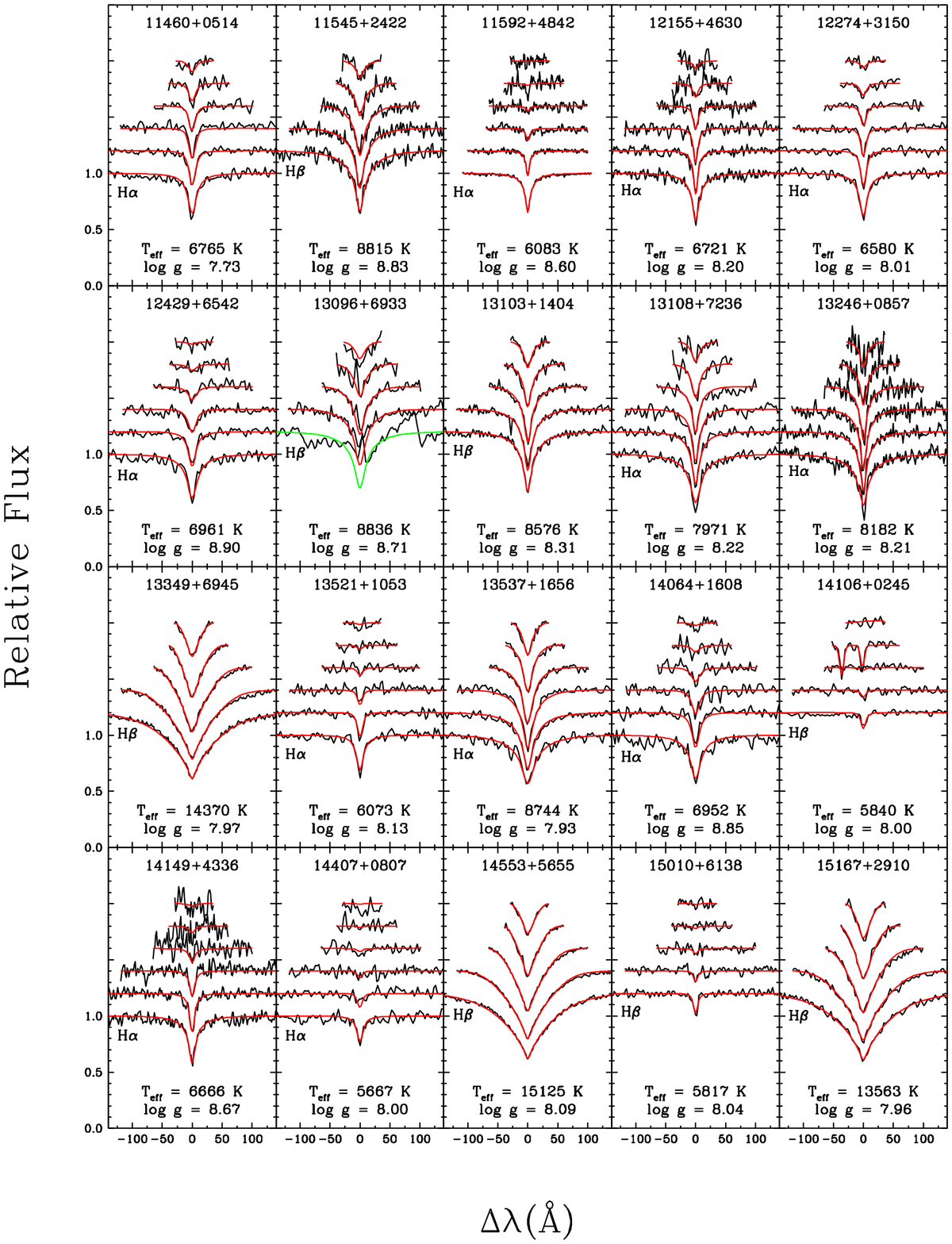}
\begin{flushright}
Figure \ref{DAspectro2}e
\end{flushright}
\end{figure}

\clearpage

\begin{figure}[p]
\plotone{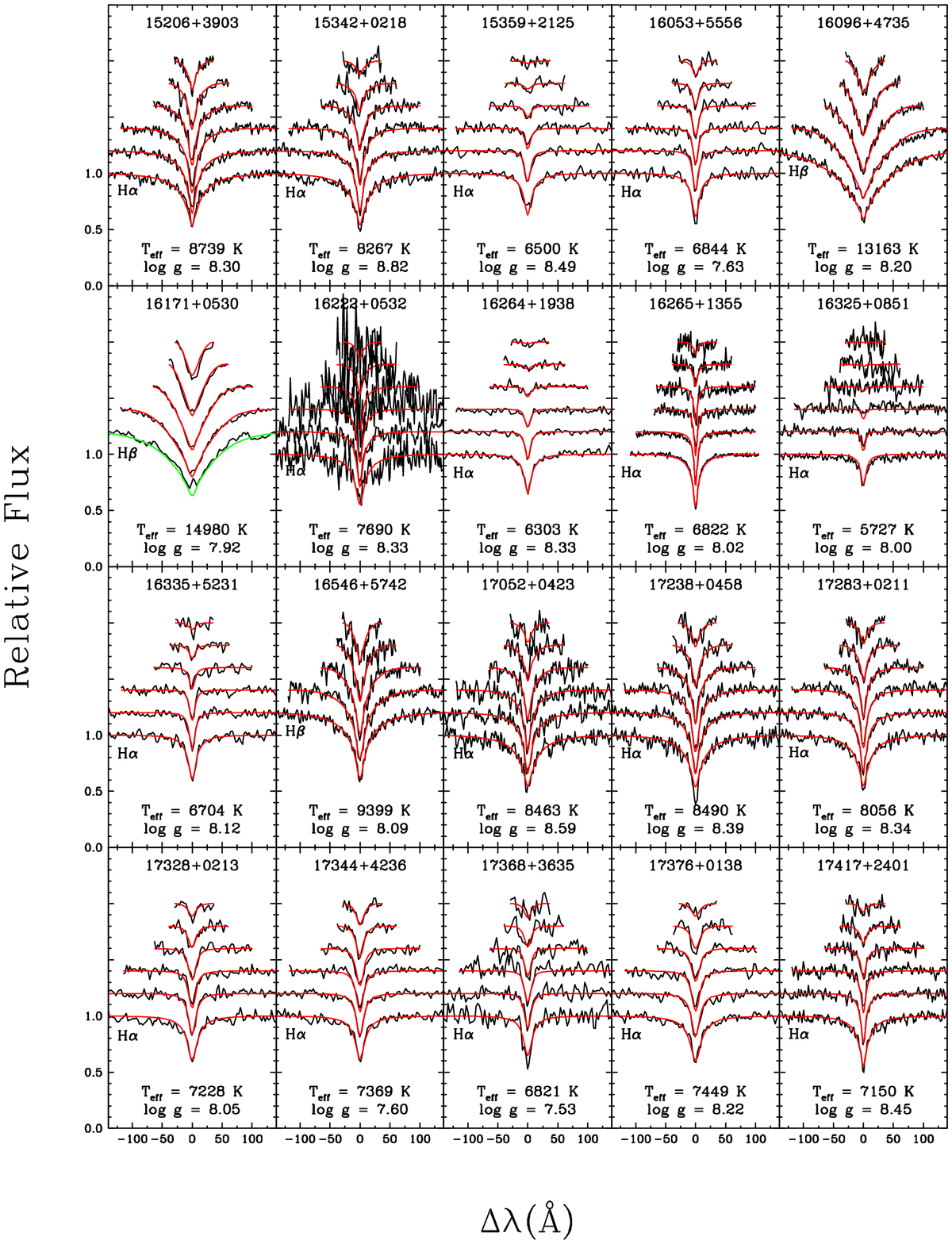}
\begin{flushright}
Figure \ref{DAspectro2}f
\end{flushright}
\end{figure}

\clearpage

\begin{figure}[p]
\plotone{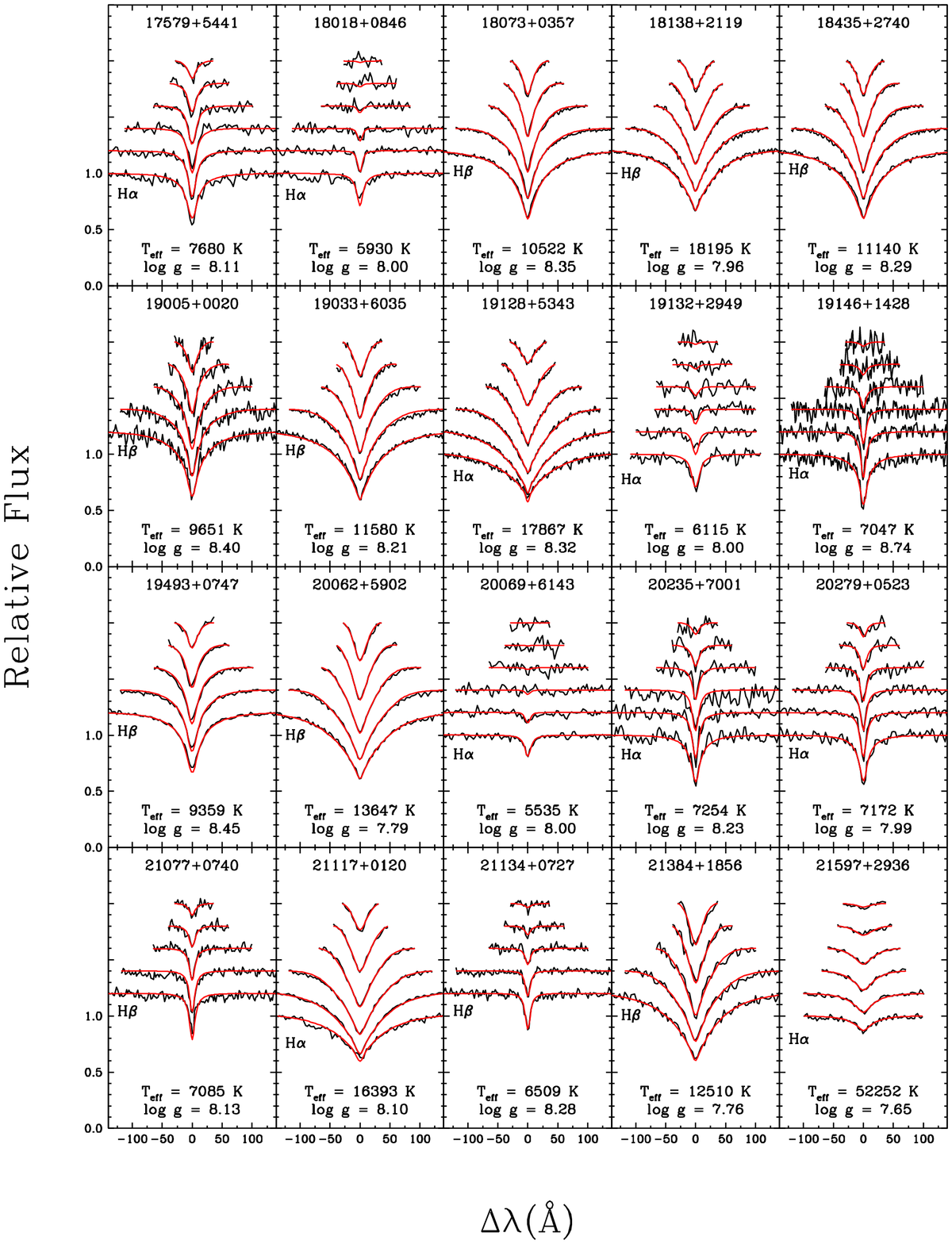}
\begin{flushright}
Figure \ref{DAspectro2}g
\end{flushright}
\end{figure}

\clearpage

\begin{figure}[p]
\plotone{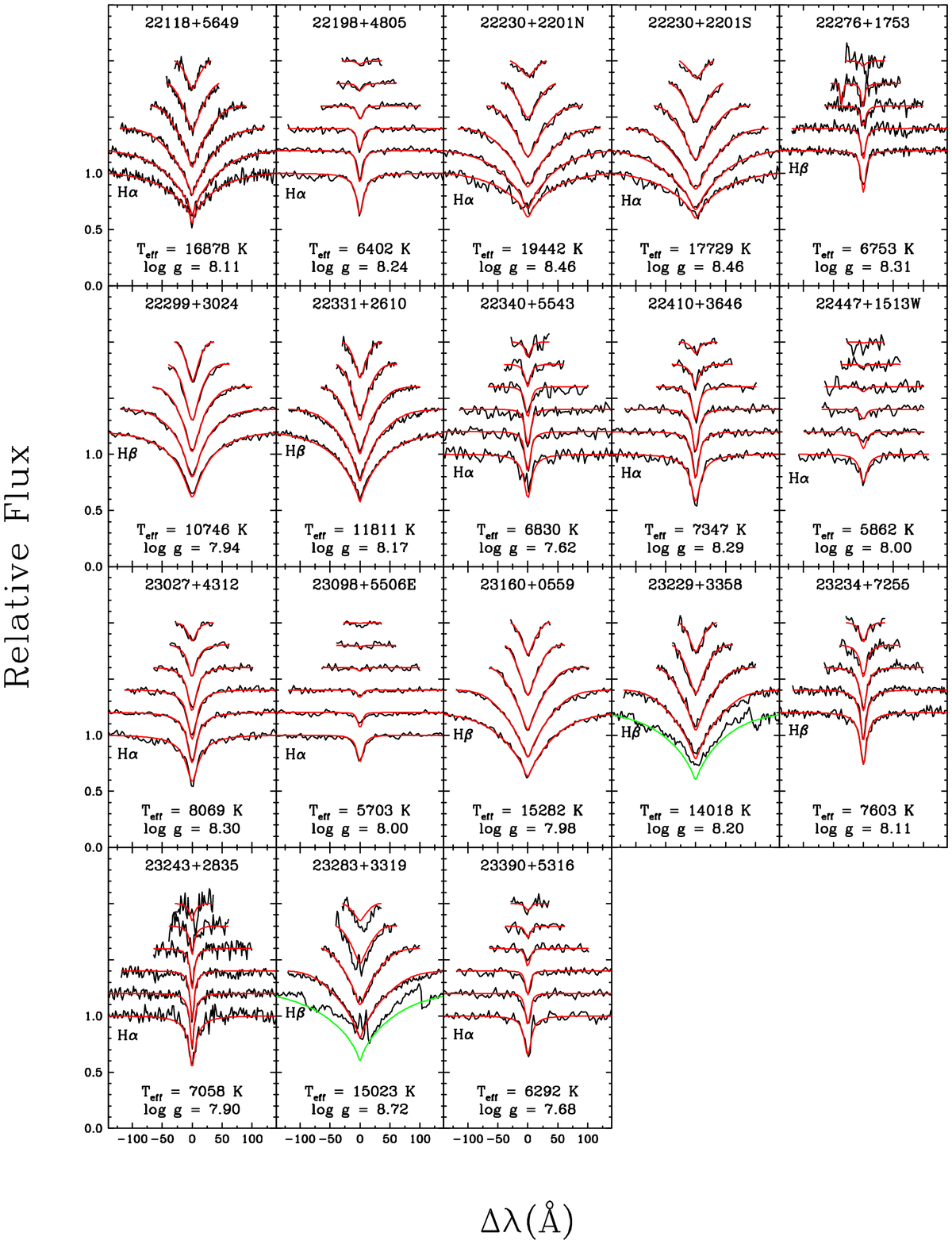}
\begin{flushright}
Figure \ref{DAspectro2}h
\end{flushright}
\end{figure}

\clearpage

\begin{figure}[p]
\plotone{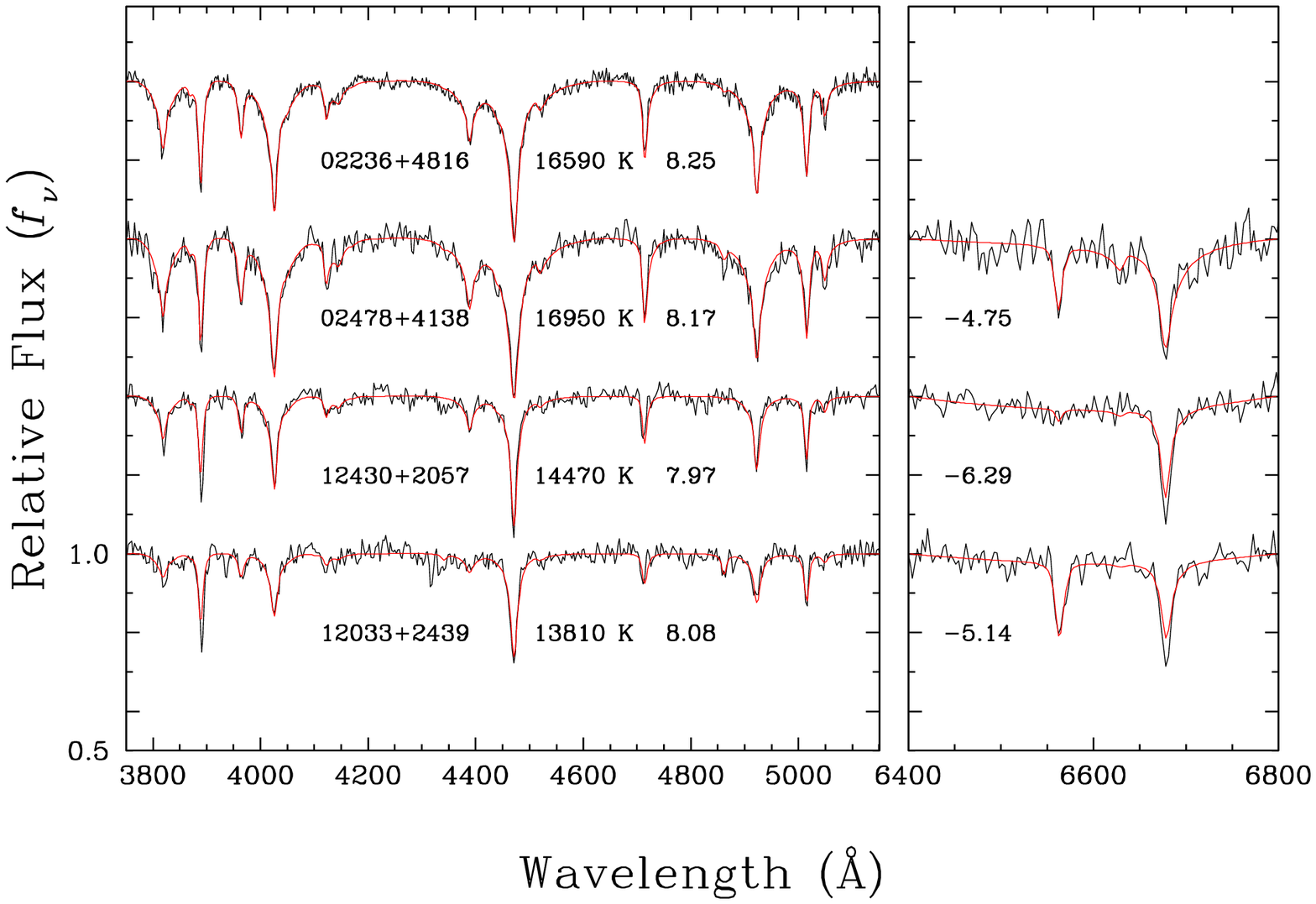}
\begin{flushright}
Figure \ref{DBfit}
\end{flushright}
\end{figure}

\clearpage

\begin{figure}[p]
\plotone{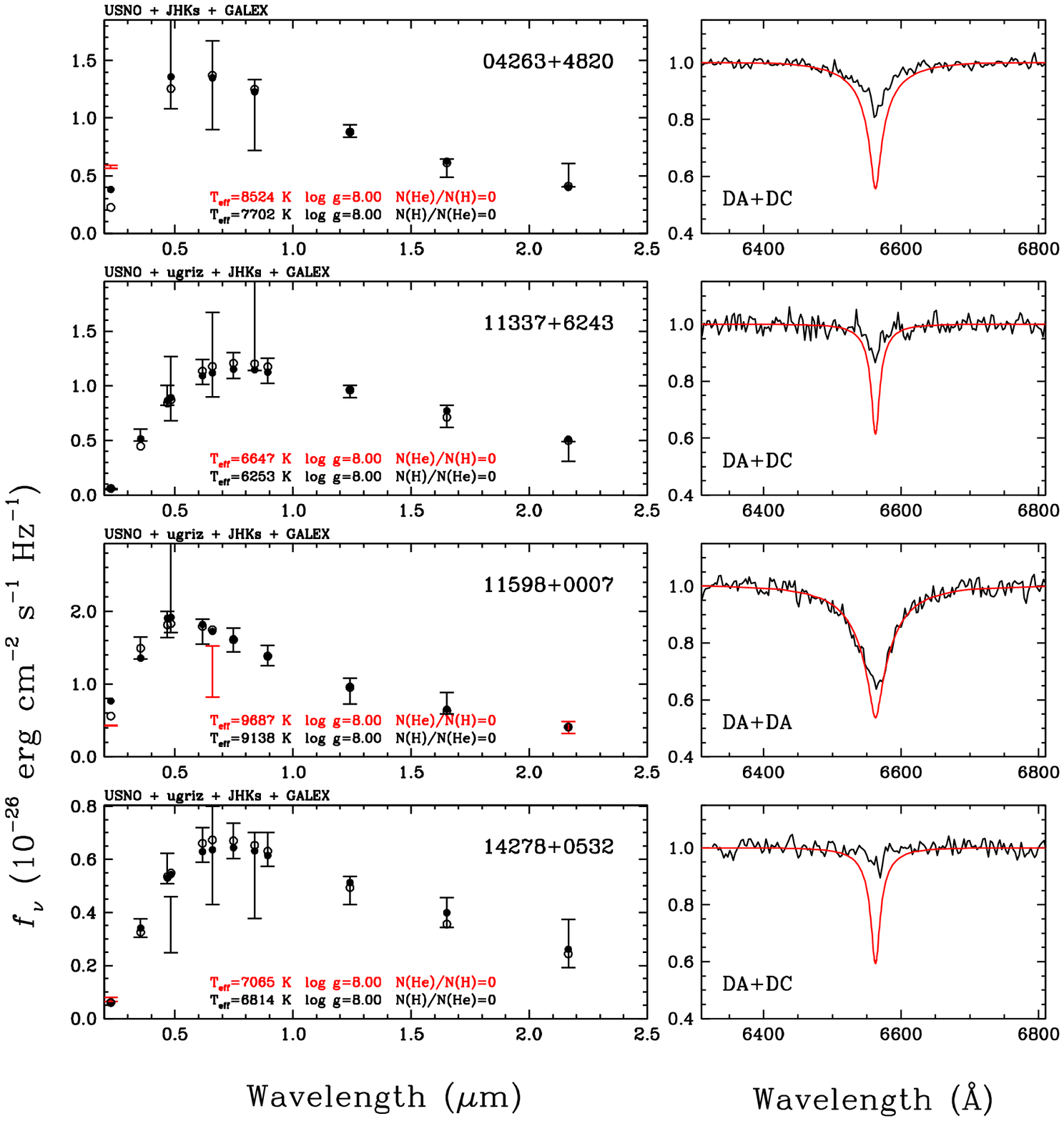}
\begin{flushright}
Figure \ref{fitsDADC}
\end{flushright}
\end{figure}

\clearpage

\begin{figure}[p]
\plotone{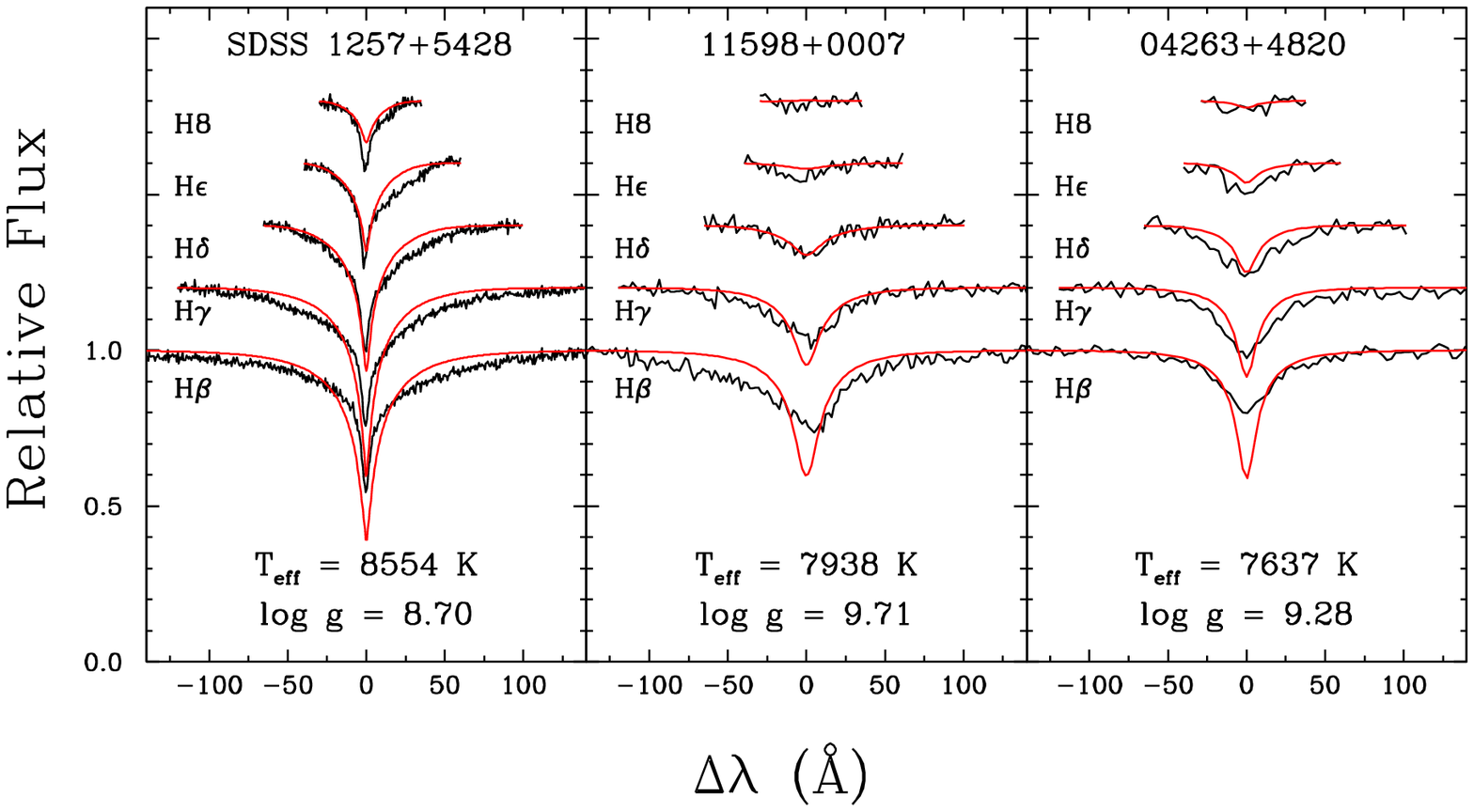}
\begin{flushright}
Figure \ref{DA+DA}
\end{flushright}
\end{figure}

\clearpage

\begin{figure}[p]
\plotone{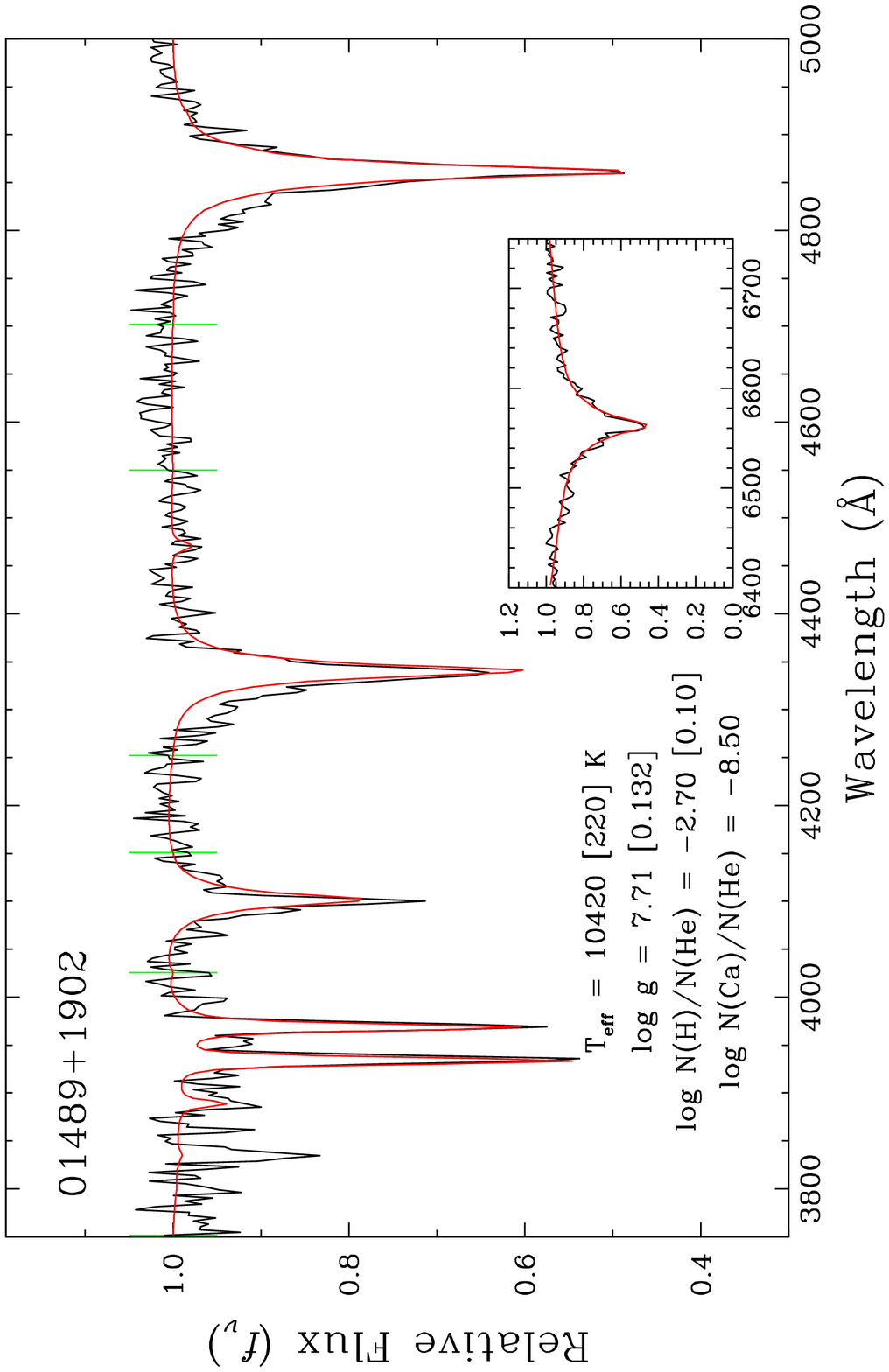}
\begin{flushright}
Figure \ref{fit_01489}
\end{flushright}
\end{figure}

\clearpage

\begin{figure}[p]
\plotone{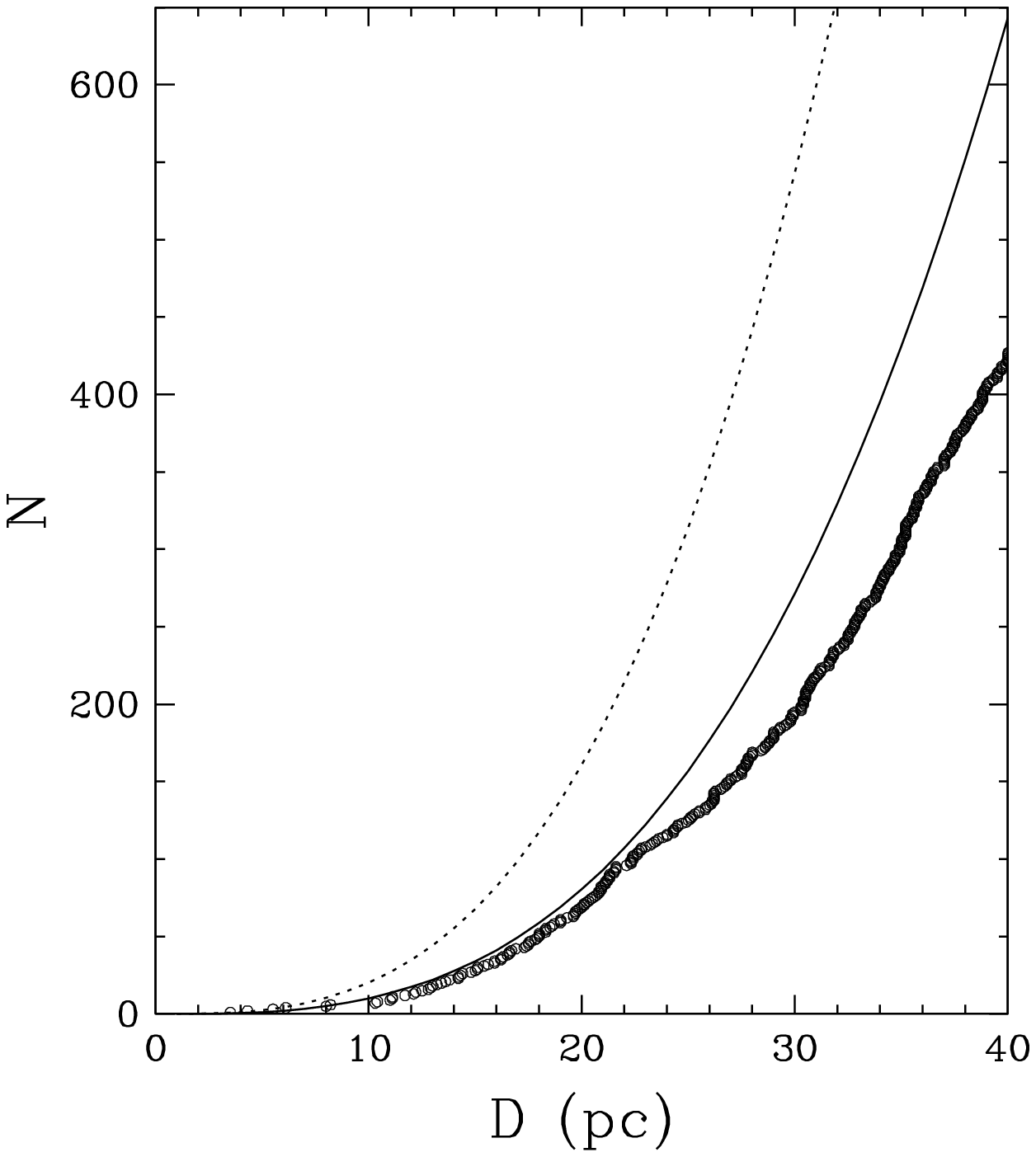}
\begin{flushright}
Figure \ref{NvsD}
\end{flushright}
\end{figure}

\clearpage

\begin{figure}[p]
\plotone{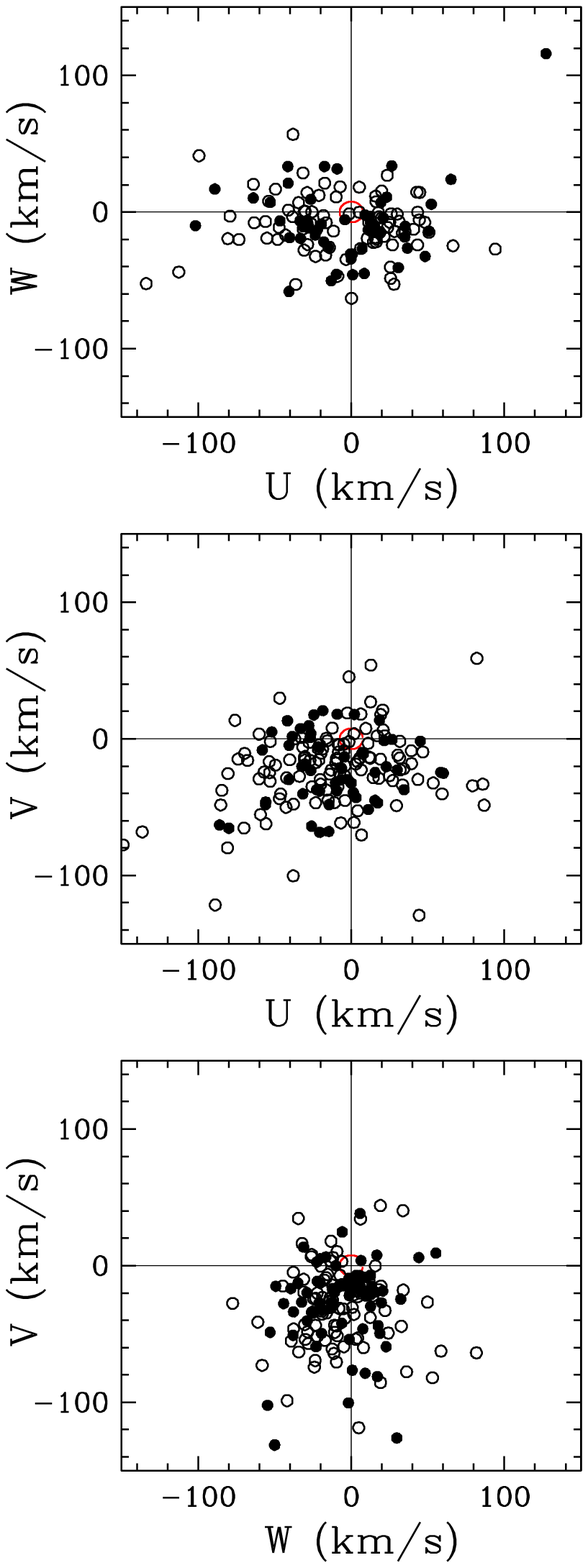}
\begin{flushright}
Figure \ref{uvw}
\end{flushright}
\end{figure}

\clearpage

\begin{figure}[p]
\plotone{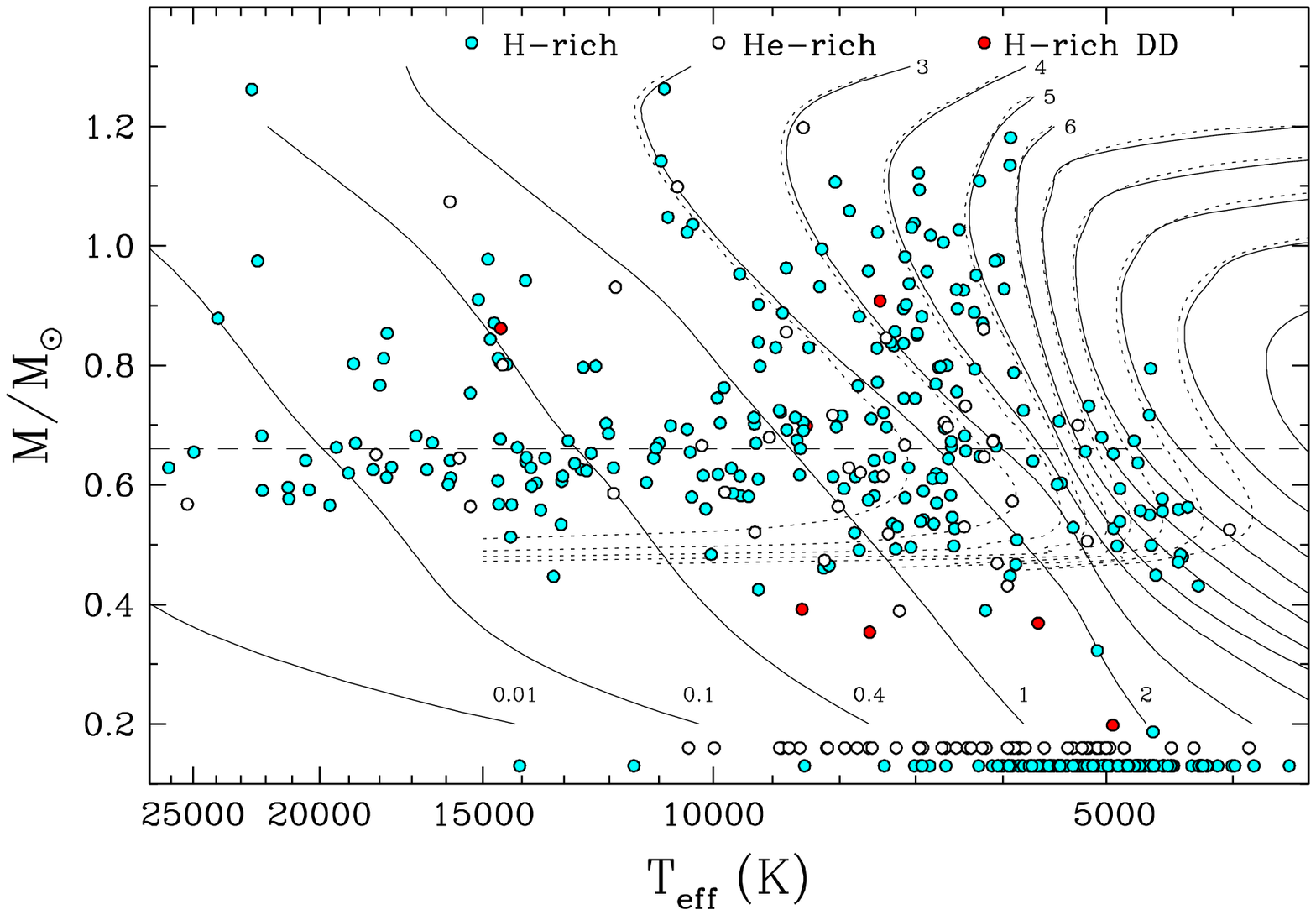}
\begin{flushright}
Figure \ref{correltm}
\end{flushright}
\end{figure}

\clearpage

\begin{figure}[p]
\plotone{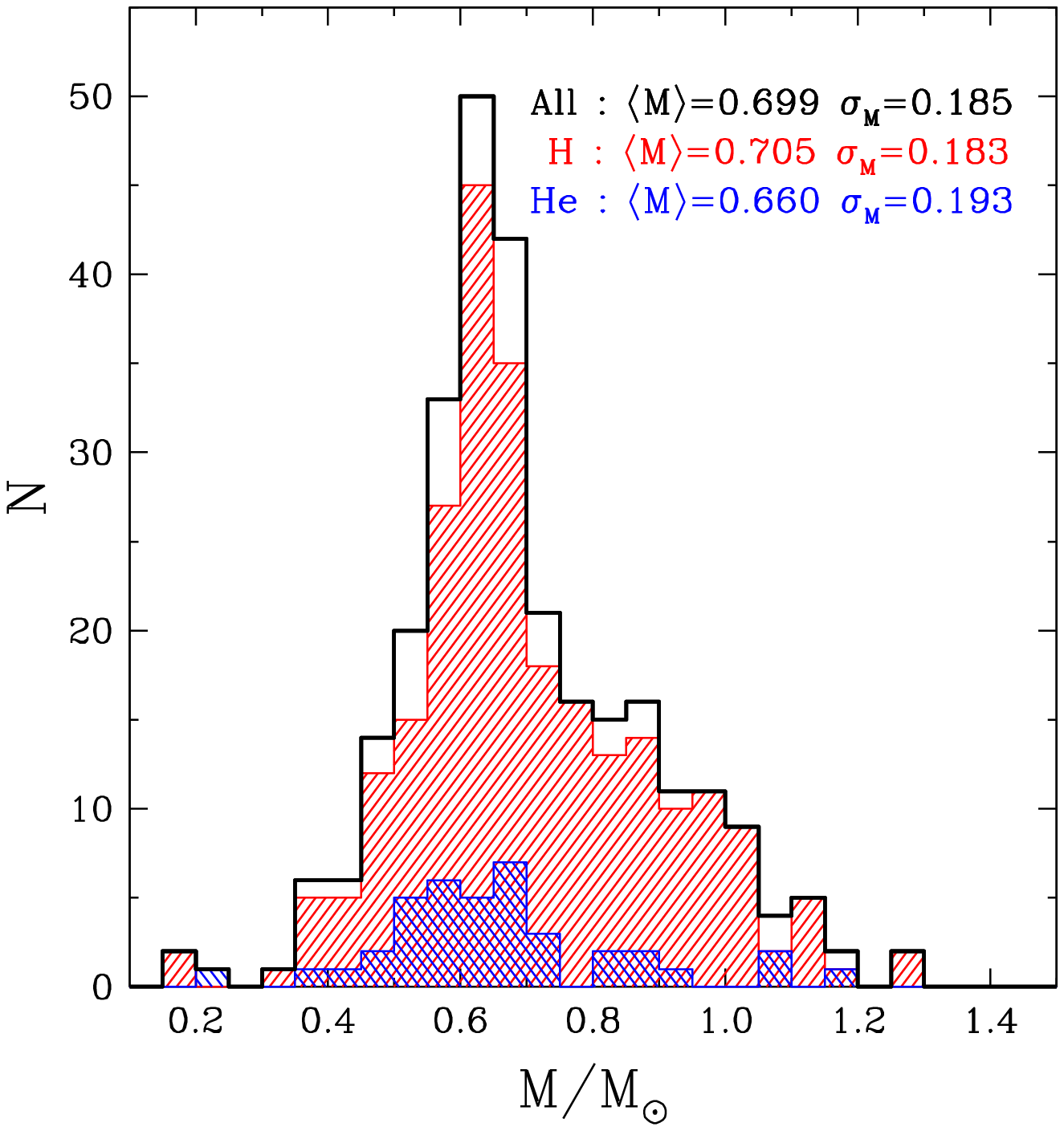}
\begin{flushright}
Figure \ref{histo_mass_spectro}
\end{flushright}
\end{figure}

\clearpage

\begin{figure}[p]
\plotone{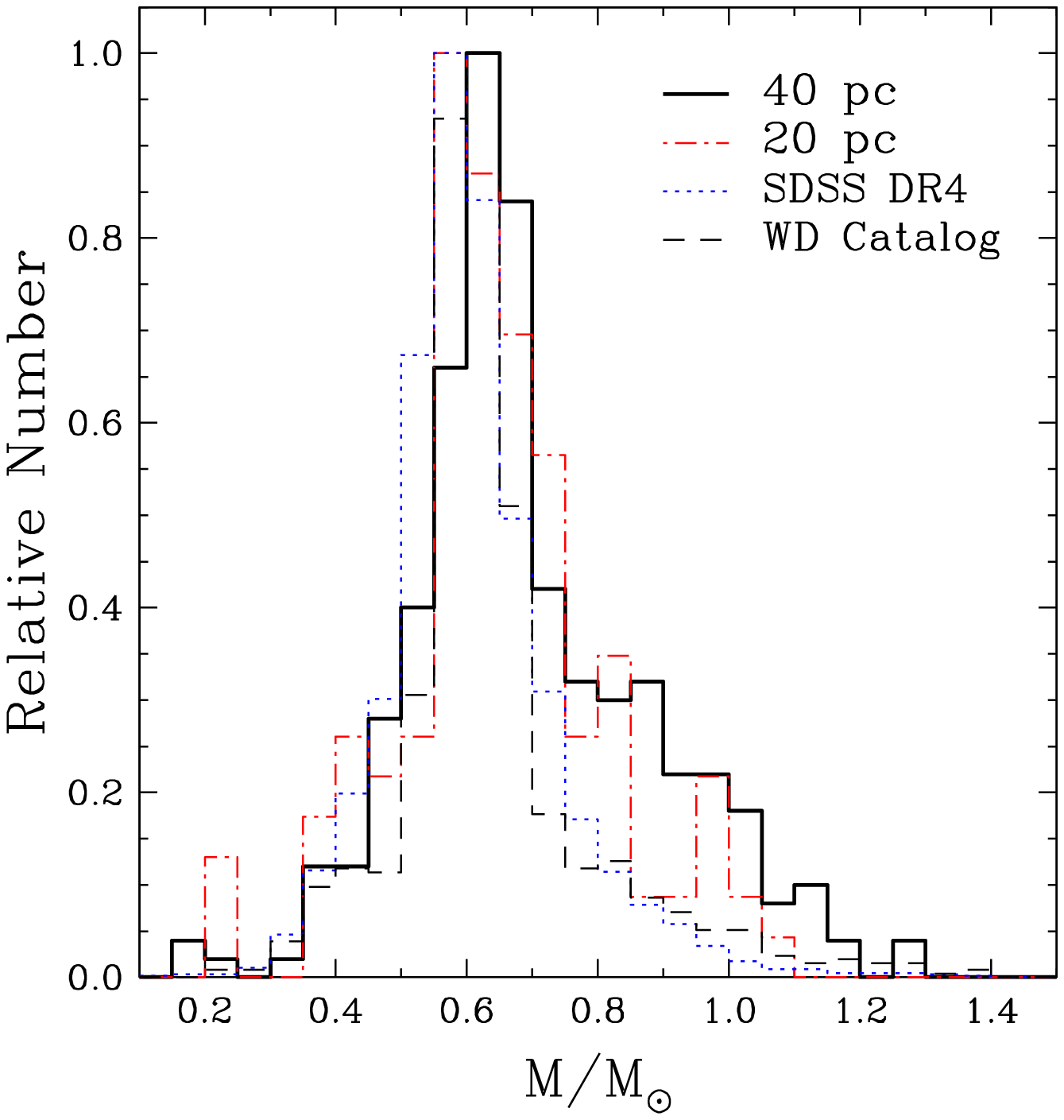}
\begin{flushright}
Figure \ref{histo_compare}
\end{flushright}
\end{figure}

\clearpage

\begin{figure}[p]
\plotone{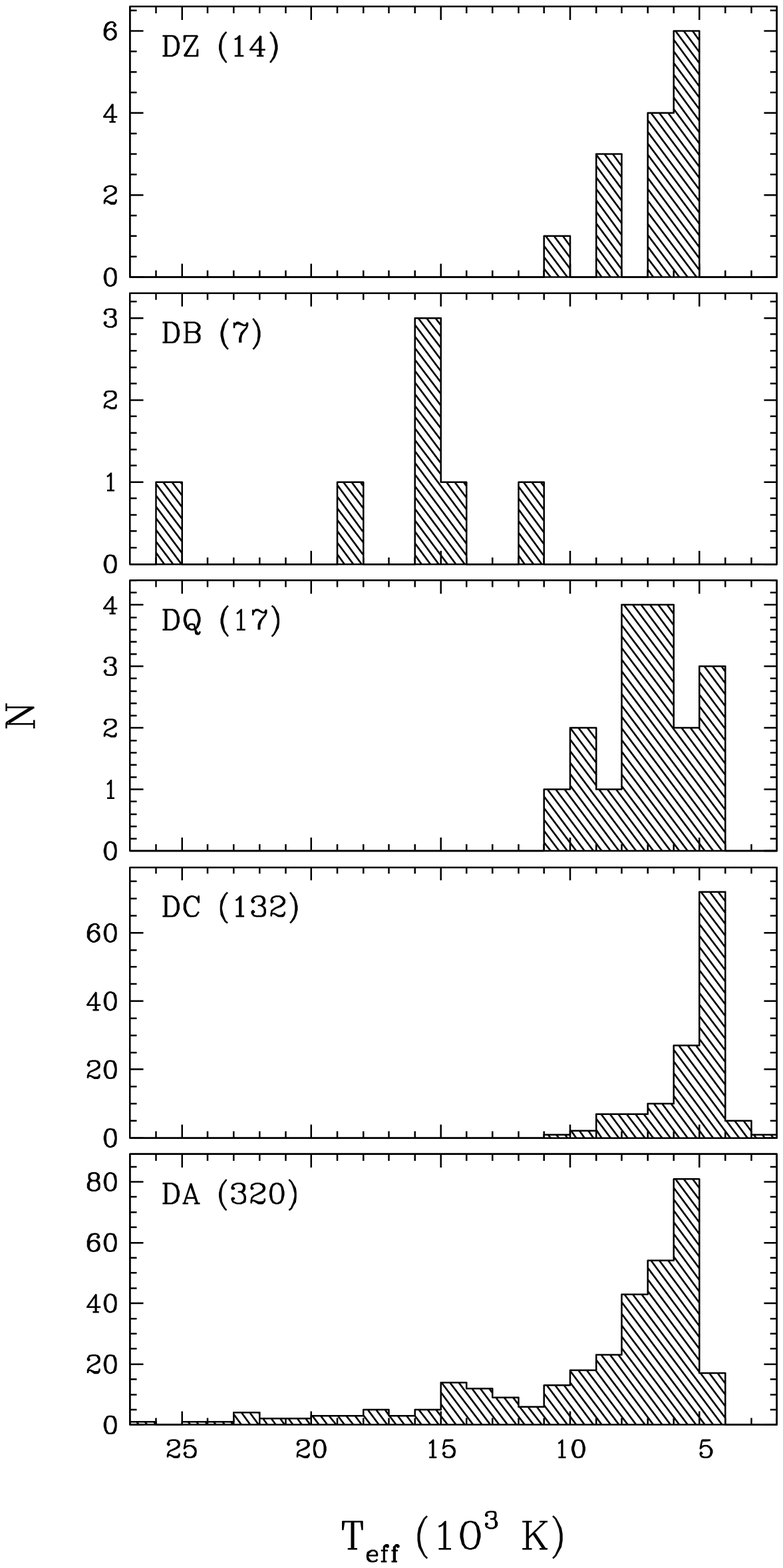}
\begin{flushright}
Figure \ref{spec_evol}
\end{flushright}
\end{figure}

\clearpage

\begin{figure}[p]
\plotone{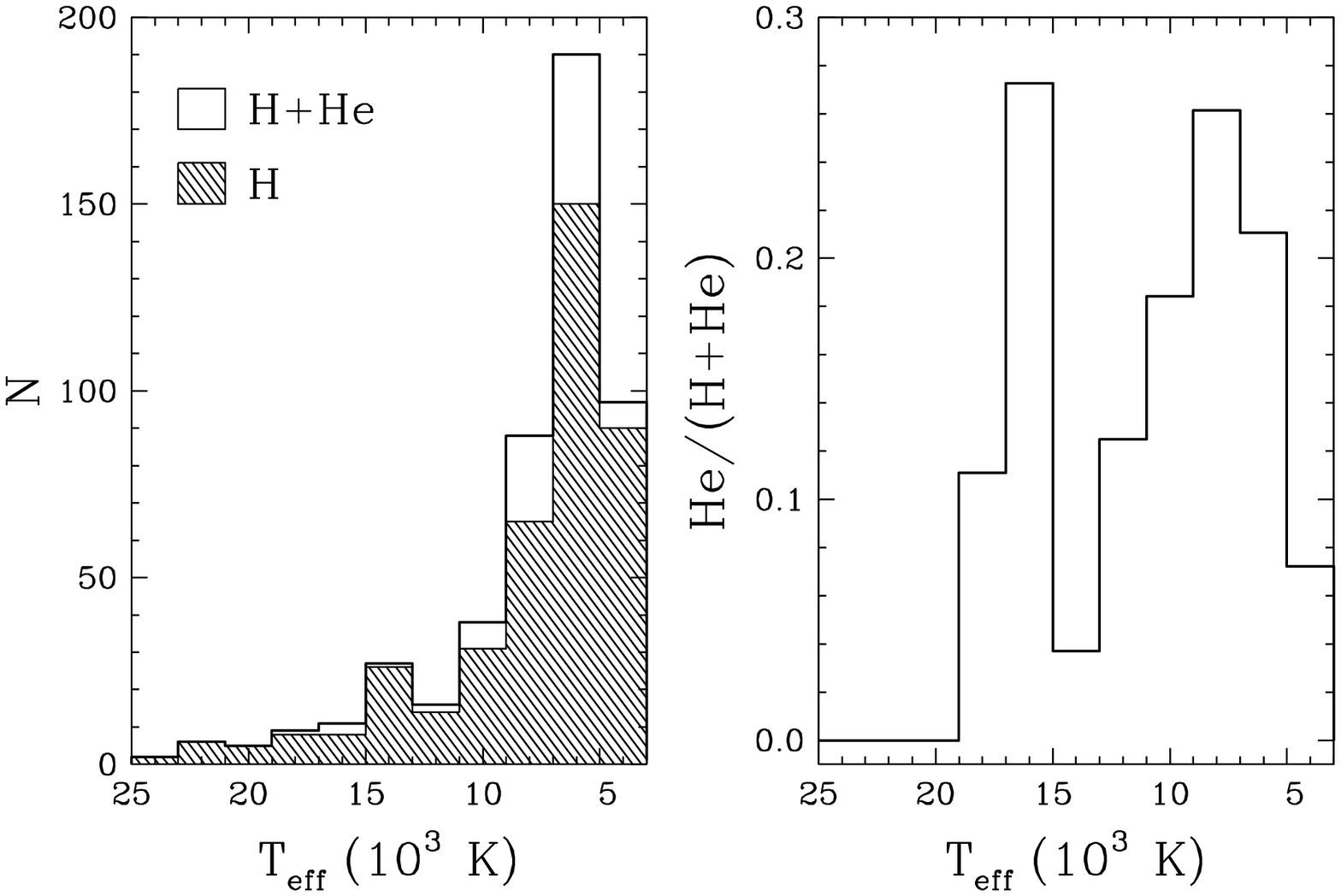}
\begin{flushright}
Figure \ref{ratio_H}
\end{flushright}
\end{figure}

\clearpage

\begin{figure}[p]
\plotone{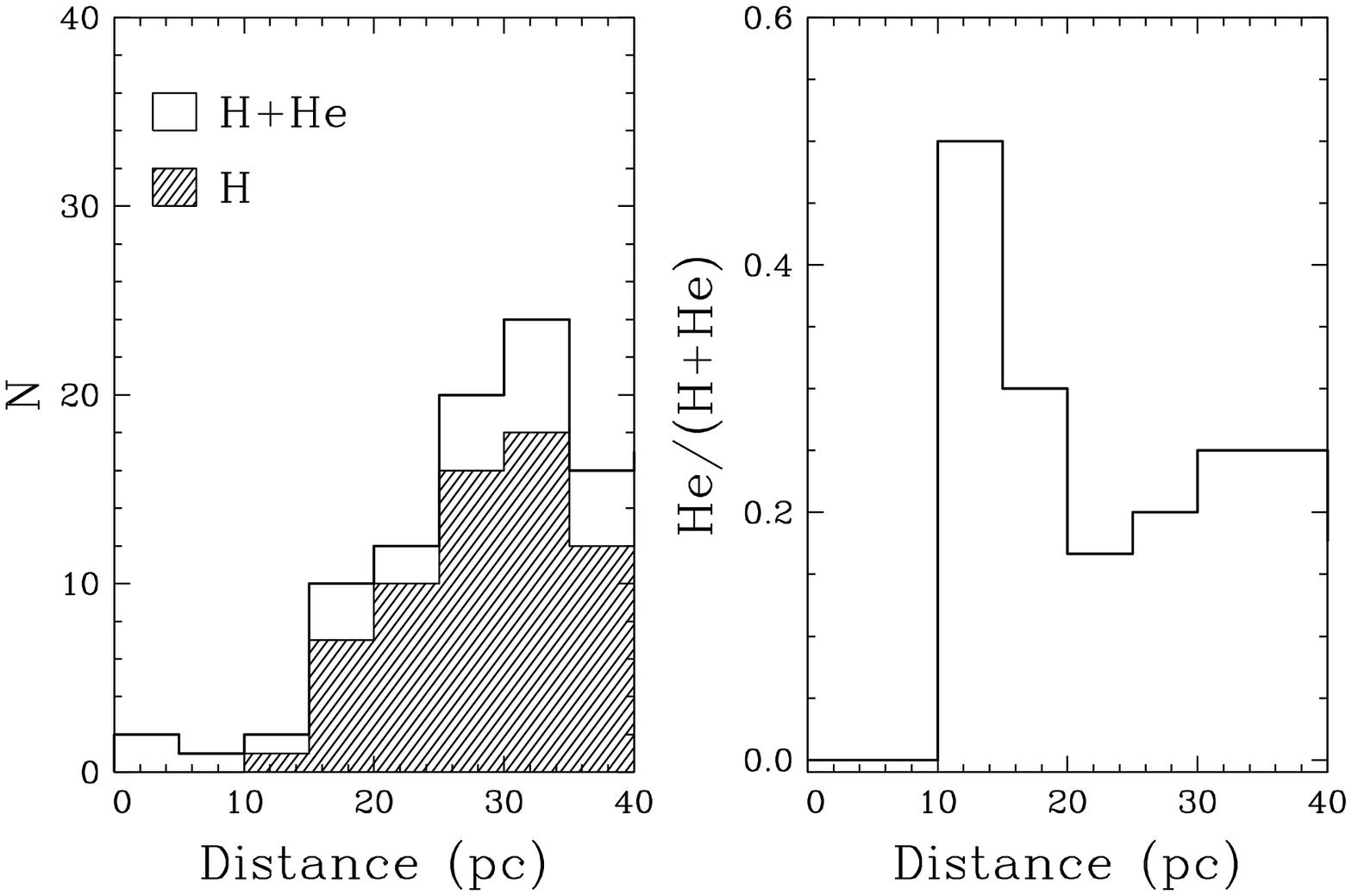}
\begin{flushright}
Figure \ref{ratio_D}
\end{flushright}
\end{figure}

\clearpage

\begin{figure}[p]
\plotone{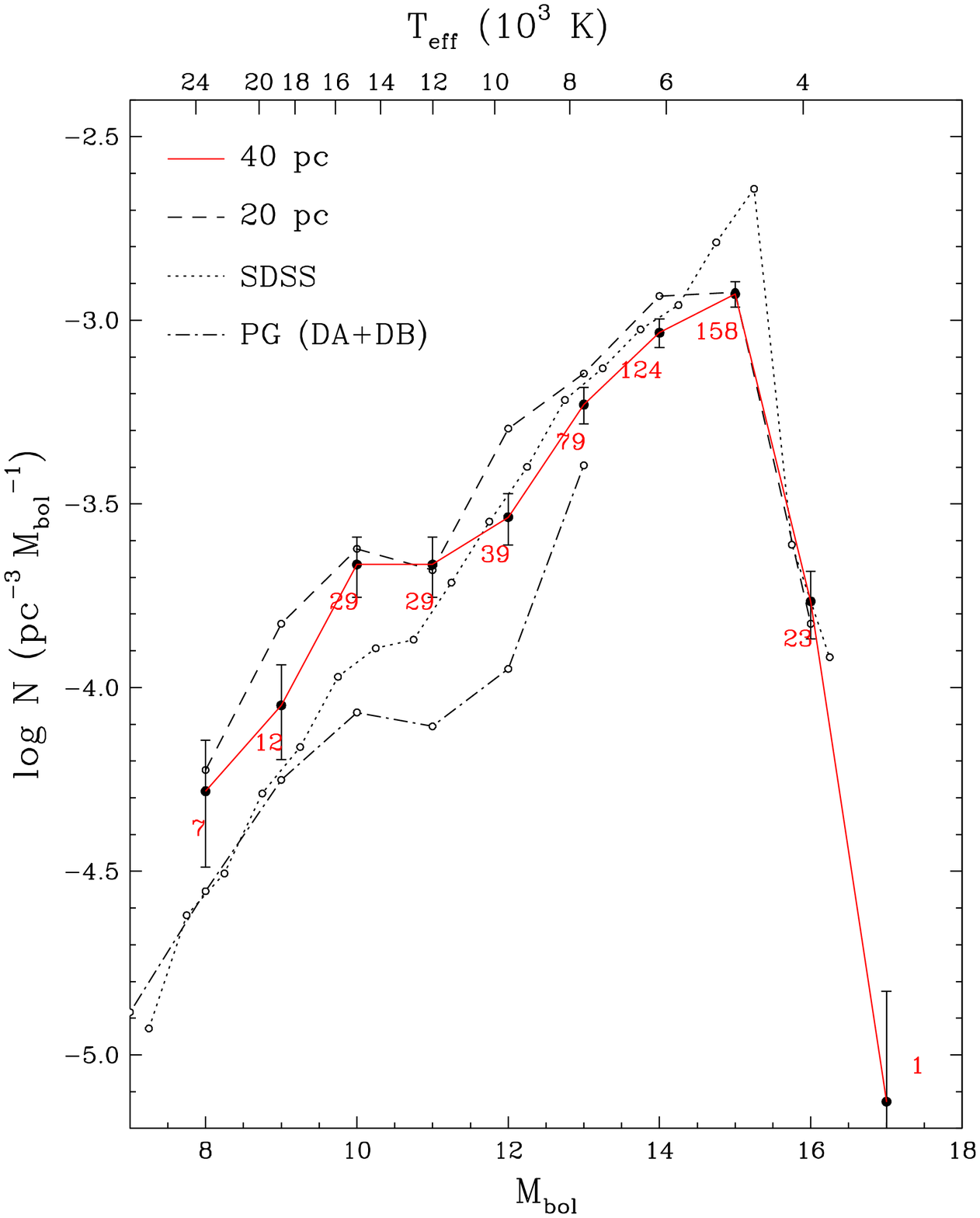}
\begin{flushright}
Figure \ref{lf}
\end{flushright}
\end{figure}

\clearpage

\begin{figure}[p]
\plotone{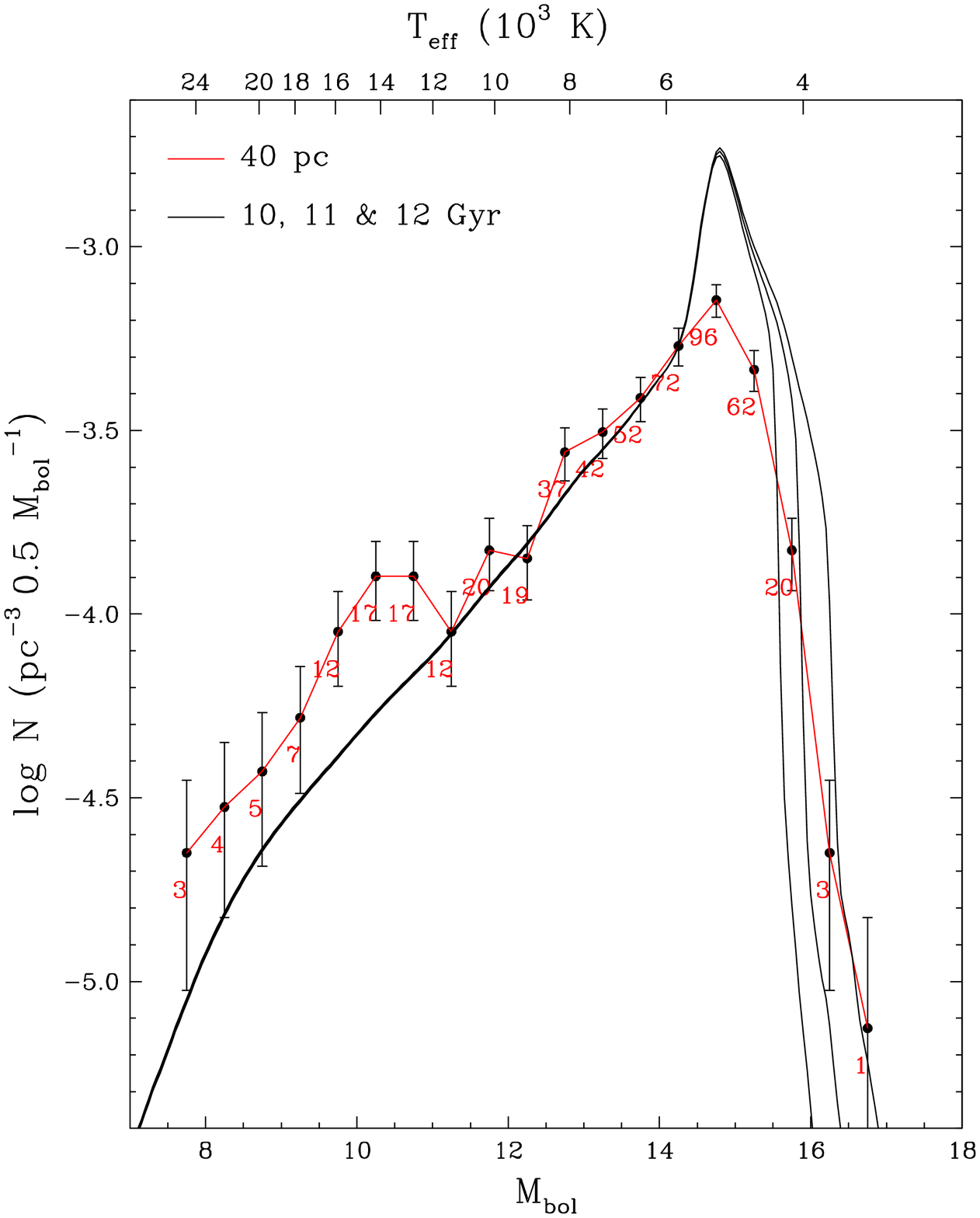}
\begin{flushright}
Figure \ref{lf2}
\end{flushright}
\end{figure}

\end{document}